\documentclass[a4paper,11pt]{article}
\pdfoutput=1 % if your are submitting a pdflatex (i.e. if you have
\usepackage{amsmath}
\usepackage{amssymb}
\usepackage{amsfonts}
\usepackage{amsthm}
\usepackage{array}
\usepackage{bm}
\usepackage{cancel}
\usepackage[usenames,dvipsnames]{color}
\usepackage{epsfig}
\usepackage{epsfig,multicol}
\usepackage{float}
\usepackage[T1]{fontenc}
\usepackage{graphicx}
\usepackage[applemac]{inputenc}
\usepackage{jheppub} 
\usepackage{latexsym}
\usepackage{mathrsfs}
\usepackage{multirow}
\usepackage{slashed}
\usepackage{subfig}
\usepackage{url}
\usepackage{wasysym}
\usepackage{xcolor}

\usepackage{hyperref}
\usepackage{xfrac}
\usepackage{tikz}
\usepackage[compat=1.1.0]{tikz-feynman}
\usepackage{color, colortbl}
\definecolor{Green}{rgb}{0,1,0}

\newcommand{\be}{\begin{equation}}
\newcommand{\ee}{\end{equation}}

\newcommand{\GeV}{\text{GeV}}
\newcommand{\TeV}{\text{TeV}}
\newcommand{\keV}{\text{keV}}

\newcolumntype{\mycol}{!{\vrule width 1pt}}
\definecolor{Grn}{rgb}{0.,0.6,0.}

\newcommand{\cmark}{$\checkmark$}
\title{\boldmath Displaced new physics at colliders and \\ the early universe before its first second}

\author[a]{Lorenzo Calibbi,}
\author[b,c]{ Francesco D'Eramo,}
\author[d,e,f]{ Sam Junius,}
\author[d,e]{\\ Laura Lopez-Honorez,}
\author[e,f]{ Alberto Mariotti}

\affiliation[a]{School of Physics, Nankai University, Tianjin 300071, China}
\affiliation[b]{Dipartimento di Fisica e Astronomia ``Galileo Galilei'', \\ Universit\`a di Padova, Via Marzolo 8, 35131 Padova, Italy}
\affiliation[c]{INFN, Sezione di Padova, Via Marzolo 8, 35131 Padova, Italy}
\affiliation[d]{Service de Physique Th\'eorique, Universit\'e Libre de Bruxelles, C.P. 225, B-1050 Brussels, Belgium}
\affiliation[e]{Theoretische Natuurkunde \& The International Solvay Institutes, \\
Vrije Universiteit Brussel, Pleinlaan 2, B-1050 Brussels, Belgium}
\affiliation[f]{Inter-University Institute for High Energies, Vrije Universiteit Brussel, \\
Pleinlaan 2, B-1050 Brussels, Belgium}

\emailAdd{calibbi@nankai.edu.cn, francesco.deramo@pd.infn.it, sam.junius@vub.be, llopezho@ulb.ac.be, alberto.mariotti@vub.be}

\abstract{Displaced vertices at colliders, arising from the production and decay of long-lived particles, probe dark matter candidates produced via freeze-in. If one assumes a standard cosmological history, these decays happen inside the detector only if the dark matter is very light because of the relic density constraint. Here, we argue how displaced events could very well point to freeze-in within a non-standard early universe history. Focusing on the cosmology of inflationary reheating, we explore the interplay between the reheating temperature and collider signatures for minimal freeze-in scenarios. Observing displaced events at the LHC would allow to set an upper bound on the reheating temperature and, in general, to gather indirect information on the early history of the universe.} 
  
\preprint{ULB-TH/21-02}

\begin{document} 
\maketitle
\flushbottom

%%%%%%%%%%%%%%%%%%%%%%%%%%%%%%%%%%%%%%%%%%%%%%%%%%%%%%%%%%%%%%%%%%
\section{Introduction}
\label{sec:intro}
%%%%%%%%%%%%%%%%%%%%%%%%%%%%%%%%%%%%%%%%%%%%%%%%%%%%%%%%%%%%%%%%%%

Detecting non-gravitational interactions of dark matter (DM) has been the target of a vast and diverse experimental effort with no conclusive evidence so far~\cite{Jungman:1995df,Bertone:2004pz,Feng:2010gw}. In this work, we exploit the interplay between two different DM probes: the measured abundance~\cite{Aghanim:2018eyx} and collider searches~\cite{Boveia:2018yeb}. In particular, we illustrate how the observation of displaced events 
at particle colliders could allow us to speculate about the early universe. 

The main characters of our study are Feebly Interacting Massive Particles (FIMPs), new stable degrees of freedom beyond the Standard Model (SM) that never achieve thermal equilibrium through the expansion history of our universe. Nevertheless, they are viable DM candidates since decays and/or scatterings of primordial bath particles can produce FIMPs that free-stream subsequently. If FIMP interactions are renormalizable then their production is most efficient at temperatures around the mass of the heaviest particle involved in the process~\cite{McDonald:2001vt,Kusenko:2006rh,Ibarra:2008kn,Hall:2009bx}. Consequently, the resulting relic density depends only on masses and couplings that we can measure in our laboratories and/or astrophysically and this ``IR-dominated'' production mechanism has been dubbed {\it freeze-in} in the literature~\cite{Hall:2009bx}. Calculations for the relic density within this framework are reliable theoretically, but they come with an important {\it caveat}: the cosmological history is assumed to be standard. Contrarily, if the interactions proceed via non-renormalizable operators then the higher the temperature the more efficient production via scattering would be: the majority of FIMPs are produced around the time of inflationary reheating~\cite{Moroi:1993mb,Rychkov:2007uq,Strumia:2010aa,Elahi:2014fsa,Eberl:2020fml}. 

FIMPs are clearly a nightmare scenario for experimental searches. All we have left today are stable particles with expected rates at conventional DM searches\,---\,barring some exceptions, see e.g.~\cite{Hambye:2018dpi}\,---\,typically too low to yield any signal.  Here, we study a peculiar experimental manifestation of several FIMP models: new heavy particles, which participate in the FIMP production in the early universe, can be pair produced at particle colliders and then decay into DM with a macroscopic decay length. The resulting signature is the one of displaced decays with missing energy in the final state~\cite{Alimena:2019zri}. 

\begin{figure}
\centering
\includegraphics[width=0.99\textwidth]{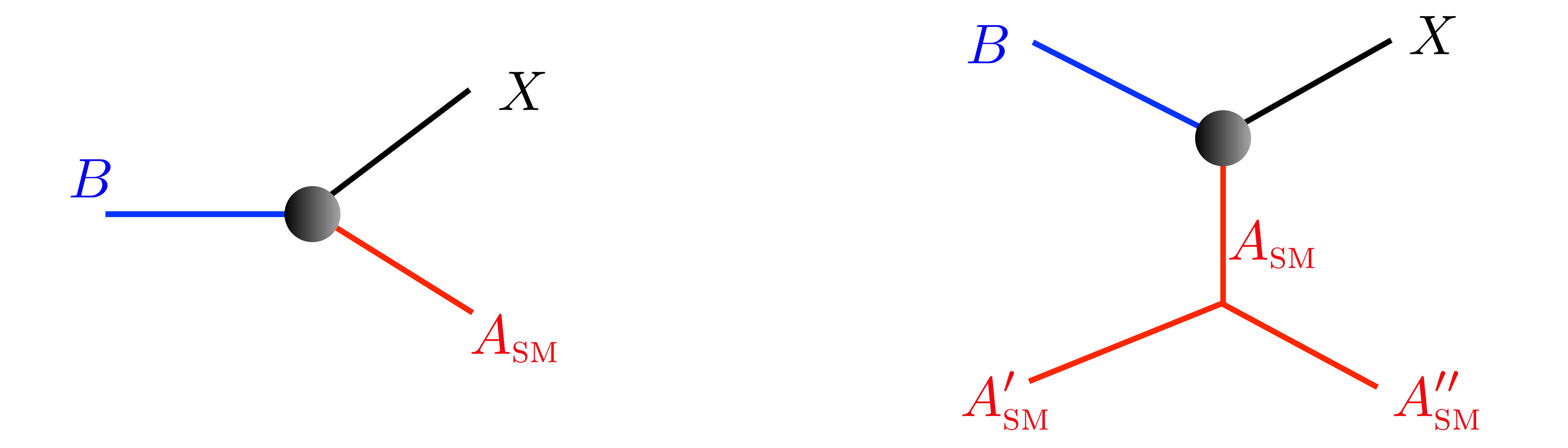} 
\caption{Basic setup of this work: a FIMP $X$ couples to the thermal bath via a cubic interaction with a SM particle $A_\textsc{sm}$ and a new BSM particle $B$. This sources $X$ production via $B$ decays (left panel) and binary collisions (one example in the right panel).}
\label{fig:FI}
\end{figure}

We investigate the scenario sketched in Figure~\ref{fig:FI}: a FIMP $X$ couples to a SM field $A_\textsc{sm}$ and a new beyond the SM (BSM) bath particle $B$ via a cubic interaction. This will be the only DM coupling relevant phenomenologically. The FIMP is a SM gauge singlet and we take the cubic interaction to be very suppressed in order to prevent thermalization. In contrast, the bath particle $B$ carries SM gauge quantum numbers and it is in thermal equilibrium with the primordial plasma at early times. Known examples for $B$ can be supersymmetric partners (gluino, wino, squarks, etc.), but we take a bottom-up approach and we do not commit to any specific realization.  Although we focus on a 3-body interaction, the generalization of our analysis to interactions involving more particles is straightforward.

Bath particles populate the early universe with $X$'s via two different classes of processes. The cubic interaction itself is responsible for $B$ decays, and it also mediates binary collisions once we account for interactions among primordial bath constituents. One example of such a collision is sketched in the right panel of Figure~\ref{fig:FI} where $A_\textsc{sm}'$ and $A_\textsc{sm}''$ are two SM particles that can differ from $A_\textsc{sm}$. Production through decay is always IR dominated whereas the latter process can be UV dominated if the FIMP interaction is non-renormalizable. 

This setup opens up a promising way to probe FIMPs: $B$ particles can be pair-produced at colliders via SM gauge interactions, and they can subsequently decay to FIMPs~\cite{Co:2015pka,Evans:2016zau,DEramo:2017ecx,Heisig:2018kfq,Calibbi:2018fqf,No:2019gvl,Belanger:2018mqt,Belanger:2018sti}. If the $B$ decay length is too short, less than approximately $100 \, \mu{\rm m}$, the signal would be indistinguishable from conventional DM searches targeting missing energy. On the contrary, if it is larger than the detector size ($\simeq 10 \, {\rm m}$) we would observe a charged track associated to the $B$ particle crossing the full detector; although this would be a spectacular manifestation of BSM physics, it has no manifest connection with the invisible universe. 

The main focus of our study are signatures that we get when the decay length lies in between these two extremes. In such a range, we can observe charged particles decaying to SM and missing energy displaced from the primary interaction vertex. 

What typical $B$ decay lengths should we expect? If we neglect the mass of $A_\textsc{sm}$, which is irrelevant to the phenomenology we study, the system in Figure~\ref{fig:FI} is described by three independent parameters: the masses of $B$ and $X$ and the decay length (or equivalently the lifetime) of $B$. For a given mass spectrum, the relic density constraint sets the $B$ decay length. Unfortunately, the typical decay length is way too large to give rise to displaced events within the detector for $B$ masses such that we can produce them at colliders~\cite{Hall:2009bx}. 

This well known observation motivates a consideration. Let us assume that we observe displaced events and that we measure the $B$ mass and decay length. In a quite optimistic scenario, let us even suppose that the decay happens via a 3-body interaction such as in Figure~\ref{fig:FI} and that we reconstruct the DM mass as well~\cite{Bae:2020dwf}. As we show in Section~\ref{sec:bound}, this would point to a cosmological disaster: the relic density of $X$ should be much larger than the observed DM abundance. Is there any way out? The answer is yes, and the reason is the {\it caveat} mentioned above: the relic density calculation assumes a standard cosmological history with freeze-in production during a radiation dominated (RD) era. However, the earliest time we can probe indirectly is the one of Big Bang Nucleosynthesis (BBN) when the temperature was $T_{\rm BBN} \simeq {\rm MeV}$ and the age of the universe was approximately one second. While it is reasonable to extrapolate the standard expansion history to earlier times and higher temperatures, this is an assumption not supported by any observation: we do not know what our universe looked like when it was younger than one second, see e.g. Ref.~\cite{Allahverdi:2020bys}. Thus displaced missing energy at colliders could indirectly provide us with some hints about the early universe: DM production could have happened during a non-standard cosmology, see for instance~Refs.~\cite{Choi:2007rh,Co:2015pka,Evans:2016zau,DEramo:2017ecx,Belanger:2018sti,Frere:2006hp,Bernal:2020yqg,Bernal:2020bjf} within the context of freeze-in. 

The standard history cannot be extrapolated back in time indefinitely. At early times, our universe undergoes through a phase of very fast (exponential) expansion with energy density dominated by one (or more) scalar field known as the inflaton. Later on, the inflaton oscillates around the minimum of its potential and our universe experiences a phase of early matter domination (MD) during which inflaton decays to SM and BSM particles populate the primordial thermal bath. Eventually, when the age of the universe is around the inflaton lifetime, the energy budget is overtaken by radiation and we enter the RD epoch that lasts until the time of the (late) matter-radiation equality ($T_{\rm eq} \simeq 1 \, {\rm eV}$). The highest temperature of such a RD epoch is conventionally called the reheating temperature, and we denote this quantity important for our study with the symbol $T_{R}$. We investigate how such an early MD era, with reheating temperature $T_R$ below the $B$ mass, alters the prediction for the $B$ decay length at colliders. While we focus for concreteness on inflationary reheating, our results hold also for the last stages of a generic early MD epoch such as the one due to supersymmetric moduli~\cite{Acharya:2009zt,Easther:2013nga,Kane:2015jia,Co:2016xti,Aparicio:2016qqb,Co:2016fln,Co:2017orl} or evaporating primordial black holes~\cite{Bernal:2020bjf}.

We revisit the prediction for the $B$ decay length during a RD epoch in Section~\ref{sec:bound}, and we discuss how it changes for an early MD. For values of the $B$ mass and decay length relevant to collider physics, and once we set the DM mass to the lowest value allowed by small scale structures constraints, we get an {\it upper bound} on the reheating temperature $T_{R}$. In Section~\ref{sec:models}, we  classify  the simplest microscopic realizations. Finally, in Section~\ref{sec:coll-results}, we first present  the searches performed at the LHC experiments for long-lived particles in Section~\ref{sec:searches} and then illustrate  in Section~\ref{sec:results} how collider searches can bound the viable parameter space for these DM scenarios. Our analysis provides a map between unconventional DM collider signatures and properties of the early universe. Within our hypothesis of modified cosmological history due to late time reheating after inflation, this feature would be the reheating temperature $T_{R}$. We defer technical details to appendices, and we conclude in Section~\ref{sec:conclusion}.

%%%%%%%%%%%%%%%%%%%%%%%%%%%%%%%%%%%%%%%%%%%%%%%%%%%%%%%%%%%%%%%%%%
\section{Cosmological implications of the $B$ decay length}
\label{sec:bound}
%%%%%%%%%%%%%%%%%%%%%%%%%%%%%%%%%%%%%%%%%%%%%%%%%%%%%%%%%%%%%%%%%%

In this section, we present numerical results for the FIMP relic density and we discuss how to understand them with the help of approximate semi-analytical solutions that capture the relevant physics. Technical details about the cosmological histories considered in our work and freeze-in production can be found in the Appendices~\ref{app:cosmology} and~\ref{app:FI}, respectively. 

\subsection{Freeze-in production via $B$ decays in a RD universe}
\label{sec:freeze-decay-standard}

Every freeze-in calculation in a RD universe relies upon two assumptions. First, the reheating temperature is set to be much higher than the mass of any of the particles participating in the DM production processes. Second, DM production through inflaton decays or initiated purely gravitationally during inflation~\cite{PhysRevD.35.2955,PhysRevD.37.3428,Fairbairn:2018bsw,Tenkanen:2019aij,Tenkanen:2019wsd} is subdominant. We begin our analysis with the investigation of $X$ freeze-in production via B decays in a RD universe and we also make these conventional assumptions.

\begin{figure}
\centering
\includegraphics[width=0.55\textwidth]{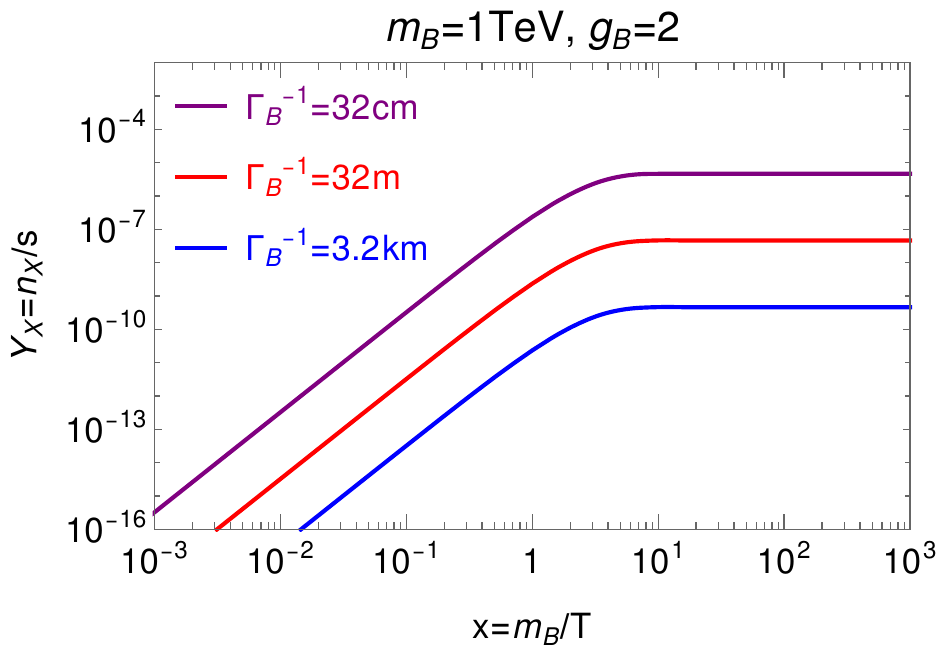} 
\caption{Freeze-in comoving number density from $B$ decays for a Weyl fermion ($g_B=2$). We fix $m_B = 1 \, {\rm TeV}$ and we consider three different decay lengths $\Gamma_B^{-1}$. Relic density requires: $m_X = 100 \, {\rm keV}$ (purple line), $m_X = 10 \, {\rm MeV}$ (red line), $m_X = 1 \, {\rm GeV}$ (blue line).}
\label{fig:StandardFI}
\end{figure}

The rigorous way to track the $X$ number density $n_X$ is by solving a Boltzmann equation, as we outline in Appendix~\ref{app:FI}, and the output of such a procedure is illustrated in Figure~\ref{fig:StandardFI}. We show numerical solutions for a fixed $B$ mass and three different values of its decay width $\Gamma_B$. The variable on the vertical axis is the so called comoving number density, and it is defined as $Y_X \equiv n_X / s$ where $s$ is the entropy density of the radiation bath. The evolution ``time variable'' we employ is the dimensionless combination $x \equiv m_B / T$ where $T$ is the temperature of the thermal bath. The comoving number density changes in a RD universe only if there are number changing processes. In other words, $Y_X$ is a convenient variable because it scales out the effect of the Hubble expansion. 

Initially, the $X$ comoving density grows with time and the slope of such a growth can be understood through a simple analytical estimate. The amount of $X$ particles produced when the thermal bath had a temperature $T$ during a Hubble doubling time, namely the time it takes for the universe to double its size, results in
\begin{equation}
Y_X(T) \simeq \mathcal{R}_{B \rightarrow A_\textsc{sm} X}(T) \, t(T) \ .
\label{eq:Yapprox}
\end{equation}
Here, $\mathcal{R}_{B \rightarrow A_\textsc{sm} X}(T)$ is the (temperature dependent) rate of particle production and $t(T)$ is the age of the universe when the temperature is $T$. We estimate the former by multiplying the $B$ rest frame decay width by an additional factor of $m_B / T$ to account for time dilatation due to the kinetic energy of the thermal bath. The latter is approximately the inverse Hubble parameter, $t(T) \simeq 1 / H(T)$, and for a RD universe we have $H(T) \sim T^2 / M_{\rm Pl}$. Thus the comoving number density scales with $x$ (i.e., with the temperature) as follows 
\begin{equation}
Y_X(x) \simeq \frac{\Gamma_{B \rightarrow A_\textsc{sm} X} \; M_{\text{Pl}}}{m_B^2} \, x^{3} \ .
\end{equation}

Figure~\ref{fig:StandardFI} makes the IR domination manifest: the lower the temperature, the more efficient DM production becomes. This is only valid as long as $B$ is present significantly in the bath since we hit the Maxwell-Boltzmann suppression once we reach temperatures around the $B$ mass. Consistently with this physical argument, we observe how the three lines in Figure~\ref{fig:StandardFI} reach a constant values for $x$ larger than a few. In what follows, we refer to the temperature at which the FIMP production peaks around the $B$ mass as $T_{FI}$, or equivalently $x_{FI}\sim{\cal O} (1)$. The $X$ comoving density freezes to the constant value
\begin{equation}
Y_X^{\infty} \simeq \frac{\Gamma_{B \rightarrow A_\textsc{sm} X} \; M_{\text{Pl}}}{m_B^2} \ .
\label{eq:YXR}
\end{equation}

A more rigorous calculation gives the expression in Eq.~\eqref{eq:YFIasymptotic} of Appendix~\ref{app:FI}. If we impose the further  constraint that the FIMP $X$ makes all the DM we get 
\begin{equation}
\tau_B \equiv \frac{1}{\Gamma_{B \rightarrow A_\textsc{sm} X}} \simeq 3.3 \times 10^3 \, {\rm m} \, \left( \frac{g_B}{2} \right) \, \left(\frac{m_X}{1\, \GeV}\right) \,  \left(\frac{1 \TeV}{m_B}\right)^{2} \ .
\label{eq:ctauRD}
\end{equation}
This relation connects the DM mass, the mass of the mother particle and its decay width. Keeping in mind that we work within the natural unit system ($\hbar = c = 1$), this is the typical decay length for $B$ decays at colliders after they have been pair-produced. Unless we consider a very light FIMP, the decay length is beyond the detector size ($\sim 10$~m) for $B$ mass values relevant to collider physics. Displaced events seem somewhat unlikely within any FIMP framework. In fact, we will argue now how their observation at colliders (i.e., $B$ decays inside the detector) could instead point to freeze-in during an early MD era.

\subsection{A motivated alternative: early MD era during inflationary reheating}

\begin{figure}
\centering
\includegraphics[width=0.55\textwidth]{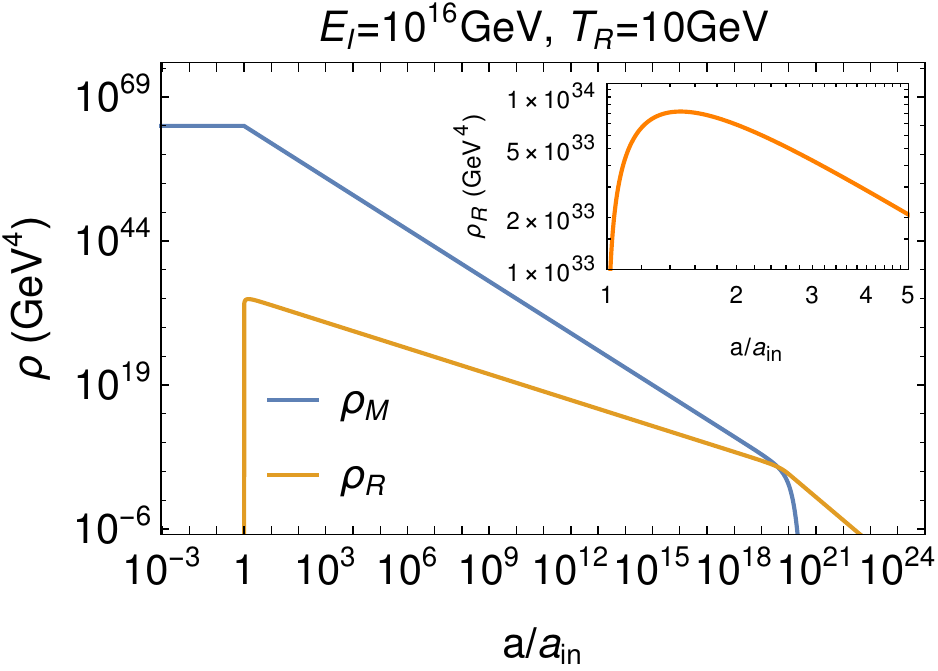} 
\caption{Inflaton (blue line) and radiation bath (orange line) energy densities as a function of the scale factor for inflationary reheating. In the top-right part of the plot we zoom in at early times and we show how the radiation bath energy density arises from inflaton decays.}
\label{fig:EarlyMD1}
\end{figure}

We provide in Appendix~\ref{app:cosmology} the Boltzmann equations describing inflationary reheating. For a specific choice of inflationary parameters, Figure~\ref{fig:EarlyMD1} shows how the inflaton and radiation energy densities depend on the scale factor $a$ describing the expanding universe. Here, we use the semi-analytical solution given in Appendix~\ref{app:cosmology} to describe the underlying physics. 

Initially, there is no radiation bath and the inflaton dominates the energy budget with a costant density $ E_I^4$. At some point, which we take corresponding to a value of the scale factor $a_{\rm in}$, inflation ends and the inflaton begins damped oscillations around its minimum. The energy density stored in these oscillations red-shifts as non-relativistic matter, $\rho_M \propto a^{-3}$, and we enter an early MD epoch. Meanwhile, inflaton decays populate a newly formed radiation bath. The inflaton decays completely when the age of the universe is approximately its lifetime, and we recover the standard RD universe. The reheating temperature $T_R$ is the highest temperature ever achieved by the radiation bath during this RD epoch.

\begin{figure}
\centering
\includegraphics[width=0.45\textwidth]{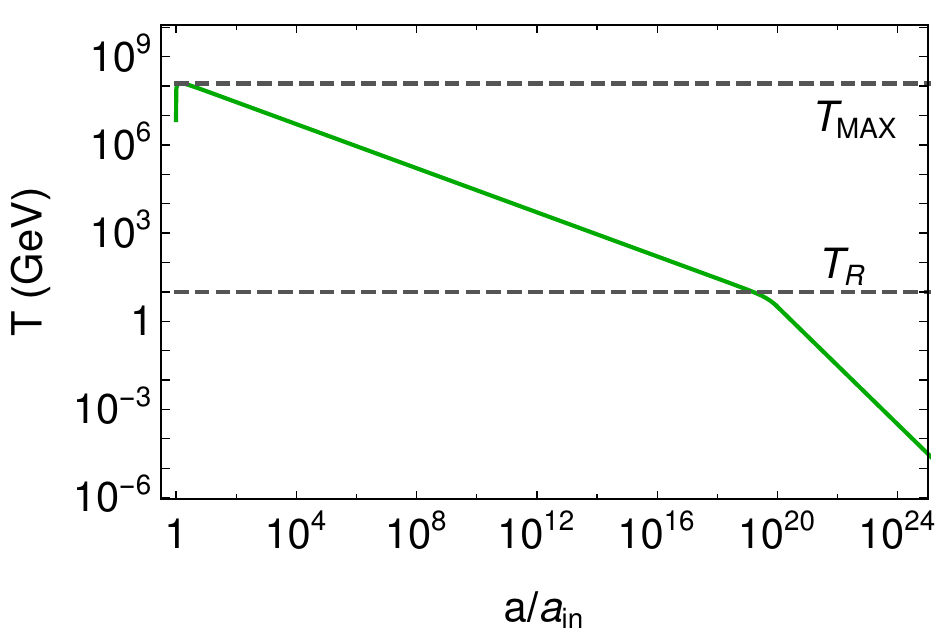} \hfill
\includegraphics[width=0.45\textwidth]{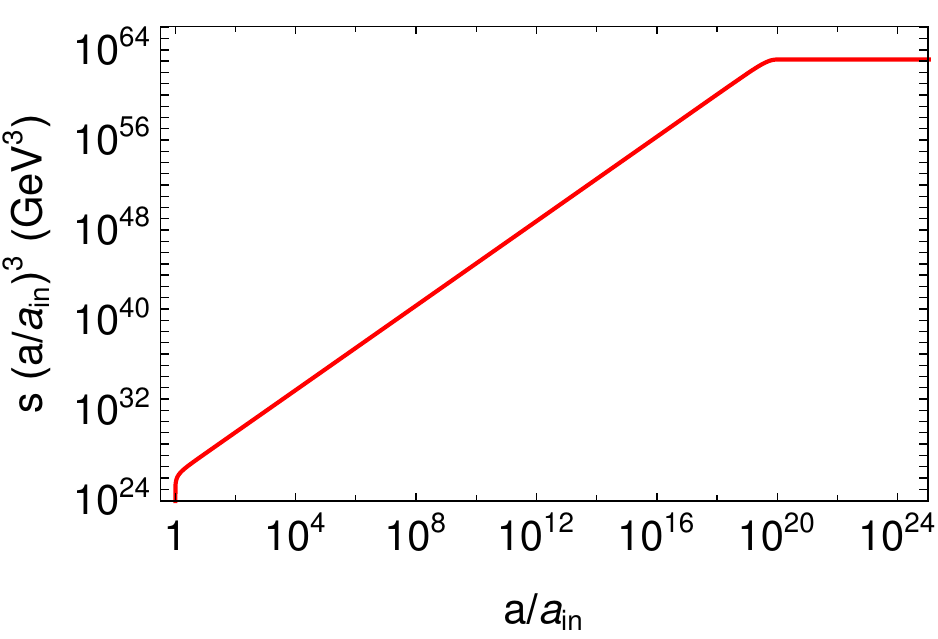} 
\caption{Radiation bath temperature (left) and the entropy in a comoving volume (right) as a function of the scale factor. We set $E_I$ and $T_R$ to the same values chosen in Figure~\ref{fig:EarlyMD1}.}
\label{fig:EarlyMD2}
\end{figure}

It is instructive to investigate bath properties during reheating. We show in Figure~\ref{fig:EarlyMD2} the evolution of two important quantities: the temperature (left panel) and the entropy in a comoving volume $S(T)= s(T) (a(T)/a_{\rm in})^3$ (right panel). For values of the scale factor $a < a_{\rm in}$, the temperature and the entropy were both zero. As soon as inflaton  oscillations began, the bath temperature had a quick growth reaching its maximum value $T_\textsc{max}$ when the scale factor did not even get twice as big, $a_\textsc{max} \simeq 1.5 \, a_{\rm in}$. Up to order one factors, the maximum temperature as a function of the inflationary parameters reads $T_\textsc{max} \simeq \left(M_{\rm Pl} \Gamma_M E_I^2 \right)^{1/4}$ with $\Gamma_M$ the inflaton decay width. Afterwards, the bath temperature decreases as follows
\begin{equation}
T(a) = T_\textsc{max} \left(\frac{a_\textsc{max}}{a}\right)^{3/8} \ ,\qquad \qquad \qquad a_\textsc{max} > a > a_R \,,
\label{eq:TaMD}
\end{equation}
where $a_R$ is the value of the scale factor when the bath temperature is $T_R \simeq \left(M_{\rm Pl} \Gamma_M\right)^{1/2}$. The decrease in Eq.~(\ref{eq:TaMD}) is slower with respect to the $T \propto a^{-1}$ of a RD universe, and this behavior persists as long as the inflaton field is around dominating the universe. The inflaton decays eventually, and we enter the RD phase for values of the scale factor $a > a_R$.

The entropy in a comoving volume grows during the reheating phase. We compute this quantity with respect to its value at the maximum temperature
\begin{equation}
\frac{S(T)}{S(T_\textsc{max})} = \frac{s(T) a^3}{s(T_\textsc{max}) a_\textsc{max}^3} =  \left(\frac{T \, a}{T_\textsc{max} \, a_\textsc{max}} \right)^3 = \left(\frac{T_\textsc{max}}{T} \right)^5 \ .
\label{eq:SvsT}
\end{equation} 
This entropy dump provides a dilution to the FIMP produced via freeze-in. Finally, it is useful to find an approximate expression for the Hubble parameter during reheating
\begin{equation}
H(T) = \frac{E_I^2}{\sqrt{3} M_{\rm Pl}} \left( \frac{a_{\rm in}}{a}\right)^{3/2} \simeq \frac{E_I^2}{M_{\rm Pl}} \left( \frac{T}{T_\textsc{max}}\right)^{4} \simeq 
\frac{1}{M_{\rm Pl}} \frac{T^4}{T_R^2} \ .
\label{eq:HearlyMD}
\end{equation}

\subsection{FIMP production  during an early MD era}
\label{sec:fi-dec-MD}

We revisit freeze-in via $B$ decays when FIMP production peaks during an early MD era, i.e.~when $m_B >T_R$. Our working assumption throughout this analysis is that the maximum temperature of the radiation bath attained during reheating, $T_\textsc{max}$, is larger than $m_B$. Thus the thermal bath generated during reheating begin its existence with a full relativistic abundance of $B$ particles. Once we ensure that $T_\textsc{max} > m_B$ the resulting FIMP relic density depends only on $T_R$. The methodology to solve numerically the Boltzmann equation for freeze-in during an early MD era can be found in Appendix~\ref{app:FI}. Here, we present numerical solutions and we understand the underlying physics thanks to analytical estimates.

\begin{figure}
\centering
\includegraphics[width=0.55\textwidth]{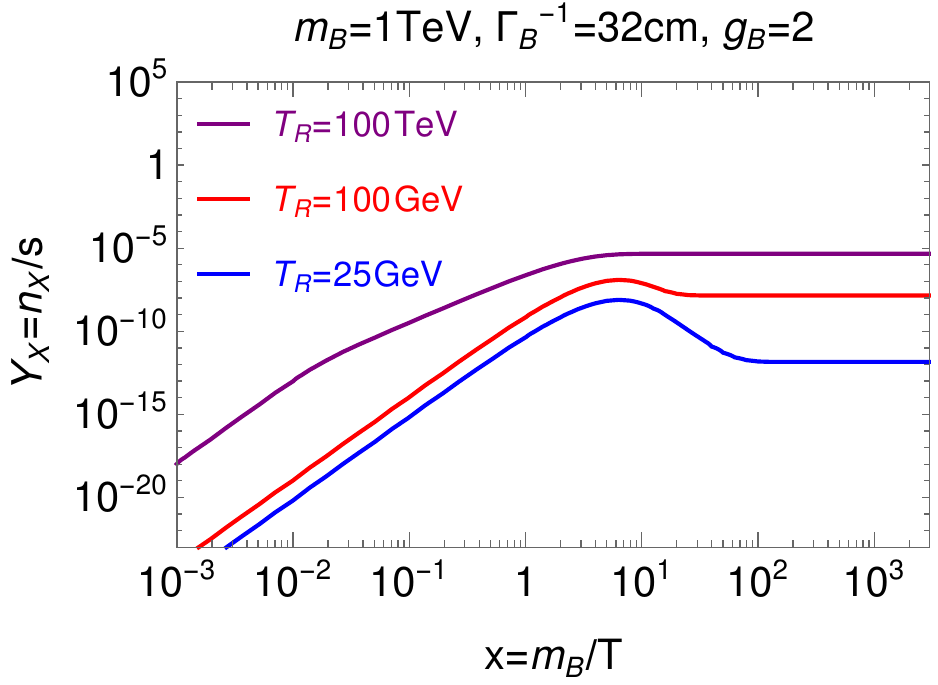}
\caption{Freeze-in comoving number density from $B$ decays for a Weyl fermion ($g_B=2$) during an early MD era. We fix $m_B = 1 \, {\rm TeV}$ and $\Gamma_{B\to A_\textsc{sm}X}^{-1}=32$~cm, and we consider three different values for the reheating temperature.  Relic density requires: $m_X = 100 \, {\rm keV}$ (purple line), $m_X = 30 \, {\rm MeV}$ (red line), $m_X = 300 \, {\rm GeV}$ (blue line).}
\label{fig:FIearlyMD}
\end{figure}

Figure~\ref{fig:FIearlyMD} shows the evolution of the DM comoving number density for fixed values of the $B$ mass and decay length, but different values of the reheating temperature. We can potentially reproduce the DM relic density upon choosing an appropriate value for $m_X$. The purple line does not present any substantial difference with respect
to the solutions in Figure~\ref{fig:StandardFI}. Indeed, the reheating temperature in this case is much larger than the $B$ mass so freeze-in happens long after inflationary reheating is over and the cosmological background is a RD universe. An estimate for the relic $Y_X^{\infty}$ is still provided by Eq.~(\ref{eq:YXR}). 

The red and the blue lines feature a quite different behavior. The DM comoving number density reaches a peak and decreases afterwards before settling down to its asymptotic value. Freeze-in production happens during an early MD epoch for both cases since $T_R < m_B$. We use the expression in Eq.~(\ref{eq:Yapprox}) to estimate the amount of FIMPs produced during a Hubble doubling time when the universe had a temperature $T$ 
\begin{equation} 
Y^{\rm prod}_X(T)  \simeq \frac{\Gamma_{B \rightarrow A_\textsc{sm} X} \, M_{\rm Pl} m_B T_R^2}{T^5} \qquad \qquad [T_\textsc{max}> T\gtrsim m_B]\,,
\label{eq:YprodMD}
\end{equation}
making use of the Hubble rate given in Eq.~\eqref{eq:HearlyMD}. The above expression approximates well the blue and red curves of Figure~\ref{fig:FIearlyMD} for temperatures larger than the $B$ mass ($x \lesssim 1$). The initial growth scales as $x^5$ instead of the $x^3$ found for a standard cosmological history. The $x^5$ and $x^3$ dependencies are also well visible in the purple curve slope when going from FIMP production in an early MD to production in a standard RD era, namely in the transition between $x < 10^{-2}$ and $10^{-2}< x <1$. Besides the different power-law behavior, the message about IR domination is confirmed and actually reinforced. 

The expression in Eq.~\eqref{eq:YprodMD} can only account for $Y_X(T)$ at $x\lesssim 1$ because $B$ hits the Maxwell-Boltzmann suppression. Unlike the case of freeze-in during RD, we cannot estimate $Y_X^\infty$ by evaluating $Y_X^{\rm prod}(T)$ at a temperature $T_{FI}$ of the order of the $B$ mass because we still have to account for the subsequent dilution of $Y_X(T)$ until the reheating time. The dilution, due to entropy released by inflaton decays, is well visible in the red and blue curves of Figure~\ref{fig:FIearlyMD} for $x> 1$ up to $x\sim m_B/T_{R}$. We quantify the dilution factor by the ratio of the entropy in a comoving volume between  a production time at a given temperature $T$ and the time when we can trust entropy conservation again (namely temperature below $T_R$)
\begin{equation}
  D(T) = \frac{S(T_R)}{S(T)} = \left(\frac{T}{T_R} \right)^5  \qquad \qquad [T_\textsc{max}> T> T_R]\,,
  \label{eq:D(T)}
\end{equation}
where we use the analytical estimate of Eq.~\eqref{eq:SvsT}. We define the ratio of the produced comoving FIMP density $Y^{\rm   prod}_X (T)$  of
Eq.~(\ref{eq:YprodMD}) and of the dilution factor $D(T)$  as 
\begin{equation}
  \tilde Y_X(T) = \frac{Y^{\rm prod}_X(T) }{D(T)} \simeq \frac{\Gamma_{B \rightarrow A_\textsc{sm} X} \, M_{\rm Pl} m_B T_R^7}{T^{10}}
  \label{eq:Ytoday}
\end{equation}
 which results from the combination of FIMP production in an early MD era at temperature $T$ and subsequent entropy dilution between $T$ and $T_R$. We notice in Eq.~(\ref{eq:Ytoday}) an extreme IR domination with a power-law scaling as $1 / T^{10}$, and such an IR domination is the reason why the resulting relic density does not depend on $T_{\rm MAX}$. 

As for the RD case, FIMP production is more efficient at the lowest possible temperature compatible with the presence of $B$ particles in the thermal bath. Thus the same argument used above would suggest that the relic FIMP comoving density is approximately $\tilde Y_X(T)$ evaluated at a temperature $T_{FI}$ that is of the order of the $B$ mass. However, Ref.~\cite{Co:2015pka} observed that FIMP production is still in action when we are in the tail of the Maxwell-Boltzmann distribution for $B$; this is just a consequence of the extreme IR domination observed in Eq.~\eqref{eq:Ytoday}. The precise calculation of the relic density would require numerically solving the Boltzmann equation, and we provide this derivation in Appendix~\ref{app:FI}. Here, we report the analytical estimate for the $X$ comoving density
 \begin{equation} Y_X^{\infty} \simeq 10^4 \times \left(
 \frac{T_R}{m_B} \right)^7 \times \frac{\Gamma_{B \rightarrow A_{\rm
       SM} X} \, M_{\rm Pl}}{m_B^2} \ .
\label{eq:YXMDfinal}
 \end{equation}
Besides a quite large enhancement factor of $ 10^4$, freeze-in during an early MD era leads  to a $(T_R / m_B)^7$ suppression to the FIMP relic density that we  observe today compared to the RD case of Eq.~(\ref{eq:YXR}). The  latter scaling is compatible with our estimate $Y_X^{\infty}\sim  \tilde Y_X(m_B)$. Consistently with~(\ref{eq:YXMDfinal}), we notice how  the red and the blue lines lead to an asymptotic relic density that is suppressed with respect to the purple line by approximately three and seven orders of magnitude, respectively.

\subsection{FIMP production from scatterings and UV sensitivity}
\label{sec:UVfreezin}

As we illustrate in Figure~\ref{fig:FI}, scatterings are an irreducible contribution to DM production within our framework. We complete our analysis by discussing this additional channel, and we show how its relative importance depends on whether the interaction between $X$ and the thermal bath is renormalizable or not. Let $d$ be the mass dimension of this operator. At high enough temperatures, larger than any mass of the particles involved in the collision, the scaling of the interaction rate with the temperature follows from dimensional analysis
\begin{equation}
\mathcal{R}_{B A^{'}_\textsc{sm} \rightarrow A^{''}_\textsc{sm} X}(T) \propto \frac{T^{2d - 7}}{\Lambda^{2d - 8}} \ ,
\label{eq:ratescattering}
\end{equation}
where $\Lambda$ is the mass scale appearing in the FIMP interaction. For non-renormalizable interactions, $d > 4$, this is the scale suppressing the operator. The amount of FIMP produced is still given by Eq.~\eqref{eq:Yapprox}, with production rate driven by scatterings as in Eq.~\eqref{eq:ratescattering}. For a RD universe, the resulting comoving number density scales with the temperature as
\begin{equation}
Y_X(T) \propto \frac{M_{\rm Pl}}{\Lambda^{2d - 8}} \, T^{2d - 9} \qquad \qquad \text{[RD epoch]}  \,.
\label{eq:YvsTscatteringRD}
\end{equation}

The value $d = 4.5$ divides two distinct regimes, and DM production is more efficient at high temperatures for $d > 4.5$. This case, where the interaction is non-renormalizable, is often referred to as UV freeze-in~\cite{Chen:2017kvz,Bernal:2019mhf,McDonald:2015ljz,Barman:2020plp}. On the contrary, for renormalizable operators scatterings dominate over decays only if a larger number of scattering processes can contribute to DM production or when the mass splitting between $B$ and $X$ is small enough to suppress the decay rate (see e.g.~Refs.~\cite{Junius:2019dci,Belanger:2018mqt,Garny:2018ali}). 

The abundance of DM produced through scatterings via non-renormalizable interactions cannot be computed in a RD universe. As it is manifest from Eq.~\eqref{eq:YvsTscatteringRD}, the production diverges as we go back in time (i.e.,~higher $T$). However, the estimate in Eq.~\eqref{eq:Yapprox} is only valid as long as FIMPs are out of equilibrium, and eventually at high enough temperatures the rate would become so large that FIMPs thermalize. We need to know to what extent we can extrapolate the RD universe back in time in order to provide a precise calculation. In other words, we need a temperature UV cutoff corresponding to the highest temperature for the RD phase, and the reheating temperature provides such a UV cutoff. 

The expression in Eq.~\eqref{eq:Yapprox} still accounts for the amount of FIMP produced in a Hubble doubling time when the temperature was $T$, and the scattering rate in Eq.~\eqref{eq:ratescattering} is still valid. However, as we have already seen in Sec.~\ref{sec:fi-dec-MD} for decays, there are two new ingredients: we need to use the Hubble parameter during reheating given in Eq.~\eqref{eq:HearlyMD}, and we need to account for dilution due to the entropy released by the inflaton decays. We find
\begin{equation}
Y_X(T) \propto \frac{M_{\rm Pl} T_R^2}{\Lambda^{2d - 8}} \, \frac{T^{2d - 11}}{D(T)} = \frac{M_{\rm Pl} T_R^7}{\Lambda^{2d - 8}} \, T^{2d - 16} \qquad \qquad \text{[MD epoch]}  \ .
\label{eq:YvsTscatteringMD}
\end{equation}

DM production via scattering for renormalizable operators (i.e.~$d \leq 4$) is IR dominated for both RD and MD. On the contrary, in the range $5 \leq d \leq 8$ DM production is dominated at small temperatures during reheating and at large temperatures during RD. In other words, during an early MD era it is mostly efficient either around $T=T_{R}$ or $T\sim m_B$, depending on the hierarchy between these two scales. For $d > 8$, DM production is UV dominated for both RD and MD. In the latter case, the resulting relic density is sensitive to $T_\textsc{max}$ and to the details of reheating. Interestingly, for $d < 8$ DM production via scattering depends only on $T_R$ and not on $T_\textsc{max}$~\cite{Co:2015pka,Garcia:2017tuj,Chen:2017kvz}.

One concrete example of non-renormalizable interactions is for a Weyl fermion singlet FIMP $\chi$. We consider the dimension 5 operator coupling $\chi$ to gluons and to a new fermion $\lambda^a$ in the adjoint representation of the $SU(3)$ color gauge group 
\begin{equation}
\mathcal{L} \supset \frac{1}{\Lambda} G_{\mu \nu }^a \lambda^a \sigma^{\mu} \bar \sigma^{\nu} \chi \ ,
\label{eq:Ldim5sec2}
\end{equation}
where $G_{\mu\nu}^a$ the gluon field strength. This is for instance the well known case of gluino-gluon-goldstino coupling in supersymmetric theories~\cite{Martin:1997ns}. Such an interaction induces DM production both via decays $\lambda \rightarrow G \, \chi$ and scatterings. The rate for the former scales proportionally to the decay width of the mother particle, $\Gamma_\lambda \simeq m_\lambda^3/\Lambda^2$, and therefore production via decays is IR dominated even for non-renormalizable operators. The operator in Eq.~\eqref{eq:Ldim5sec2} induces also DM production through t-channel scattering processes. Considering for example the t-channel exchange of a gluon, we have the matrix element
\begin{equation}
	|\mathcal{M}|^2 \sim \frac{g_s^2}{\Lambda^2} \frac{st(s+t)}{(t-\Pi(T)^2)^2} \, ,
\label{eq:Mt}
\end{equation}
where $g_s$ is the strong coupling, $s$ and $t$ are Mandelstam variables, and $\Pi(T)$ is a thermal mass that we insert to regulate the t-channel IR divergence.

\begin{figure}
\centering
\includegraphics[width=0.44\textwidth]{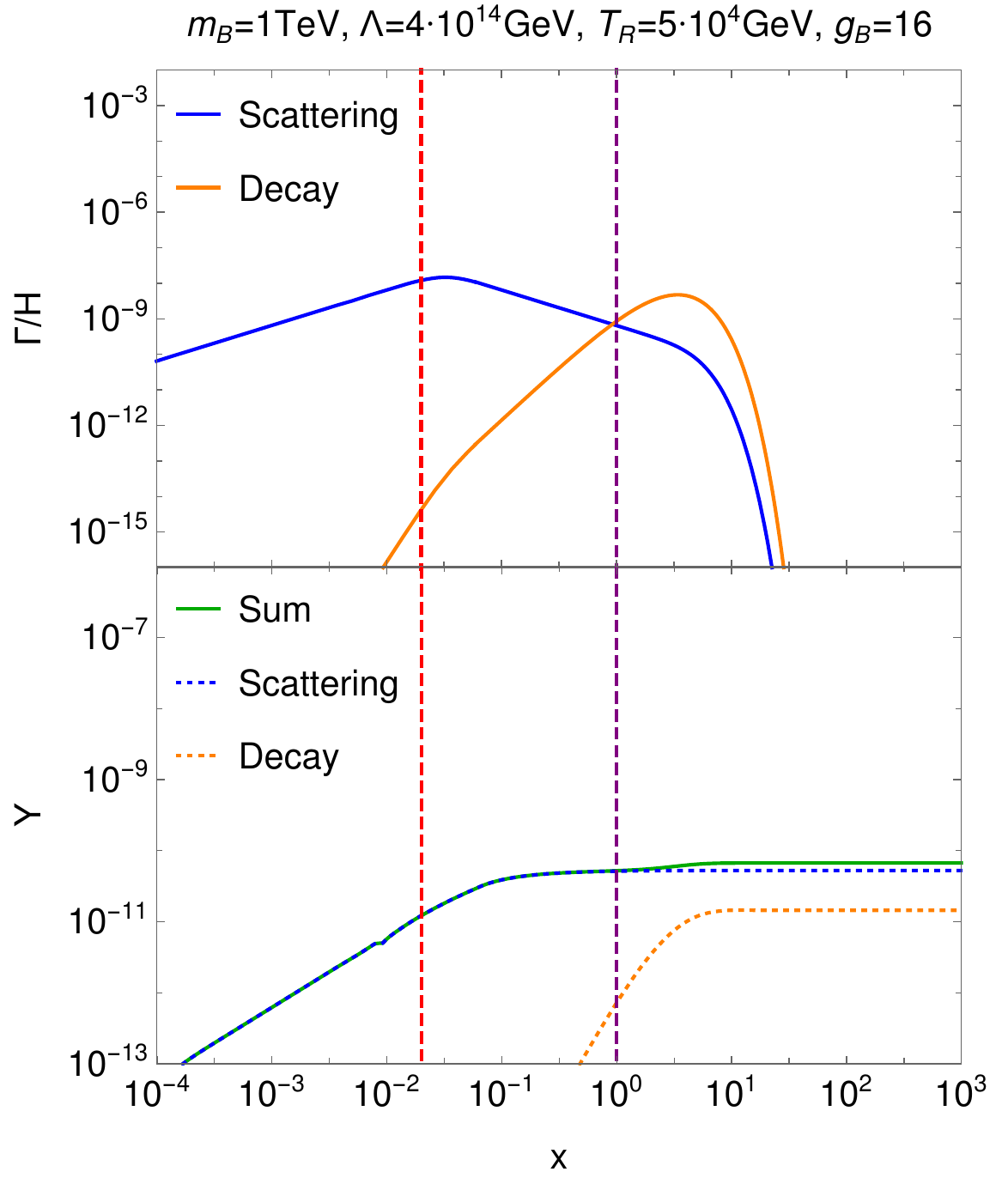} \hfill
\includegraphics[width=0.44\textwidth]{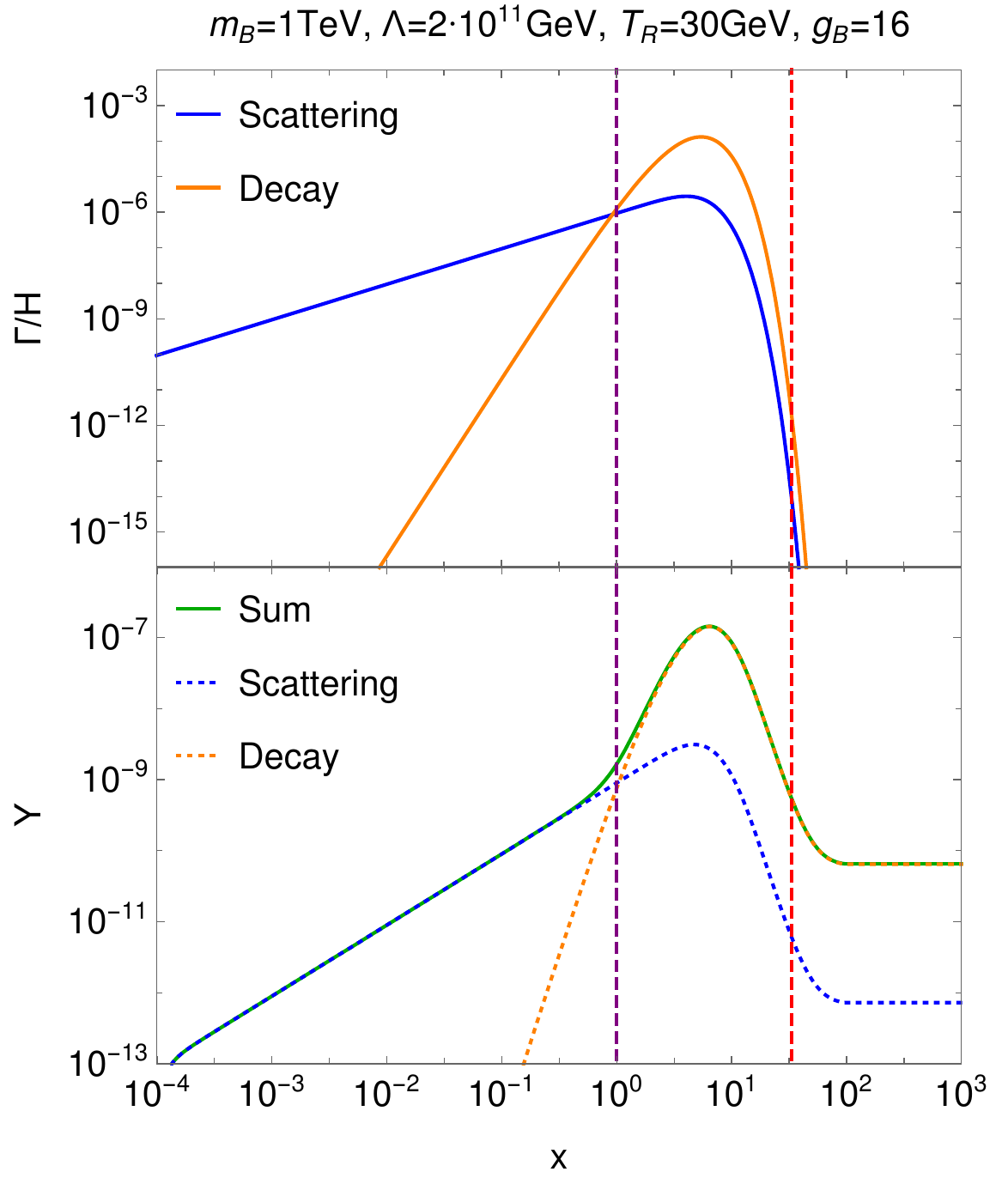} 
\caption{FIMP production via decays (orange) and scatterings (blue). The top panel depicts the ratio between interaction and Hubble rates while the bottom panel shows the total DM yield. We show two benchmarks with $T_{R}>m_B$ (left) and $T_{R}<m_B$ (right). The dashed vertical lines denote $T=T_{R}$ (red) and $T = m_B$ (purple). For these parameters, the observed DM abundance is reproduced for $m_{\rm DM} = 10$~GeV.}
\label{fig:rates}
\end{figure}

Taking $\Pi (T)= g_s T$ for illustration, Figure~\ref{fig:rates} presents numerical results for the operator in Eq.~\eqref{eq:Ldim5sec2}. Fixing the value of $m_B$, we consider two different choices for the combination $(\Lambda, T_R)$ in the two panels. For both cases, we take $\Lambda$ to be much larger than the weak scale\footnote{This is appealing since it could explain the tiny FIMP coupling from a hierarchy of mass scales.} and we choose numerical values such that we reproduce similar asymptotic DM comoving densities. The reheating temperature is always below the new physics scale, $T_R < \Lambda$, in order to provide consistency with the contact interaction. As argued above, we do not need to specify the value of $T_\textsc{max}$ for $d = 5$ but it is implicit that we assume the hierarchy $T_\textsc{max} < \Lambda$ for the same reason. The left and right panels have reheating temperature larger and smaller than the $B$ mass, respectively. We show the evolution of the production rates with respect to the Hubble parameter (upper part) and the DM comoving density (lower part) as a function of $x = m_B / T$. We keep the two contributions, decays (orange lines) and scatterings (blue lines). Vertical dashed lines identify the two key temperatures when freeze-in production is mostly efficient around $x_{FI}\sim {\cal O}( 1)$ (the purple lines are shown for $x=1$ as a guide for the eye) and when inflationary reheating is over $x= m_B/T_R$ (red lines). As expected, the scattering rate for the case $T_R > m_B$ is mostly efficient around $T_R$ whereas the decay rate is maximized when the temperature is around $m_B$. The relative height of the peaks is set by the reheating temperature: the higher $T_{R}$, the higher the contributions from the scattering processes. Consequently, the comoving density features two plateaus corresponding to the two production channels. In contrast, both scattering and decay rates peak around $T \approx m_B$ for the case $T_R < m_B$. For the specific interaction of Eq.~\eqref{eq:Ldim5sec2}, the scattering rate is smaller than the decay rate around the peak and therefore scatterings provide a negligible contribution to the total DM abundance. Thus it does matter whether the operator mediating the decay is renormalizable or not because the resulting scattering production may have a relatively large importance and get UV dominated~\cite{Moroi:1993mb,Rychkov:2007uq,Strumia:2010aa,Elahi:2014fsa}.

\subsection{Learning about the early universe from displaced events}
\label{sec:BoundTR}

The main message from the analysis above is the one concerning the $B$ decay length. The FIMP relic density is suppressed for freeze-in during an early MD, and consequently we need larger couplings than for freeze-in in RD. Thus if we want to account for the observed DM density we need a larger interaction rate, or in other words we need a shorter $B$ decay length compared to Eq.~(\ref{eq:ctauRD}): displaced events at colliders within the detectors is a rather typical prediction of freeze-in during an early MD epoch. Assuming that we can measure the masses from reconstructing the collider event in combination with information from measurements of the $B$ production cross section\,---\,see e.g.~Ref.~\cite{Bae:2020dwf} for new methods to read off the mass spectrum involved in displaced events\,---\,the observed decay length for $B \to A_\textsc{sm} X$ will directly pinpoint the reheating temperature $T_R$ if we require that $X$ makes all the observed DM. If we are less demanding and we require that $X$ is just a subdominant
component, it would at least provide an upper bound on $T_R$. 

Reconstructing the complete spectrum at collider is rather challenging so we should explore the less optimistic scenario where we cannot reconstruct the DM mass. FIMPs cannot be too light otherwise they would behave similarly to thermal warm DM (WDM) and erase small scales structures in contradiction with observations. Data from the Lyman-$\alpha$ forest put bounds on how warm these relics can be~\cite{Viel:2005qj,Viel:2013apy,Yeche:2017upn,Irsic:2017ixq}. In particular, Ref.~\cite{Irsic:2017ixq} put bounds on the WDM mass by implementing the velocity distribution into Boltzmann codes to get the linear transfer function and input the latter into dedicated hydrodynamical simulations. As a result, we have the bounds $m_\textsc{wdm}\gtrsim 3.5 \, {\rm keV}$ (conservative) and $m_\textsc{wdm}\gtrsim 5.3 \, {\rm keV}$ (stringent), corresponding to different assumptions on the thermal history of the universe. Refs.~\cite{Boulebnane:2017fxw,Heeck:2017xbu,Kamada:2019kpe,McDonald:2015ljz,Ballesteros:2020adh,DEramo:2020gpr} used the results of this analysis for WDM to obtain mass bounds on FIMPs, produced by decays or scatterings in a RD era, by comparing the associated linear transfer functions.\footnote{More accurate tools in the form of Lyman-$\alpha$ likelyhoods~\cite{Archidiacono:2019wdp} have not yet been applied to freeze-in.} For the case of decays, assuming that the final state particles masses are negligible compared to initial state particle masses, Refs.~\cite{Boulebnane:2017fxw,Ballesteros:2020adh,DEramo:2020gpr} have obtained bounds in the range $m_{X}\gtrsim 7-16 \, {\rm keV}$. The strongest bound, obtained by Ref.~\cite{DEramo:2020gpr} by employing the area criterium introduced in~\cite{Murgia:2017lwo}, results in $m_X \gtrsim 16 \, {\rm keV}$, and in the next sections we will refer to it as a ``stringent'' bound. As a guide for the eye, in the next sections, we will make use of the bound $m_{X}\gtrsim 10 \, {\rm keV}$ as a ``conservative'' Lyman-$\alpha$ bound on FIMPs. It is worth keeping in mind that the freeze-in production via scatterings or decays during an early MD epoch might also affect these bound, see e.g.~\cite{McDonald:2015ljz}.

Even if we cannot reconstruct the DM mass, for a known $B$ mass and decay length we can still determine a specific value of the reheating temperature $T_R$ once we set $m_X$ to the lowest value allowed by Lyman-$\alpha$ forest data. The DM mass cannot get any lower than that, and for such a smallest allowed value we need the smallest amount of dilution
after freeze-in because $\rho_X \propto m_X$. And the smallest amount of dilution corresponds to the largest value of $T_R$; this procedure also provides an upper bound on $T_R$.

Finally, if we consider a non-renormalizable interaction DM production can be controlled by scatterings. If this is the case, the amount of DM grows with $T_R$. If we fix the DM mass to the smallest value compatible with the Lyman-$\alpha$ bound, for a given $m_B$ and $\tau_B$ we can derive an upper bound on $T_{R}$, which is valid for any other (higher) DM mass.  Thus, even when the dominant  production channel would be freeze-in through scatterings, instead of decays, colliders could indirectly provide hints about cosmology.

To summarize, displaced events at colliders could indirectly provide constraints on the cosmological history of the early universe.

%%%%%%%%%%%%%%%%%%%%%%%%%%%%%%%%%%%%%%%%%%%%%%%%%%%%%%%%%%%%%%%%%%
\section{Minimal FIMP frameworks}
\label{sec:models}
%%%%%%%%%%%%%%%%%%%%%%%%%%%%%%%%%%%%%%%%%%%%%%%%%%%%%%%%%%%%%%%%%%
%
\renewcommand{\arraystretch}{1.0}
\begin{table}[t]
	\centering
	\begin{tabular}{ | c | c | c | c | c | }
		\hline
		$\boldsymbol{A_\textsc{sm}}$ & \textbf{Spin DM} & \textbf{Spin B} & \textbf{Interaction} & \textbf{Label} \\ 
		\hline \hline
		\multirow{2}{1cm}{\centering$\psi_\textsc{sm}$} & 0 & 1/2 & $\bar{\psi}_\textsc{sm} \Psi_B \phi $ & $\mathcal{F}_{\psi_\textsc{sm}\phi}$ \\
		& 1/2 & 0 &  $\bar{\psi}_\textsc{sm} \chi \Phi_B $ &$\mathcal{S}_{\psi_\textsc{sm}\chi}$ \\  
		\hline
		$F^{\mu\nu}$ & 1/2 & 1/2 & $\bar{\Psi}_B \sigma_{\mu\nu} \chi F^{\mu\nu} $ & $\mathcal{F}_{F\chi}$ \\
		\hline
		\multirow{2}{1cm}{\centering$H$} & 0 & 0 & $H^\dagger \Phi_B \phi$ & $\mathcal{S}_{H\phi}$ \\
		& 1/2 & 1/2 & $\bar{\Psi}_B \chi H$ & $\mathcal{F}_{H\chi}$ \\
		\hline
	\end{tabular}
	\caption{Classification of the simplest possible operators featuring a cubic interaction of the type of Figure~\ref{fig:FI}. The DM particle is a SM singlet denoted by $\phi$ if it is a scalar, $\chi$ if it is a fermion. The bath field ($\Psi_B$ for fermions, $\Phi_B$ for scalars) has therefore the same quantum numbers as the corresponding SM field $A_\textsc{sm}$.
	See the text for further details. 	}
	\label{tab:classification}
\end{table}

The main goal of our analysis is to emphasize the interplay between early universe DM production and displaced events with missing energy at colliders. As argued in Section~\ref{sec:BoundTR}, a positive signal for this kind of events at the LHC would allow us to gain information about both the DM model parameter space and the cosmological history, even in the pessimistic case where the mass of the missing energy carrier can not be reconstructed. 

We exploit this synergy considering a set of well defined scenarios giving rise to the cubic interaction shown in Figure~\ref{fig:FI}. Restricting ourselves to BSM particles of spin $0$ and $1/2$,\footnote{For a FI model with instead spin-1 DM produced by the decay of a long-lived vectorlike lepton, see~\cite{Delaunay:2020vdb}.}  and assuming DM to be a singlet under the SM gauge group, the resulting list is compactly reported in Table~\ref{tab:classification}.  These ``simplified models'' are minimal in the sense that they feature the lowest dimensional operators (with $d\leq 5$), containing one SM field and two extra fields of the dark sector, giving rise to the three-body interaction. Hence for each model only two fields need to be added to the SM field content. Furthermore, we impose an unbroken $Z_2$ symmetry under which the SM fields are even and the new fields are odd; this new parity, together with the fact that we always assume the DM field to be the lightest of the dark sector fields, ensures DM stability and restricts the three-field interactions to the form shown in Figure~\ref{fig:FI}.  In the first column of Table~\ref{tab:classification}, we consider the finite set of options for the SM particle $A_\textsc{sm}$: (i) a SM fermion field $\psi_\textsc{sm}$, that is, a left-handed or right-handed lepton or quark; (ii) a SM gauge boson (photon, $W$, $Z$, gluon), following from an interaction involving the field strength $F^{\mu\nu}$ of $U(1)_Y$, or $SU(2)_L$, or $SU(3)_c$; (iii) the Higgs boson $H$. The second and third columns contain all the corresponding possible choices for the spins of $X$ and $B$, respectively. The gauge quantum numbers of $B$ under the SM gauge group should be equal to the ones of $A_\textsc{sm}$ since the DM field is a gauge singlet. Notice how the case when $A_\textsc{sm}$ is the $U(1)_Y$ hypercharge gauge boson is of no phenomenological relevance at colliders because $B$ would be a complete SM singlet for gauge invariance. In the fourth column, we display the Lagrangian term relevant for freeze-in DM production and for the decay of $B$ into DM at colliders.  We write these interactions denoting a fermionic  DM particle (bath particle) as $\chi$ ($\Psi_B$) and a scalar DM particle (bath particle) as $\phi$ ($\Phi_B$). In the last column, we give a label to each minimal model: the first letter indicates whether the bath particle $B$ is a fermion ($\cal F$) or a scalar ($\cal S$), the second letter indicate the SM field involved, while the third letter indicate the nature of DM.

It is worth noticing that all the interactions of Table~\ref{tab:classification} are renormalizable except for the ones of the models $\mathcal{F}_{F\chi}$ which are of dimension $d=5$. The interaction term $\mathcal{F}_{F\chi}$ resembles the gluino-gluon-goldstino coupling or Wino-$W$-goldstino interactions in supersymmetry,~cf.~e.g.~\cite{Martin:1997ns}.\footnote{One could in principle add analogous models featuring scalar DM and a scalar bath particle $B$ in an adjoint representation of $SU(2)_L$ or $SU(3)_c$ with couplings to the field strengths given by dimension-6 operators of the kind $\partial_\mu\Phi_B \partial_\nu\phi F^{\mu\nu}$. However, as one can show by integrating by parts, such an interaction is   proportional to the gauge boson masses, thus vanishing in the $SU(3)_c$ case and for the unbroken phase of $SU(2)_L$ (up to  thermal corrections). We defer to future work a detailed study of this peculiar scenario.}  Among the renormalizable interactions, three classes have $d=4$.  Some instances of models involving interactions of this kind have already been studied within the context of freeze-in in e.g.~\cite{Garny:2018ali} for $\mathcal{F}_{\psi_{\rm SM}\phi}$, in \cite{Ibarra:2008kn,Belanger:2018sti} for $\mathcal{S}_{\psi_{\rm SM}\chi}$ and in~\cite{Calibbi:2018fqf,No:2019gvl} for $\mathcal{F}_{H\chi}$.
   Finally, $\mathcal{S}_{H\phi}$ is a dimension-three operator involving scalars only.

This simple scheme encodes all possible cubic interactions involving one FIMP, another BSM field and one SM particle (restricting to spins 0 and 1/2 for the BSM sector).  They amount to a total of $35$ simplified models (counting all possible combinations of chirality and flavor of the SM fermions, and the three different choices for the gauge boson field strength $F^{\mu\nu}$).  We will focus on a representative subset of these simplified models which can illustrate the general interplay between collider phenomenology and early universe cosmology.

%%%%%%%%%%%%%%%%%%%
\section{Collider meets Cosmology}
\label{sec:coll-results}
%%%%%%%%%%%%%%%%%%

In recent years there has been an important effort of the LHC community to explore the sensitivity of the LHC to BSM signatures involving displaced decays and/or long-lived particles (together denoted as LLP searches)~\cite{Alimena:2019zri}. As argued in the introduction, LLP searches can test the nature of DM produced through freeze-in and they can also indirectly probe the cosmological parameters involved. In particular, the LHC LLP searches listed in Table~\ref{tab:searches} can directly probe the minimal frameworks presented in Section~\ref{sec:models} as detailed in the summary Table~\ref{tab:models-searches}. Within the context of an early MD era, we can obtain the DM relic density as a function of $T_{R}$ and confront the viable parameter space to the relevant collider searches. Below, we first underline the relevant features of the selected LLP searches in Section~\ref{sec:searches}. In Section~\ref{sec:results}, we discuss how each of the theoretical models can be probed by experimental constraints on general grounds, and then illustrate the procedure for a representative set of minimal  scenarios including leptophilic and topphilic  as well as the non-renormalizable case of the singlet-triplet model in Sections~\ref{sec:leptophilic-scenario},~\ref{sec:topphilic-scenario} and~\ref{sec:non-renorm-oper} respectively. More details on the collider searches and our recastings can be found in Appendix~\ref{sec:Ap-LHC}.
 
%%%%%%%%%%%%%%%%%%
\subsection{Collider searches for long-lived particles}
\label{sec:searches}
%%%%%%%%%%%%%%%%%%

\renewcommand{\arraystretch}{1.2}
\begin{table}[t]
	\centering
	\begin{tabular}{ | c | c | c | c | c | }
		\hline
		\multirow{2}{1.5cm}{\textbf{Signature}} &  \multirow{2}{2.1 cm}{\centering \textbf{Exp.~\&~Ref.}} & \multirow{2}{1.5cm}{\centering 
		$\boldsymbol{\mathcal{L}}$} &   \textbf{Maximal} & \multirow{2}{1.5cm}{\centering \textbf{Label}} \\ 
		& & &  \textbf{sensitivity} & \\
		\hline
		\hline
		R-hadrons  & CMS~\cite{CMS:2016ybj} &  $12.9~\text{fb}^{-1}$ & \multirow{2}{1.8cm}{\centering $c\tau\gtrsim 10$~m} & RH\\
		Heavy stable charged particle & ATLAS~\cite{Aaboud:2019trc} & $36.1~\text{fb}^{-1} $ &  & HSCP\\
                \hline
		\multirow{2}{4cm}{\centering Disappearing tracks}  &ATLAS~\cite{Aaboud:2017mpt}&  $36.1~\text{fb}^{-1}$ &  $c\tau \approx 30$~cm & \multirow{2}{2cm}{\centering DT} \\
		& CMS~\cite{CMS:2018ail,Sirunyan:2020pjd} &  $140~\text{fb}^{-1} $& $c\tau \approx 60$~cm & \\
                \hline
		\multirow{3}{4cm}{\centering Displaced leptons} & \multirow{1}{1.5cm}{\centering CMS~\cite{Khachatryan:2014mea} }$^{\,\dagger}$  &  $19.7~\text{fb}^{-1}$ & \multirow{2}{1.8cm}{\centering $ c\tau\approx 2$~cm}&\multirow{3}{2cm}{\centering DL} \\
		& \centering CMS~\cite{CMS:2016isf}&  $2.6~\text{fb}^{-1} $& & \\
                 & \centering ATLAS~\cite{Aad:2020bay}&  $139~\text{fb}^{-1}$ & $c\tau\approx 5$~cm& \\
                \hline
		Displaced vertices + MET & ATLAS~\cite{Aaboud:2017iio} &  $32.8~\text{fb}^{-1}$ & $c\tau \approx 3$~cm & DV+MET \\
		\hline
		Delayed jets + MET  &CMS~\cite{Sirunyan:2019gut} &  $137~\text{fb}^{-1}$ & $c\tau \approx 1-3$~m & DJ+MET \\
		\hline
		Displaced vertices + $\mu$ &ATLAS~\cite{Aad:2020srt} &  $136~\text{fb}^{-1}$ & $c\tau \approx 3$~cm& DV+$\mu$ \\
			\hline
		Displaced dilepton vertices &ATLAS~\cite{Aad:2019tcc} & $32.8~\text{fb}^{-1}$ & $c\tau \approx 1-3$~cm& DLV \\
		\hline
		Delayed photons & CMS~\cite{Sirunyan:2019wau} &  $77.4~\text{fb}^{-1}$ &  $c\tau \approx 1$~m & D$\gamma$ \\
		\hline
	\end{tabular}
	\caption{Searches for long-lived particles (LLP) at the LHC experiments performed on $\sqrt{s} = $ 13 TeV run data (with the exception of the 8 TeV search denoted
          by$^{\,\dagger}$). See Section~\ref{sec:searches} and Appendix~\ref{sec:Ap-LHC} for details. The maximal sensitivities in the proper decay length refer to the models employed in the original Refs.~of the second column to interpret the searches. } 
	\label{tab:searches}
\end{table}

\begin{figure}[t]
	\centering
	\includegraphics[width=1.\textwidth]{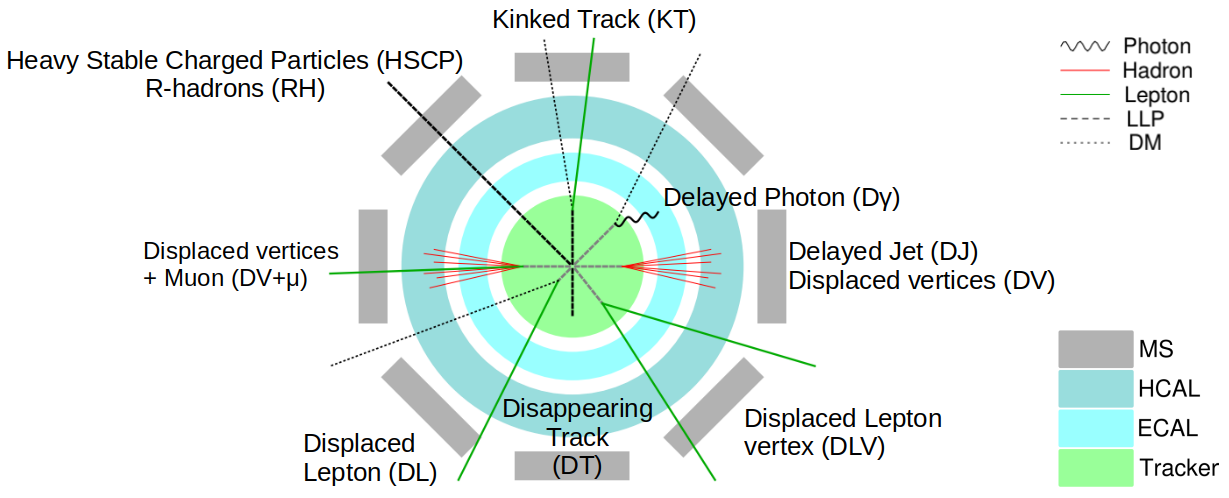}
		\caption{Schematic illustration of the LLP searches considered in Table~\ref{tab:searches} and Table~\ref{tab:models-searches} and described in Appendix~\ref{sec:Ap-LHC}. The green area represents the tracker, the blue (cyan) region denoted by HCAL (ECAL) refers to the hadronic (electromagnetic) calorimeter, the gray boxes represent the muon spectrometer (MS). Dotted lines refer to DM, dashed lines refer to the LLP and they are darker for searches that require  a charged track associated to it.}
                  \label{fig:searches}
\end{figure}

We collect in Table~\ref{tab:searches} a selection of existing LLP searches performed by ATLAS and CMS employing data from collisions with center of mass (c.o.m.)~energy of $13$ TeV, and we assign a label to each of them. The selection is dictated by the attempt to have a full coverage over the possible SM final states with LLP signatures, and, when available, searches including missing transverse energy (MET) are preferred. In  Appendix~\ref{sec:def-other} we comment about other possibly relevant LHC searches not considered in our study. Table~\ref{tab:searches} shows how ATLAS and CMS are spanning already a quite wide landscape of possible LLP-type signatures, involving leptons, gauge bosons, charged tracks, and also MET. As pictures can sometimes summarize information better than a long text we illustrate these searches in Figure~\ref{fig:searches} while we defer to Appendix~\ref{sec:Ap-LHC} a brief description of each experimental analysis. Details on our recastings are also provided in length in the same appendix. 

A large MET component is clearly a smoking gun signature of DM. Only two of these experimental analyses explicitly require missing energy in the final state.  Nevertheless, the other searches are often sensitive to our simplified DM models due to a rather inclusive event selection.  We remark that missing energy is not typically required in LLP searches since the SM background is already very much suppressed by the displacement selection.  However, the presence of a large MET in the targeted BSM signature could also be exploited for triggering the relevant events. Hence for the purpose of detecting DM models with LLP signatures, the MET requirement could be beneficial.

\subsection{Early matter dominated era and collider signatures}
\label{sec:results}

We can confront the minimal simplified models of Table~\ref{tab:classification} to the series of LHC searches listed in Table~\ref{tab:searches} and discussed in Appendix~\ref{sec:Ap-LHC}. The result of this comparison is illustrated in Table~\ref{tab:models-searches}. We distinguish models involving the first and second generation fermions ($\ell$ and $q$) from models involving the third generation ($\tau$ and $t$), because of their different collider signatures. We can exploit all the available collider searches focusing on models of the classes $\mathcal{F}_{W\chi}$, $\mathcal{F}_{\ell\phi}$ \& $\mathcal{S}_{\ell\chi}$ and $\mathcal{F}_{t\phi}$ \& $\mathcal{S}_{t\chi}$. These classes are highlighted in green.

Another characteristic signature that can arise from the type of models under study is the one of the so-called ``kinked tracks'' (KT) included in the last column of Table~\ref{tab:models-searches}. This signature corresponds to a long-lived charged mediator\,\footnote{From now on, we will often refer to the bath particle $B$ as the ``mediator''.} $B$ decaying inside the detector into DM and an electron or muon. The track due to the daughter particle appears as the continuation of the track of the mediator, however there is a kink at the point where the decay takes place. No dedicated and fully experimental analysis has been performed addressing this type of signature. We nevertheless include KT in Table~\ref{tab:models-searches} to highlight the range of physics targets that such a search would have; the reader interested in models of which kinked tracks would be a key signature can find an early discussion in Ref.~\cite{Calibbi:2014pza}. Moreover, constraints on KT signatures can be obtained by re-interpreting the disappearing track  searches along the lines of Ref.~\cite{Jung:2015boa}, as we will do in the rest of this section.

For our phenomenological discussion, we select within the model classes of Table~\ref{tab:classification} the following three that, we argue, represent an exhaustive set of simplified FIMP models:  
 \begin{itemize}
 \item $\mathcal{S}_{\ell_R\chi}$, that features a scalar mediator with the quantum numbers of a right-handed lepton, coupling to DM through a renormalizable operator;
 \item  $\mathcal{F}_{t_R\phi}$, another example of renormalizable cubic interaction but involving strongly-interacting fields, and the third generation (namely, a right-handed top);
 \item $\mathcal{F}_{W\chi}$, an explicit case of freeze-in controlled by a non-renormalizable operator, involving the $SU(2)_L$ field strength, hence a fermion mediator in a triplet representation.
 \end{itemize}
Our selection includes renormalizable interactions (with first/second and third generation) and a non-renormalizable interaction. We perform a detailed analysis of the collider implication for early-universe cosmology for these models. As shown in Section~\ref{sec:bound}, DM freeze-in production for renormalizable interactions will be IR dominated. In addition, for simplicity, we will assume  that the DM only couples to one of the lepton/quark flavors at a  time. Notice that even for a mediator coupling to multiple families we do not expect any relevant constraint from flavor violation because of the smallness of the couplings considered, see e.g.~\cite{Kopp:2014tsa} for a leptophilic scenario of such kind.

\renewcommand{\arraystretch}{1.1}
\begin{table}[t]
	\centering
	\begin{tabular}{|c||c|c|c|c|c|c|c|c|c||c|c|c|}
		\hline
		& DV & DJ & DV & & & & & & &  \\
		\textbf{Label} & + & + & + & DL & DLV & D$\gamma$ & DT & RH & HSCP &  KT \\
		& MET & MET & $\mu$ & & & & & & &  \\
		\hline
		\hline
		\rowcolor{Green}
		$\mathcal{F}_{\ell\phi} \ \& \ \mathcal{S}_{\ell\chi}$ &  &  & & \cmark & &  &  &  & \cmark  & \cmark \\
		\hline
		$\mathcal{F}_{\tau\phi} \ \& \ \mathcal{S}_{\tau\chi}$ & \cmark & \cmark &  & \cmark & & & & & \cmark  & \cmark \\
		\hline
		$\mathcal{F}_{q\phi} \ \& \ \mathcal{S}_{q\chi}$ & \cmark & \cmark & & & & & & \cmark &  &  \\
		\hline
		\rowcolor{Green}
		$\mathcal{F}_{t\phi} \ \& \ \mathcal{S}_{t\chi}$ & \cmark & \cmark & \cmark & \cmark & & & & \cmark &  & \\
		\hline
		$\mathcal{F}_{G\chi}$ & \cmark & \cmark & & & & & & \cmark &  & \\
		\hline
		\rowcolor{Green}
		$\mathcal{F}_{W\chi}$ & \cmark & \cmark & \cmark & \cmark & \cmark & \cmark & \cmark & &  & \cmark \\
		\hline
		$\mathcal{S}_{H\phi} \ \& \ \mathcal{F}_{H\chi}$ & \cmark & \cmark & \cmark & \cmark & \cmark & & \cmark & &  & \cmark \\
		\hline
	\end{tabular}
	\caption{Sensitivity of the LHC searches, shown in Table~\ref{tab:searches} and discussed in Appendix~\ref{sec:Ap-LHC},
          to our simplified models as labeled in Table~\ref{tab:classification}. The fermion $\psi_\textsc{sm}$ of Table~\ref{tab:classification} can be either a light
          lepton $\ell\in \{ e,\mu \}$, a tau lepton $\tau$, a light quark $q\in \{ u,d,s,c,b \}$ or a top quark $t$. Fermions can be either left-handed ($L$) or right-handed ($R$), a
          choice that has little impact on the final state but it can affect LHC production of the mediator $\Phi_B/\Psi_B$. $\mathcal{F}_{W\chi}$ and $\mathcal{F}_{G\chi}$ refer
          to models of the class $\mathcal{F}_{F\chi}$ where we consider the $SU(2)_L$ (coupling to electroweak gauge bosons) and the $SU(3)_c$ (coupling to gluons) field strength, respectively. The models we study in this work belong to the classes highlighted in light green.}
        \label{tab:models-searches}
\end{table} 
 
The interplay between LHC and cosmology has been considered by Ref.~\cite{Belanger:2018sti} for models similar to our category $\mathcal{F}_{\ell\phi}$ (thus with scalar instead of
fermion DM) and $\mathcal{F}_{q\phi}$ (hence with first and second
generation quarks involved, instead of the top).\footnote{Another difference is that we do not make the instantaneous reheating approximation. This would not change dramatically the overall picture presented in Ref.~\cite{Belanger:2018sti} for this class of models (see Appendix~\ref{sec:comparison}).}  Also, the model $\mathcal{S}_{t_R\chi}$ was already
studied in the context of freeze-in in~\cite{Garny:2018ali}, but
without the special emphasis on the role of $T_{R}$ and on displaced
events that we put here. In addition, R-hadron searches applied to a
model $\mathcal{F}_{G\chi}$ have been considered
in~\cite{Davoli:2018mau} but in the context of freeze-out through
coannihilations with feeble conversion processes (see also~\cite{Brooijmans:2020yij} in the case of $\mathcal{S}_{b\chi}$).

\subsubsection{Leptophilic scenario, $\mathcal{S}_{\ell_R\chi}$}
\label{sec:leptophilic-scenario}

We augment the SM field content by a new charged scalar $\Phi_B$ with the quantum numbers of a right-handed lepton and a Majorana fermion singlet $\chi$: this is the $\mathcal{S}_{\ell_R\chi}$ simplified model. Both extra fields are odd under an unbroken $Z_2$ symmetry. For simplicity and to ease the comparison of our results with a similar scalar DM model discussed in Ref.~\cite{Belanger:2018sti}, we choose to couple the new fields to right-handed muons only, i.e.~$\ell_R=\mu_R$. The Lagrangian reads
\begin{equation}
{\cal L}_{\mathcal{S}_{\ell_R\chi}} \supset \frac{1}{2} \bar{\chi} \gamma^{\mu} \partial_{\mu}
\chi-\frac{m_{\chi}}{2} \bar{\chi} \chi+ (D_{\mu}\Phi_B)^\dagger
\ D^{\mu}\Phi_B -m_\phi^2 |\Phi_B|^2 \ - \ \lambda_{\chi} \Phi_B \bar \chi
\mu_R \ + \ h.c.\,,
\label{eq:lagr}
\end{equation}
where $\lambda_\chi$ is a dimensionless Yukawa coupling.
Within this model, there are two types of
processes that contribute to the DM abundance via the freeze-in mechanism: (i) the
decay of the mediator and (ii) the scattering processes converting the
mediator into DM.\footnote{For t-channel processes involving massive bosons and a lepton in the propagator, t-channel divergences should be regulated by thermal corrections. We expect that the regulated process contribute similarly to other scatterings, with a few percent correction in the abundance that we neglect  in the leptophilic case.}

Notice that no symmetry prevents the new scalar $\Phi_B$ to couple to the SM Higgs as
\begin{equation}
\label{eq:Lhiggs}
{\cal L}_{H} = -\lambda_H H^{\dag} H \Phi_B^{\dag} \Phi_B\,,
\end{equation}
 In general, this quartic interaction will be radiatively generated at the level $\lambda_H\gtrsim 10^{-2}$, through e.g.~a box loop  involving electroweak gauge bosons.  This interaction also contributes to the DM abundance through scattering processes involving $H$. However, we checked that its contribution is subleading, at most few percent for $\lambda_H \sim O(1)$. We can thus safely neglect it, and in what follows we set the quartic coupling to $\lambda_H=10^{-2}$.

\paragraph{Collider and cosmological constraints.} The mediator $\Phi_B$ can be pair produced at collider due to its electroweak (EW) gauge interactions. The production cross section at next to leading order (NLO) can be obtained by interpolating the tables available at~\cite{slepton-xsec}. We have checked that the NLO cross section is well reproduced by the leading order (LO) cross section obtained with {\tt MadGraph5\_aMC@NLO}~\cite{Alwall:2014hca} times a K-factor of approximatively~1.3. Depending on its lifetime, the mediator $\Phi_B$ can lead to displaced leptons~\cite{Khachatryan:2014mea,CMS:2016isf}, disappearing/kinked tracks~\cite{Aaboud:2017mpt,CMS:2018ail} and heavy stable charged particles~\cite{Khachatryan:2015lla,CMS:2016ybj} (see Table~\ref{tab:models-searches}).

\begin{figure}
	\centering
	\subfloat[]{\includegraphics[scale=0.48]{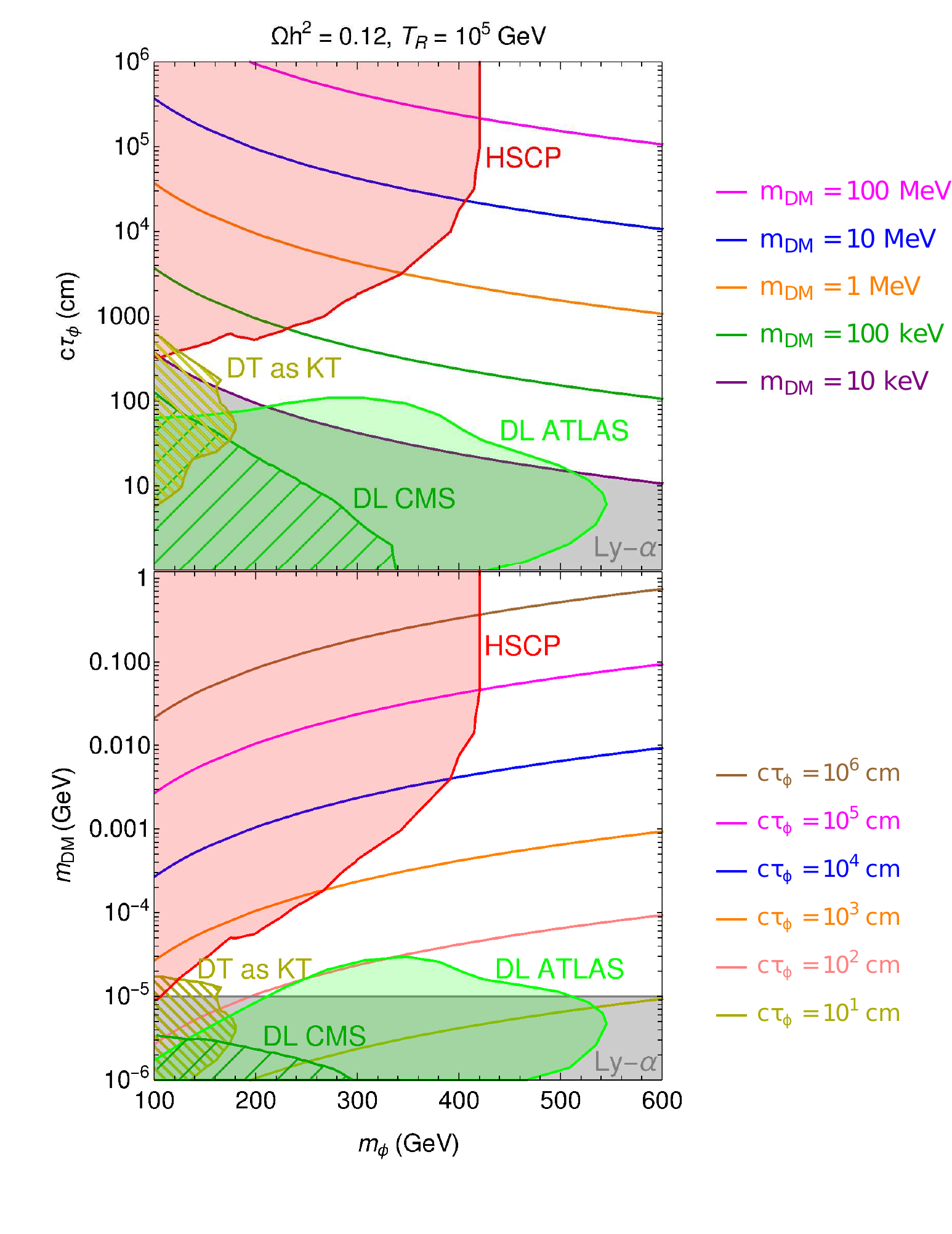}}
	\subfloat[]{\includegraphics[scale=0.48]{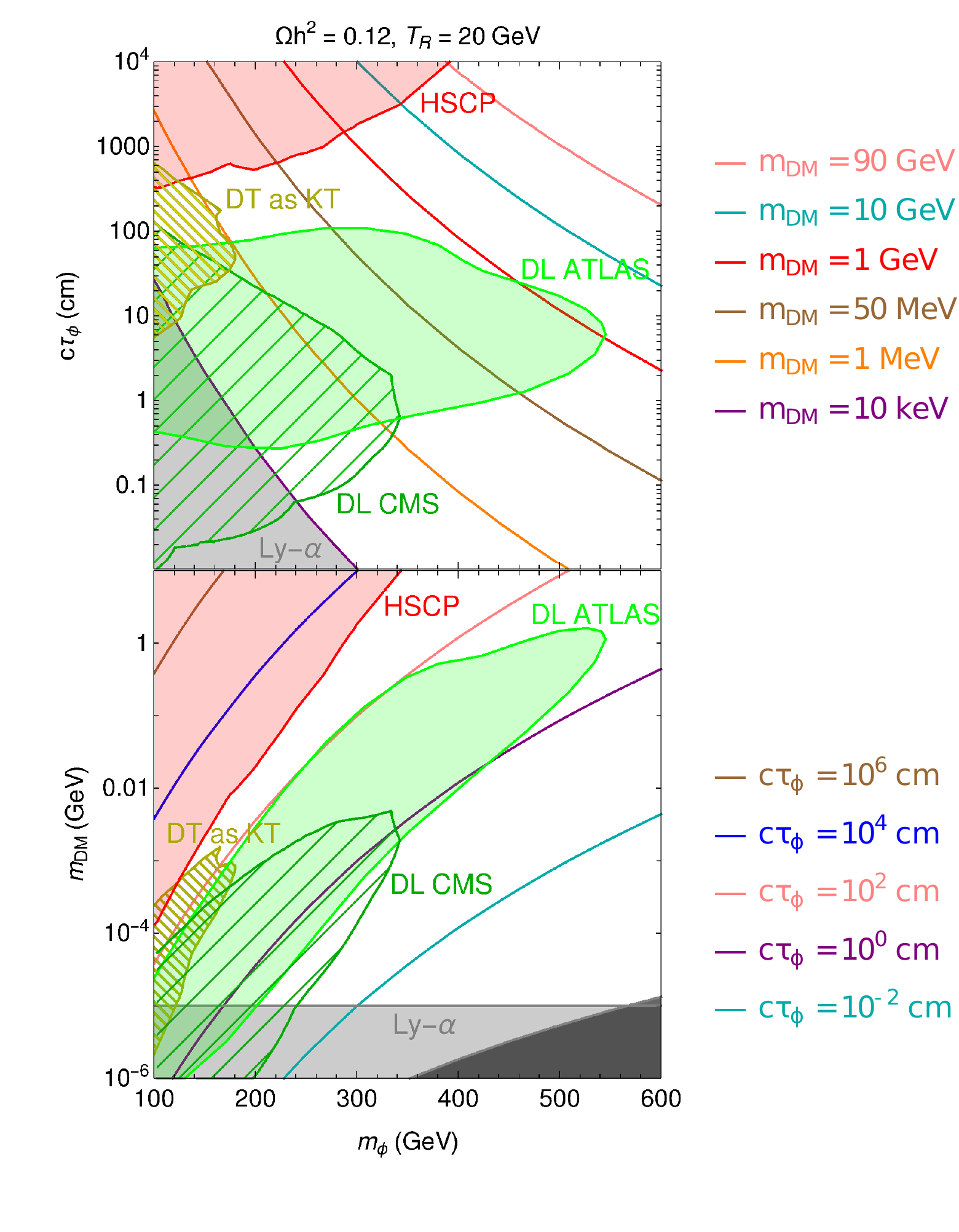}}
	\caption{Leptophilic DM with $T_{R} = 10^5~\GeV$ (left) and $T_{R} = 20~\GeV$ (right). In the top panels, solid lines in the plane $(m_\phi,\,c\tau_\phi)$ reproduce the observed DM relic density for the associated $m_{\rm DM} = m_\chi$. Similarly, in the bottom panel we show relic density contours for the proper mediator decay length in the plane $(m_\phi,\,m_{\rm DM})$. We identify regions excluded by HSCP searches from CMS and ATLAS (HSCP - red) and by displaced same-flavor lepton searches by ATLAS (DL ATLAS - uniform green). The possible reach from displaced same-flavor lepton by CMS is shown with a green loosely hatched region while our recasting of disappearing track searches in the case of a kinked track signature (DT as KT) is shown as a dark yellow densely hatched region. We denote with a light gray area regions excluded by the Lyman-$\alpha$ bound $m_{\rm DM}\gtrsim 10$~keV, see text for details. Finally, in the dark gray region DM thermalize in the early universe and therefore cannot be produced via freeze-in.}
	\label{fig:lepto2}
\end{figure}

\begin{figure}
	\centering
	\subfloat[]{\includegraphics[scale=0.48]{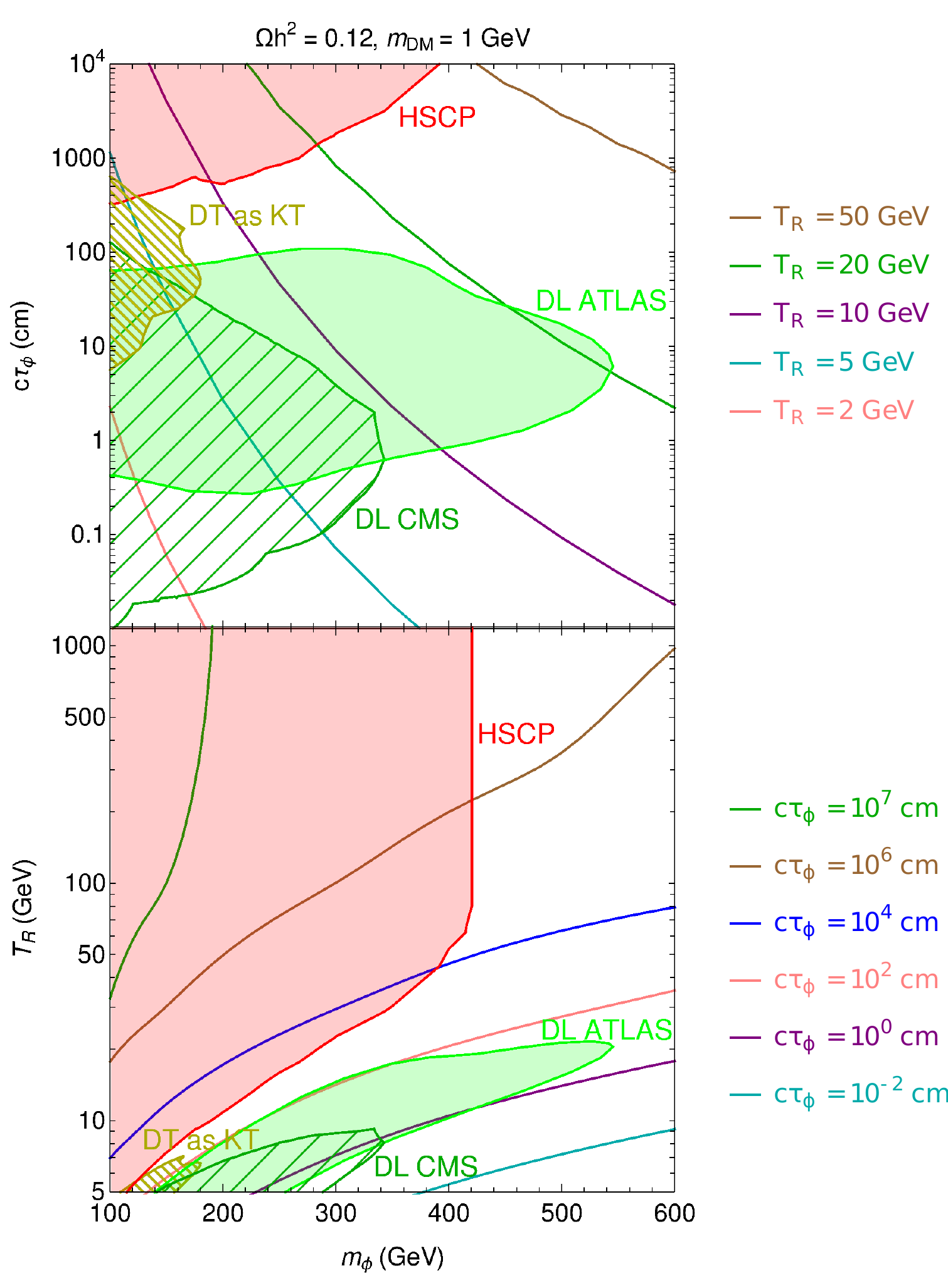}}
	\hfill
	\subfloat[]{\includegraphics[scale=0.48]{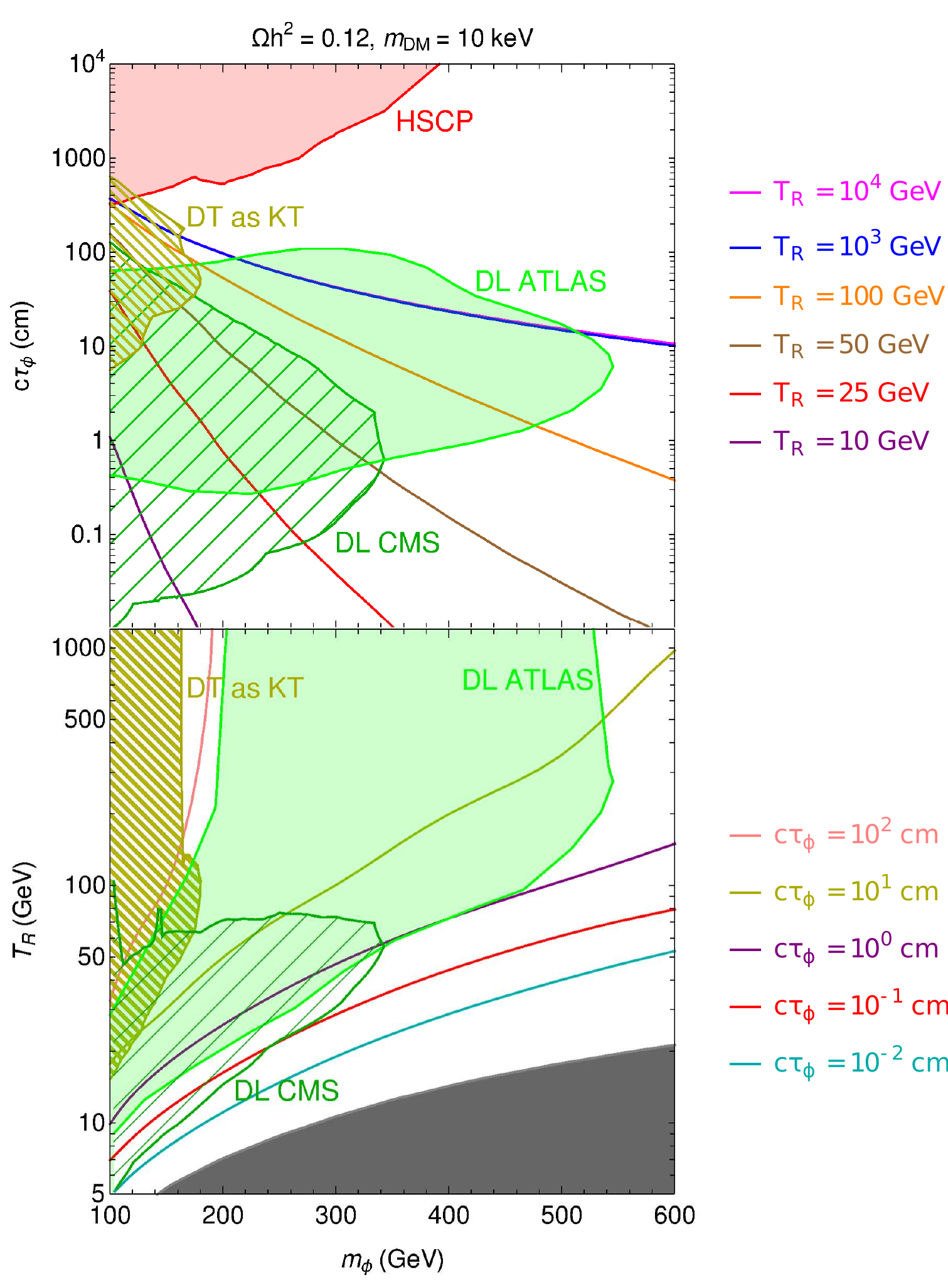}}
	\caption{Leptophilic DM with $m_\chi = 1~\GeV$ (left) and
          $m_\chi = 10~\keV$ (right). The upper panel shows
          constraints in the $(m_\phi,c\tau_\phi)$ plane while the
          lower panel identifies the viable parameter space in the
          $(m_\phi,T_{R})$ plane. Solid lines reproduce the observed
          relic density, and the color code is the same as in
          Figure~\ref{fig:lepto2}. The value of $T_R$
          associated to a given $(m_\phi,c\tau_\phi)$ in the right panel
          essentially provides an upper bound on $T_R$ for any dark
          matter mass as $m_\chi \gtrsim 10~\keV$ due to small scale
          structures constraints, see the text for details.}
	\label{fig:lepto}
\end{figure}

For long lifetimes, corresponding to a proper decay length exceeding the meter, we make use of CMS and ATLAS searches on HSCPs in Table~\ref{tab:searches}, cf.~Appendix~\ref{sec:def-HSCP} for details of the recasting method.  The corresponding regions excluded at 95\% confidence level (CL)  appear in red in Figures~\ref{fig:lepto2} and~\ref{fig:lepto}.

 For intermediate lifetimes, when the decay of the charged mediators occur inside the tracker, one expects kinked tracks. Disappearing track searches can have some sensitivity to this type of signatures, that we estimate in the following.  First, the charged track of the mother particle (the mediator) and the one of the daughter particle (the muon) should have a sufficient angular separation such that the track of the mother particle is interpreted as a disappearing track.  In our analysis we conservatively require that the angular separation of the daughter track with respect to the mother track should be larger than $\Delta R > 0.1$.  Second, the DT searches include a lepton veto, and in this DM model the second part of the charged track is a non-soft muon, since the mass splitting between the DM and mediator is large $\Delta m \gg m_\mu$.  If the muon is reconstructed, the event would be discarded.  However, by the same requirement above on $\Delta R$, the emitted muon does not generically point towards the primary vertex and hence we assume that it is not properly reconstructed and the lepton veto does not apply.\footnote{We thank Steven Lowette for discussions on this point.}  We refer to the Appendix~\ref{sec:def-DT} for more details. In the figures of this section we will denote as ``DT as KT'' the sensitivity lines obtained by reinterpreting the DT search for a KT signature through the procedure explained above.  For this purpose we have used the CMS search~\cite{Sirunyan:2020pjd} since it is the one with the largest luminosity. The area that would be excluded by KT following our reinterpretation appears as a dark-yellow densely-hatched region in Figures~\ref{fig:lepto2} and~\ref{fig:lepto}.  As can be seen, our reinterpretation demonstrates that a dedicated KT would nicely complement the other LLP searches for $c \tau_\phi\sim 1$~m.
  
For shorter lifetimes, we employ the very recent ATLAS search for displaced leptons~\cite{Aad:2020bay}. In the auxiliary material provided for this analysis, one can find the case of oppositely charged dimuons final state that can be applied directly to our model, leading to the exclusion region in Figures~\ref{fig:lepto2} and~\ref{fig:lepto} colored with light green (tagged DL ATLAS).
  
For even shorter lifetime, we exploit the CMS displaced lepton analysis~\cite{Khachatryan:2014mea,CMS:2016isf} to estimate the possible reach of an analogous search but targeting a different flavor configuration. The CMS analysis looks for events with a displaced $e\mu$ pair, while our final states consists of a pair of displaced muons since DM couples to muons only. The CMS results in~\cite{Khachatryan:2014mea,CMS:2016isf} can not then be straightforwardly used for our leptophilic model.  However, the CMS collaboration has provided the efficiency tables for displaced electrons and muons derived in these searches~\cite{1317640}. We optimistically apply the same efficiencies to a di-muon final states to obtain the sensitivity lines of an hypothetical experimental search, analogous to~\cite{Khachatryan:2014mea,CMS:2016isf}, but performed instead for a $\mu^+ \mu^-$ final state. The area that would be excluded by our reinterpretation of the CMS DL searches appear as a green loosely hatched region in Figures~\ref{fig:lepto2} and~\ref{fig:lepto} (tag DL CMS). More details about our procedure are provided in Appendix~\ref{sec:def-DL}.\footnote{We are grateful to Freya Blekman for discussions.} It is interesting to observe that displaced lepton searches of ATLAS and CMS covers slightly different regions of displacement.

In addition to collider constraints, the Lyman-$\alpha$ observations mentioned in Section~\ref{sec:BoundTR} impose a conservative bound of $m_X\gtrsim 10$~keV (and a stringent bound of $m_X\gtrsim 16$~keV) and further limit the viable parameter space for DM. In the left panel of Figure~\ref{fig:lepto2}, the light gray region is excluded by the conservative Lyman-$\alpha$ constraint. In the right panel of Figure~\ref{fig:lepto2}, we also show our conservative $m_X\gtrsim 10$~keV Lyman-$\alpha$ bound with a light gray area as a guide for the eye. In fact, a dedicated analysis of FI as non-cold DM arising from an early MD era is still missing in the literature. Finally, in some regions of the parameter space, DM may eventually achieve thermodynamical equilibrium and the freeze-in computations presented in Section~\ref{sec:bound} do not hold. We shade the corresponding parameter space regions in dark gray in Figures~\ref{fig:lepto2} and~\ref{fig:lepto}.

\paragraph{Results for fixed $T_{R}$.} The parameter space under investigation, including also the early MD cosmology, is four-dimensional and it is described by the mediator mass $m_\phi$ and proper decay length $c \tau_\phi$, the DM mass $m_{\rm DM}$ and the reheating temperature $T_R$. In Figure~\ref{fig:lepto2}, we show slices that identify the cosmological history, or equivalently we fix the value of $T_{R}$.

In the left panel of Figure~\ref{fig:lepto2}, we consider a standard cosmological scenario with a rather high reheating temperature, $T_{R}= 10^5$~GeV $\gg T_{FI}$. We show in the upper part with different colored lines, each one correspondent to a different value of the DM mass, the relation between the mediator mass and its decay length necessary to account for all the observed DM abundance. The colored areas tagged with the labels HSCP, KT, DL ATLAS and CMS are excluded or can be probed by the corresponding collider searches as discussed above. The only relevant searches for this case are those for charged tracks, such as HSCP, because mediator will typically cross the detector entirely. LLP searches targeting displaced vertices can essentially not test the viable DM space. Although the detection of a charged track associated to $\Phi_B$ would be a spectacular signal of new physics, it would not let us learn much about the possible connection to DM or the physics of the early universe. A similar conclusion can be reached looking at the bottom part, where the colored curves are associated to different decay lengths and we visualize the $(m_\phi,m_{\rm DM})$ plane. 

In the right panel of Figure~\ref{fig:lepto2}, we illustrate how this situation can be turned into a more promising one when considering freeze-in occurring during an early MD era with $T_{R}= 20$~GeV. The upper part shows how displaced searches can probe DM with masses $m_{\rm DM}\lesssim 1$~GeV. From the bottom panel, it appears that the decay length of the mediator ranges from the order of a millimeter to hundreds of meters while still accommodating the correct relic abundance when $m_\text{DM}> 10$~keV and 100 GeV$<m_\phi<$600~GeV. Leptophilic FIMPs can thus be probed at colliders through HSCP, KT and DL signatures.  The HSCP searches probe charged mediator masses up to $m_\phi\sim 400$~GeV and $c\tau_\phi \gtrsim 5$~m. The recent DL ATLAS search has a very strong reach, excluding mediator masses up to $\sim~500$~GeV for $c\tau_\phi \sim 10$~cm, and covering a large portion of the interesting displacements, i.e.  0.3~cm~$\lesssim c\tau_\phi \lesssim$~80~cm.  For even shorter decay lengths, the DL CMS search, under the assumption that it can be performed for same-flavor leptons,
as explained above, provide further sensitivity.  For intermediated lifetimes, it is interesting to observe how a KT search (or any search sensitive to such signature) can close the gap between HSCP and displaced leptons searches.

\paragraph{Results for fixed $m_{\rm DM}$.}
Measuring the DM mass from a displaced event is quite challenging. However, we can still learn something about the early universe, and in particular extract an upper bound on $T_{R}$, from a displaced event. We illustrate this point, and exemplify the discussion in Section~\ref{sec:BoundTR}, with a comparison in Figure~\ref{fig:lepto} between a 10 keV DM scenario (right panel) and a case with heavier FIMP (left panel). In the upper parts, colored lines identify in the
($c\tau_\phi$, $m_\phi$) plane the values of $T_{R}$ reproducing the observed DM abundance. Likewise, in the lower parts they give the values of $c\tau_\phi$ (again reproducing the DM relic density) in the ($T_R$, $m_\phi$) plane. The colored areas show the regions probed by the collider searches as discussed above.

In the top-left panel of Figure~\ref{fig:lepto}, the case for $m_\chi = 1$~GeV, all continuous curves illustrate that freeze-in occurs in an early MD epoch, i.e.~$T_{R}<T_{FI}$. In order to get a lower mediator lifetime for a given $m_\phi$ we have to reduce $T_{R}$, as also shown in the bottom-left panel. Reducing $c \tau_\phi$ corresponds to effectively enhancing the coupling constant $\lambda_{\chi}$ in Eq.~\eqref{eq:lagr}, which in turn increases the DM production rate and gives a larger freeze-in contribution. Therefore we need the early MD era to last longer in order to provide more dilution. The same panel also show that, for a given value of the mediator lifetime, larger $m_\phi$ imply larger values of $T_{R}$. Indeed, by increasing $m_\phi$ we suppress $Y_X^{\infty}$ and we need a shorter duration of the early MD epoch to provide less dilution, see Eq.~\eqref{eq:YXMDfinal}.

Similar conclusions can be drawn from the bottom-left panel of Figure~\ref{fig:lepto}, where we plot contours of fixed mediator lifetime accounting for the correct relic abundance in the plane $(m_\phi,T_{R})$. In the upper part of the plot, the contours become vertical since we recover freeze-in in a RD era $T_R\gg T_{FI}$. Focusing on the bottom curves, an increase in the reheating temperature implies an increase in the decay length, and then a decrease in $\lambda_{\chi}$. A larger reheating temperature (with $T_{R}< T_{FI}$) means less entropy dilution between $T_{FI}$ and $T_R$, and in order to compensate for this effect there should be less DM produced during freeze-in, i.e.~the decay width should decrease.

The picture changes when we consider the lowest value of the DM mass compatible with the Lyman-$\alpha$ constraint, $m_\chi$ = 10 $\keV$. Since the relic density is proportional to the DM mass, less dilution is needed to give displaced-vertex signatures at the LHC. Equivalently, the relic density can be reproduced for larger values of $T_R$ as it is well visible comparing the two panels of Figure~\ref{fig:lepto}.  Furthermore, the colored curves in the top-right panel  {\it do not} always satisfy $T_{R}< T_{FI}$, i.e.~both cases of freeze-in occurring during an early MD era and a RD era are possible. Indeed, the contours for $T_{R} = 10^4~\text{GeV}$ and $10^3~\text{GeV}$ are superposed. In the bottom panels, the contours at fixed values of $c\tau_\phi$ become vertical for $T_{R}\gg m_\phi$.

The contours in the right panel of Figure~\ref{fig:lepto} are associated to lightest FIMP allowed by Lyman-$\alpha$ constraints, and they inform us on the maximal $T_{R}$ allowed compatible with freeze-in production (in particular, not larger than $\Omega_\chi h^2=0.12$). For instance, suppose that a long-lived charged scalar decaying into a muon and MET is observed and its mass and decay length are measured to be $m_\phi \approx 500$~GeV, $c\tau_\phi\approx 1$~cm. In such a case, even without a measurement of the DM mass, Figure~\ref{fig:lepto} tells us that this particle can be a freeze-in mediator compatible with the observed DM relic density only if $T_R \lesssim 100$~GeV. Such information would have a profound impact on our understanding of the early universe (for instance inflation, baryogenesis, etc). Given that for heavier DM the upper bound on the reheating temperature only becomes more stringent, the discussion of the following models will focus on the $m_\text{DM}=10~\keV$ case only.

As a final remark, we comment on the possible impact of future experiments. 
First, assuming that a LLP search at the LHC remains background free,
we can estimate the reach of HL-LHC (3000 fb$^{-1}$) by (linearly) rescaling its sensitivity on the mediator production cross section with the luminosity. 
For instance, studying the cross section dependence on the mediator mass $m_\phi$, we can estimate that the DL ATLAS analysis will probe masses up to $\approx 900$ GeV. This will considerably extend the explored parameter space of the model also to regions where lower $T_{R}$ upper bounds could be inferred (see Figure \ref{fig:lepto} top right).

On the other hand, we observe that future dedicated detectors targeting larger values of $c\tau_{\phi}$ (such as MATHUSLA~\cite{Curtin:2018mvb} or CODEX-b~\cite{Aielli:2019ivi}) will not constitute a very sensitive probe for this specific simplified model.
First, the long-lived particle is electromagnetically charged and hence the regions of large $c {\tau}_{\phi}$ will be covered efficiently by HSCP searches at HL-LHC. 
Second, large $c {\tau}_{\phi}$ corresponds to very small couplings and hence to scenarios where dilution is not necessary in order to obtain the correct relic abundance. 
As a consequence, even in the event of a discovery, it will not be possible to set an absolute upper bound on $T_R$ in the region of large $c {\tau}_{\phi}$, as it is visible from the ``overlapping'' contours in the top-right panel of Figure \ref{fig:lepto}.

The above discussion changes significantly in the case of models with non-renormalizable interactions and UV sensitivity, as we will discuss at the end of Section \ref{sec:non-renorm-oper}, where future facilities dedicated to LLP signatures would be beneficial in exploring further the parameter space and the connection with $T_R$.

\subsubsection{Topphilic scenario, $\mathcal{F}_{t_R\phi}$}
\label{sec:topphilic-scenario}
The second simplified model we study is the one with a real scalar DM particle $\phi$ and a vectorlike fermion mediator $\Psi_B$ with the quantum numbers of right-handed up-type quarks. We consider a ``topphilic'' scenario where BSM particles couple only to the top
\begin{equation}
 {\cal L}_{\mathcal{F}_{t_R\phi}}  \supset \partial_{\mu}\phi
 \ \partial^{\mu}\phi -\frac{m_{\phi}^2}{2} \phi^2 + \frac{1}{2} \bar{\Psi}_B \gamma^{\mu} D_{\mu}
 \Psi_B -m_\psi \bar \Psi_B \Psi_B \ - \ \lambda_{\phi}  \bar \Psi_B  t_R \phi\ + \ h.c.\,,
 \label{eq:lagr-top}
 \end{equation}
where $\lambda_\phi$ is a dimensionless Yukawa coupling. As usual, both the DM field and the mediator are odd under an unbroken $Z_2$ symmetry, and we neglect possible extra contributions to DM production from a DM-Higgs portal coupling~\cite{Lebedev:2019ton,Heeba:2018wtf} so as to focus on the cubic interaction sketched in Figure~\ref{fig:FI}. Decay and scattering processes can both produce DM in this scenario. Moreover, the higher (color) degrees of freedom and the larger gauge coupling are such that scattering processes are expected to play a more important role than in our previous leptophilic model, see also the discussion in~\cite{Garny:2018ali}. As a consequence, for fixed values of $m_\phi$ and $T_R$, one will need a longer lifetime in order to account for the relic abundance in this topphilic scenario than in models of the class ${\cal  F}_{\ell\phi}$ (or ${\cal S}_{\ell\chi}$ studied above).
\begin{figure}
	\centering
	\subfloat[]{\includegraphics[width=0.48\textwidth]{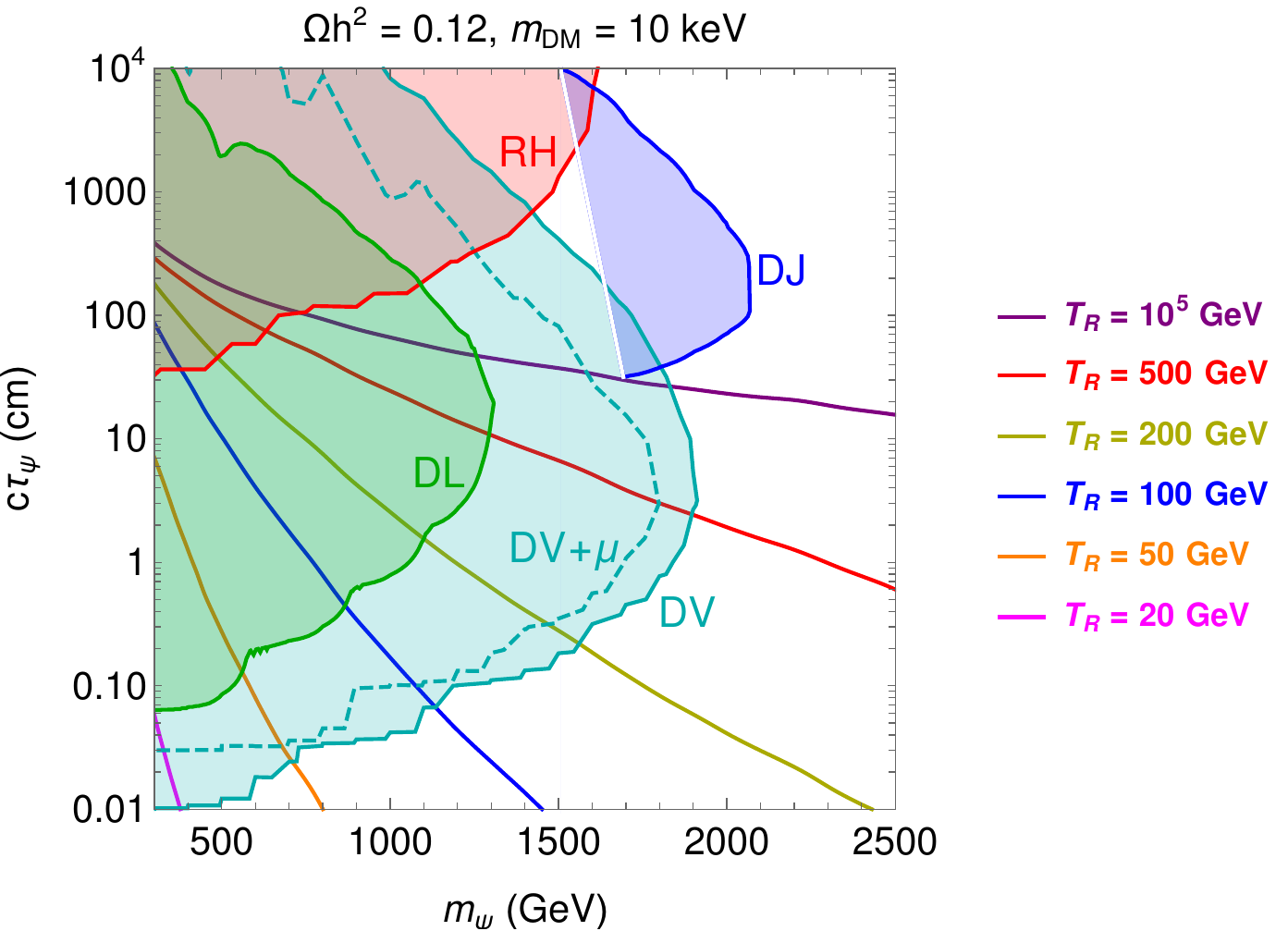}}
	\hfill
	\subfloat[]{\includegraphics[width=0.48\textwidth]{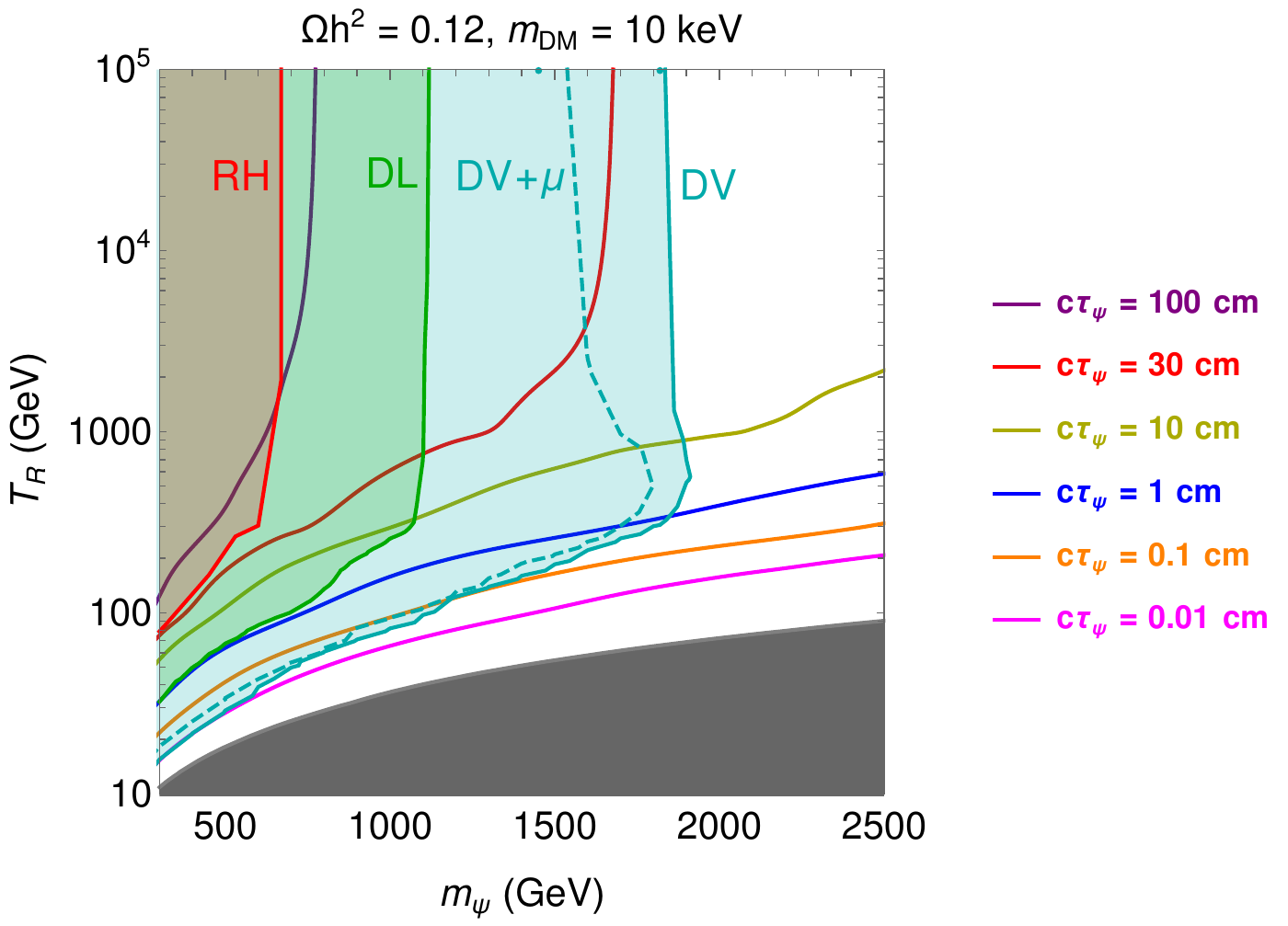}}
	\caption{Tophilic DM with $m_\chi = 10~\keV$. Contours for
          fixed value of $T_R$ account for the DM relic abundance in
          the $(m_\psi,\,c\tau_\psi)$ plane (left), and contours of
          fixed $c\tau_\psi$ do the same in the $(m_\psi,\,T_R)$ plane
          (right). The value of $T_R$
            associated to a given $(m_\psi,c\tau_\psi)$ provides
            an upper bound on $T_R$ for any dark matter mass as
            $m_\chi \gtrsim 10~\keV$ due to small scale structures
            constraints. The blue shaded area is excluded by the
          DV+MET search, the red area by R-hadron searches, the blue
          dashed area by DV+$\mu$ search, the dark blue area by the DJ
          search, and the green area by the DL search. }
	\label{fig:top}
\end{figure}

In Figure~\ref{fig:top} we consider the topphilic scenario for a DM mass of 10 keV, a value that saturates the conservative Lyman-$\alpha$ constraint and allows to extract an upper bound on $T_R$, as illustrated at the end of Section~\ref{sec:leptophilic-scenario}. In the left panel, colored lines identifies the value of $T_R$ needed to reproduce the relic density in the $(m_\psi,\,c\tau_\psi)$ plane. Likewise, colored lines in the right panel show the values of $c \tau_\psi$ needed in the $(m_\psi, T_R)$ plane. Comparing contours of Figure~\ref{fig:top} with those of the right panel of Figure~\ref{fig:lepto}, we can see that longer lifetimes can be reached for fixed value of the mediator mass and $T_R$ in models ${\cal F}_{t_R\phi}$. Except from the LHC searches, this is the main difference between the leptophilic and topphilic scenarios. The dependence on the masses and temperature scaling are the same since for both cases all processes producing DM are renormalizable and thus IR-dominated.

\paragraph{Collider signatures and constraints.} Looking at Table~\ref{tab:models-searches}, one can infer that the topphilic model can lead to a rich variety of signatures at the LHC. The mediator decays into top whose decays can in turn give both fully hadronic and semi-leptonic signatures. In order to assess the most relevant searches, we first obtain the production cross section of the colored vectorlike fermion mediator $\Psi_B$ at the LHC multiplying the LO result of {\tt MadGraph5\_aMC@NLO} by a flat K-factor of 1.6. This K-factor reproduces well the ratio of LO to NNLO cross-section for squark pair production tabulated in~\cite{tR-xsec} and we assume that the same K-factor can be applied to the case of colored vectorlike fermion mediator.
 
A long-lived $\Psi_B$ form an R-hadron (RH) before decaying~\cite{Buchkremer:2012dn}. If the decay occurs outside the detector, the R-hadron leaves a highly ionized track. The region excluded at 95\% CL by RH searches according to our recasting is shown as a red area in Figure~\ref{fig:top}. To obtain these results, we employed the public code {\tt SModelS}~\cite{Ambrogi:2018ujg,Heisig:2018kfq} and the one obtained from the ``LLP recasting repository''~\cite{LLPrepos}, and applied the analysis done for a stop mediator in both CMS~\cite{CMS:2016ybj} and ATLAS~\cite{Aaboud:2019trc} searches (see Section~\ref{sec:def-HSCP}) to our vectorlike top mediator as the spin of the produced particle is not expected to affect the sensitivity~\cite{Buchkremer:2012dn}.\footnote{We assume that the RH has the same lifetime as the heavy vectorlike fermion $\Psi_B$. This is supported by e.g.~\cite{Zyla:2020zbs,Lenz:2015dra} where the lifetime of hadrons containing heavy quarks, in particular $b$, is discussed.}
 
For shorter decay lengths, the decay happens before the R-hadron leaves the detector with a top quark in the final state that itself decays dominantly into a $bW$ pair. This in turn can decay hadronically to form jets, giving rise to displaced vertices (DV) or delayed jet (DJ) signatures. For details about these searches and our implementations we refer to, respectively, Appendix~\ref{sec:def-DV} and ~\ref{sec:def-DJ}. Concerning the DV analysis, let us emphasize that the algorithm built to reconstruct the events in~\cite{Aaboud:2017iio} combines displaced vertices that are located within 1~mm from each other. This is of particular relevance when the mediator couples to a top quark which dominantly decays to a $b$-quark and a $W$ bosons. In the latter  case, two displaced vertices possibly arise from the decay of the long-lived mediator since jets originating from a b-quark are displaced themselves (with a proper decay length of $\mathcal{O}(0.5)$~mm). Treating this properly is beyond the scope of this work. Therefore, we conservatively take only the tracks originating from the $W$ into account for the reconstruction of the displaced vertex. The region excluded by the DV+MET search is shown with a light blue region while the region excluded by the DJ+MET is shown in dark blue.  We see that the DJ+MET analysis can probe larger $c\tau_\psi$ than DV+MET, see discussion in Appendix~\ref{sec:def-DJ}. We also display with a dashed blue line the area probed by DV+$\mu$ (cf.~Appendix~\ref{sec:def-DVmu}) which is similar to the one of DV searches. This captures the topologies where one of the top quark decays leptonically to a muon. The sensitivity is comparable to the DV searches since the suppression of the signal due to the leptonic branching ratio of the $W$ is approximately compensated by the higher integrated luminosity employed in the analysis.

Alternatively, the two tops can both decay leptonically inside the detector leading to a displaced lepton pair. This signature can be covered by the DL searches~\cite{Khachatryan:2014mea,CMS:2016isf,Aad:2020bay}. We focus on the very recent ATLAS analysis~\cite{Aad:2020bay} which employs the largest set of data at $\sqrt{s}=13$ TeV. In the Appendix~\ref{sec:def-DL} we discuss the details of our recasting procedure, whose uncertainty is significant for some values of the LLP mass.

Nevertheless our analysis provides an indicative estimate of the LHC reach on the leptonic channel,
which we display as a uniform green region in Figure~\ref{fig:top}.
Even if the signal yield is limited by the leptonic branching ratio of the $W$ boson,
the search has a certain coverage of the parameter space, which is however superseded by the displaced jets + MET search
for all ranges of $c\tau_{\psi}$.

LLP searches for the topphilic ${\cal F}_{t_R\phi}$ model probe a larger parameter space than in the leptophilic ${\cal S}_{\ell_R\chi}$ of Section~\ref{sec:leptophilic-scenario}. The main reason is a larger LHC production cross section because the mediator is colored, and fermions display a higher production cross section than scalars. We see that RH searches probe the mediator masses up to $m_\psi \approx 1.6$ TeV for $c\tau_\psi\gtrsim$ 30 cm. The DV searches cover a large range of parameter space corresponding to masses up to $m_\psi \approx 1.9$ TeV and $c\tau_\psi\gtrsim$ 0.1 mm, partially overlapping with RH searches. Complementary constraints are obtained
making use of DJ searches that cover 1.5 TeV~$\lesssim m_\psi \lesssim$ 2.1 TeV and 30 cm $\lesssim c\tau_\psi \lesssim 100$~m. As in the case of the leptophilic model considered above, the plots in Figure~\ref{fig:top} show how a measurement of an LLP mass and lifetime could be employed to bound the reheating temperature.

Also for this model, the impact of HL-LHC searches will considerably extend the mass reach and hence the coverage of regions corresponding to  low $T_R$ bounds. By naively scaling the sensitivity of the DV+MET search on the production cross section of the colored mediator with the luminosity, we estimate that HL-LHC could reach $m_\psi \approx 2.8$ TeV.

\subsubsection{Non-renormalizable operators: the singlet-triplet model,  $\mathcal{F}_{W\chi}$}
\label{sec:non-renorm-oper}
The two examples considered so far feature renormalizable FIMP interactions and therefore DM production is dominated by physics in the IR. We now discuss the $\mathcal{F}_{W\chi}$ model where interactions are non-renormalizable and scattering processes contributing to DM production are UV-dominated. Looking at Table~\ref{tab:models-searches}, this model appears to be sensitive to a large set of existing LLP searches. The two BSM fields of the $\mathcal{F}_{W\chi}$ model are an extra singlet fermion DM and an $SU(2)_L$ triplet fermion mediator that we denote with
\begin{equation}
	\chi_S,	\qquad \chi_T= 
	\begin{pmatrix}
	\chi_T^0/\sqrt{2} & \chi_T^+ \\
	\chi_T^- & - \chi_T^0/\sqrt{2}
	\end{pmatrix}
	\,,
\end{equation}
and with masses $m_S$ and $m_T$ respectively.  Both new fields are odd under an unbroken $Z_2$ symmetry and couple to the EW gauge bosons through the Lagrangian:
\begin{eqnarray}
  \mathcal{L} &\supset& - \frac{m_S}{2} \bar\chi_S \chi_S - \frac{m_T}{2} \text{Tr}\left[\bar\chi_T \chi_T \right] + \frac{1}{2} \text{Tr}\left[ \bar\chi_T i \cancel{D}_\mu \chi_T \right] % \nonumber\\ &&
  +\frac{1}{\Lambda} (W^a_{\mu \nu} \bar \chi_S \sigma^{\mu \nu} \chi_T^a + \text{h.c.}),
\label{eq:unphysL}
\end{eqnarray}
where $W^a_{\mu \nu}$ is the $SU(2)_L$ field strength and $\Lambda$ is the scale of new dynamics responsible for generating the higher dimensional operator. More gauge invariant interactions exist at dimension five (for instance $H \chi_T H \chi_S$, see for details Appendix~\ref{sec:13}). However, our goal is to explore interactions of the kind depicted in Figure~\ref{fig:FI}, and we will assume that other possible dimension-five operators are suppressed, possibly by a higher mass scale, to the extent that they can be neglected for what concerns both DM production and LHC phenomenology. We only consider reheating scenarios with $T_{\textsc{max}} < \Lambda$ in order for the effective description in Eq.~\eqref{eq:unphysL} to be valid during DM production. The singlet-triplet model has been also considered in Ref.~\cite{Filimonova:2018qdc} in the context of freeze-out DM production with compressed spectra.

We work in the $m_T\gg m_S$ regime and we neglect any Higgs portal coupling. As a result, singlet/triplet mixing will have minor influence and the lightest neutral fermion under the $Z_2$ symmetry, i.e.~the DM candidate, is essentially the singlet fermion $\chi \simeq \chi_S$ with mass $m_\text{DM}\simeq m_S$. The mediator multiplet has a neutral component that is essentially the neutral component of the triplet $\Psi_B^0\simeq \chi^0_T$ with mass $m_T$. The charged components, corresponding to the charged components of the triplet $\Psi_B^{\pm}=\chi_T^{\pm}$, have a slightly higher mass $m_C= m_T+\Delta m$ with a splitting induced by EW quantum corrections $\Delta m \simeq 160$~MeV, see Appendix~\ref{sec:13} for details. As a result, the mass
hierarchy of the new fields in our analysis is always: $m_C\gtrsim m_T\gg m_\text{DM}=m_S$.
\paragraph{DM production in the early universe.} The processes reported in Table~\ref{tab:processes} of Appendix~\ref{sec:13}, decays and scatterings, contribute to DM production. As discussed in Section~\ref{sec:UVfreezin}, scattering mediated by non-renormalizable operators are UV dominated. The t-channel exchange of a photon in $\Psi_B^0e^-\to \chi e^-$ gives rise to a transition matrix $\propto (g/\Lambda)^2$ (where $g$ is the $SU(2)_L$ gauge coupling) that has a form similar to that shown in Eq.~(\ref{eq:Mt}) given that all particles involved are much lighter than the mediator particle $\Psi_B^0$.\footnote{We regulated the IR divergence of the diagram with a photon thermal mass, $m_\gamma \sim eT$, in the propagator.} 

%%%%%%%%%%%%%%
\begin{figure}
%missing
  \centering
	\subfloat[]{\includegraphics[width=0.41\textwidth]{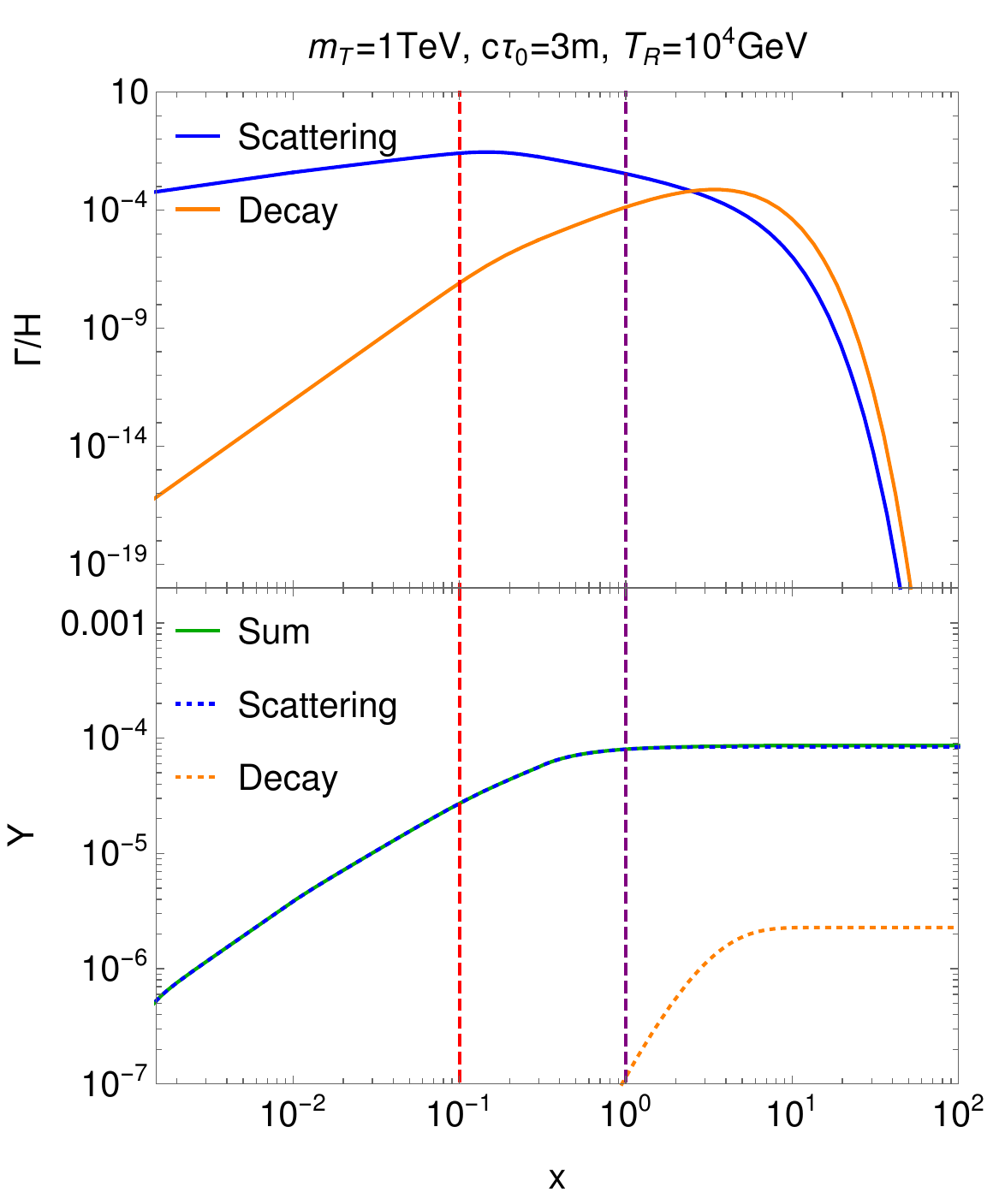}}
	%\hfill
	\qquad
	\qquad
	\subfloat[]{\includegraphics[width=0.41\textwidth]{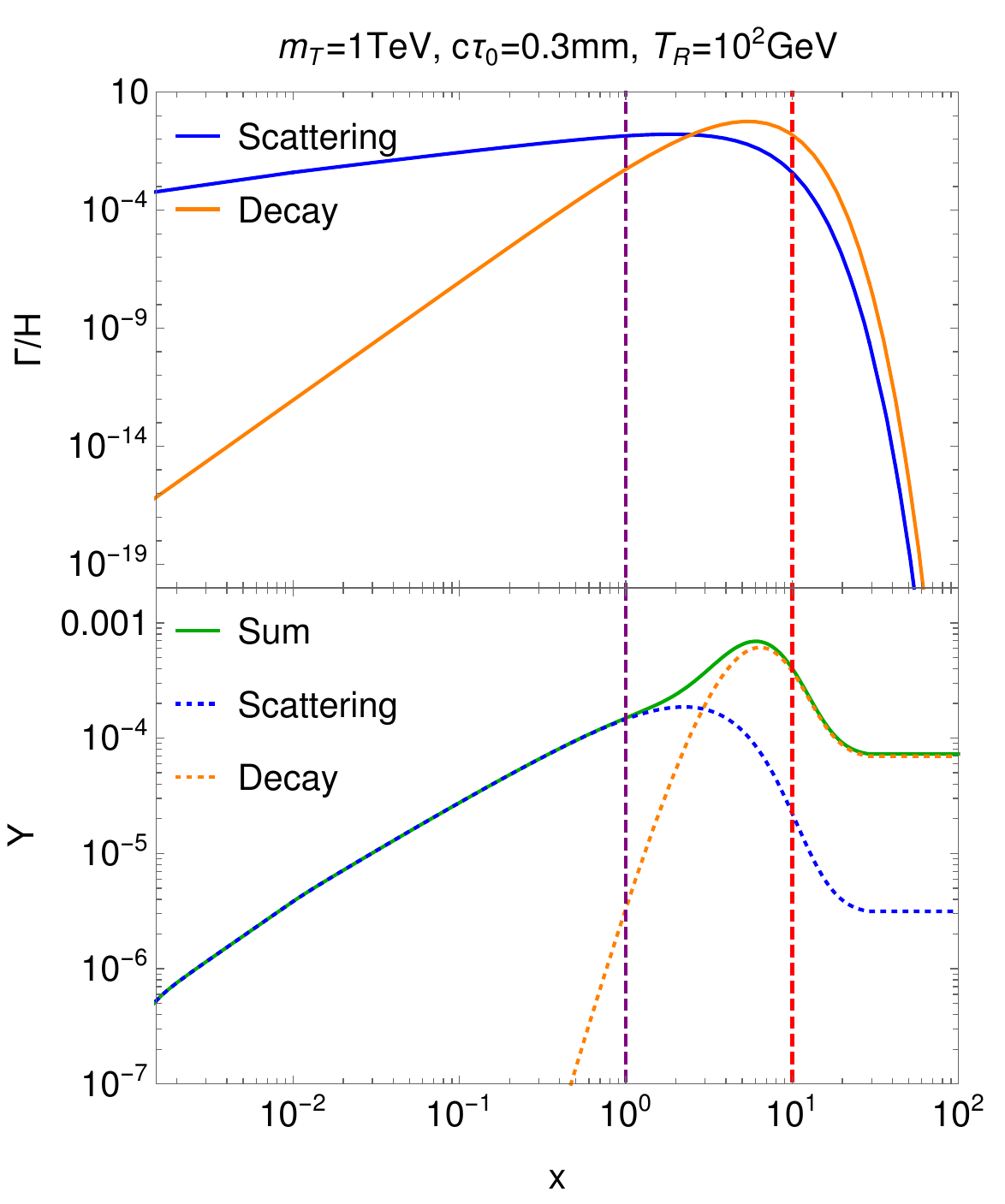}}
	\caption{Singlet-triplet model. Above: production rate/Hubble ratio for decay (blue) and scattering (orange). Below: DM yield. We consider $T_R>T_{FI}$ (left) and $T_{R}<T_{FI}$ (right). The dashed lines denote $T=T_{R}$ (red) and $T=m_T$ (purple) as a proxy for $T_{FI}$. For these parameters, the observed DM abundance is reproduced for $m_{\rm DM} = 10$~keV.}
        \label{fig:rates13}
\end{figure}
%%%%%%%%%%%%%%

Similarly to Figure~\ref{fig:rates}, we compare in the top panels of Figure~\ref{fig:rates13} the production rate from scatterings (in blue) to the one of the decays (in orange) normalized by the Hubble rate as a function of $x=m_T/T$. In the bottom panels, we plot the contributions of decay and scatterings (orange and blue dotted, respectively) to the total DM yield (continuous green). We fix $m_T=1$ TeV, and the prefactor  $1/\Lambda$ reproduces the relic density constraint for $m_\text{DM}=10$~keV; this gives a lifetime for the neutral mediator $\Psi_B^0$ of $c\tau_0=3$~m when $T_{R}\gg T_{FI}$ and $c\tau_0=0.3$~mm when $T_{R}< T_{FI}$. The main contribution to DM production in the left panel comes from scatterings consistently with our discussion in Section~\ref{sec:bound}. A larger reheating temperature implies an enhanced DM production, leading ultimately to a smaller DM-mediator coupling (i.e.~a longer mediator lifetime) needed to account for the relic abundance. In contrast, in the right panel of Figure~\ref{fig:rates13} we recover the yield dilution post freeze-in for $x>x_{FI}$ observed both for renormalizable and non-renormalizable operators in Figures~\ref{fig:FIearlyMD} and~\ref{fig:rates}. In this second case scattering processes play a subleading role.

\begin{figure}[t]
	\centering
	\subfloat{\includegraphics[width=0.48\textwidth]{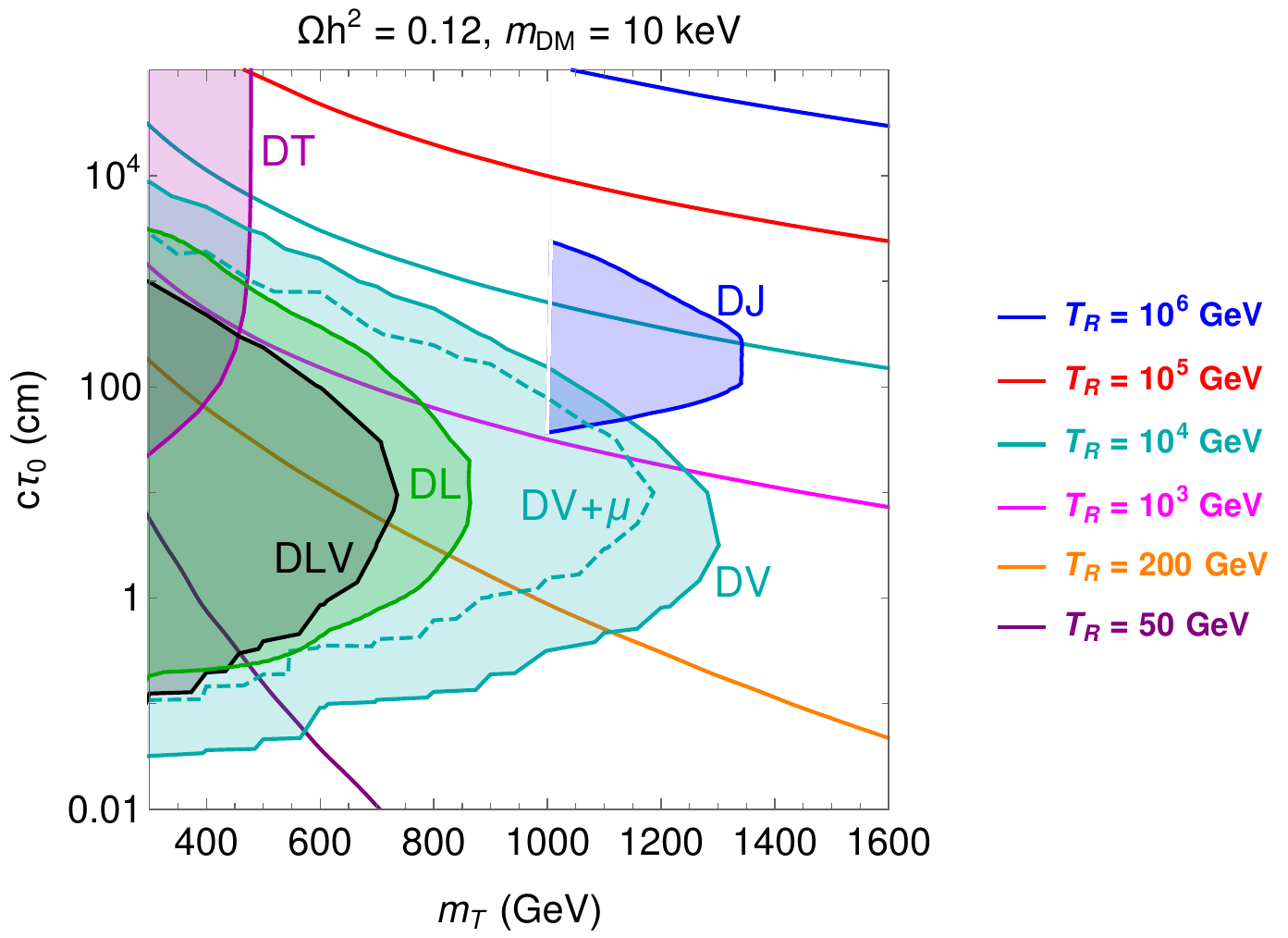}}
	\hfill	
	\subfloat{\includegraphics[width=0.48\textwidth]{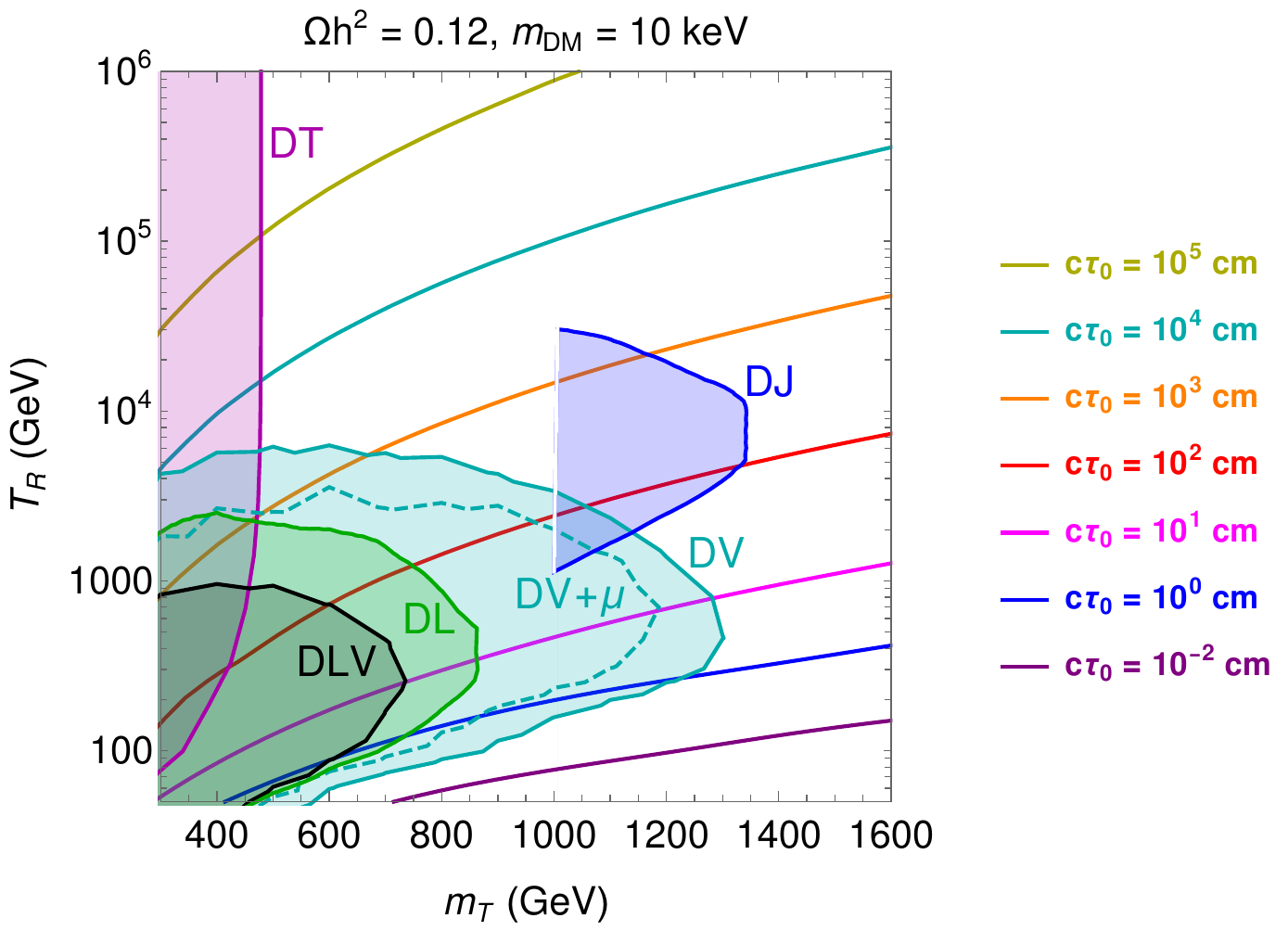}}
	\caption{Contours of the value of $T_R$ accounting for the whole DM relic abundance in the  $(m_T,\,c\tau_0)$
          plane (left), contours of the neutral mediator lifetime $c\tau_0$ in the $(m_T,\,T_R)$ plane (right), for the singlet-triplet DM model with $m_\chi=10$~keV. The value of $T_R$
            associated to a given $(m_T,c\tau_0)$ provides
            an upper bound on $T_R$ for any dark matter mass as
            $m_\chi \gtrsim 10~\keV$ due to small scale structures
            constraints. The colored areas are excluded by LHC searches for displaced vertices + MET (blue), displaced vertices + $\mu$ (dashed blue), delayed jets (dark blue), displaced lepton vertices (black), displaced leptons (green), and the disappearing tracks (purple). 
              }
	\label{fig:ExclRegion13}
\end{figure}

We consider again $m_\text{DM}=10$~keV, and we show in Figure~\ref{fig:ExclRegion13} contours of the values of $T_{R}$ giving the correct relic density in the  $(m_T,\,c\tau_0)$ plane (left panel) and $c\tau_0$ contours satisfying the same constraint in the $(m_T,\,T_R)$ plane (right panel). In the left panel, we see how higher $T_{R}$ induces larger values of the proper mediator lifetime without reaching the saturation effect at large $T_{R}$ that was observed in the case of renormalizable operators (see the top-right panel of Figure~\ref{fig:lepto} and Figure~\ref{fig:top}). Furthermore, $c\tau_0$ contours in the right panel never become $T_R$ independent. These are direct implications of the UV-dominated scatterings at work when $T_{R}\gg T_{FI}$ for non-renormalizable operators, requiring smaller couplings (increasing $c \tau_0$) when increasing~$T_{R}$.

\begin{figure}[t]
%missing
  \centering
	\subfloat[]{\includegraphics[width=0.45\textwidth]{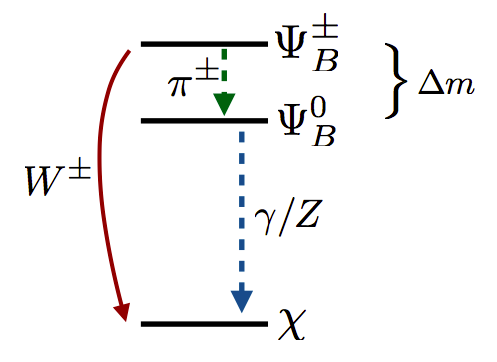}}
	\hfill	
	\subfloat[]{\includegraphics[width=0.5\textwidth]{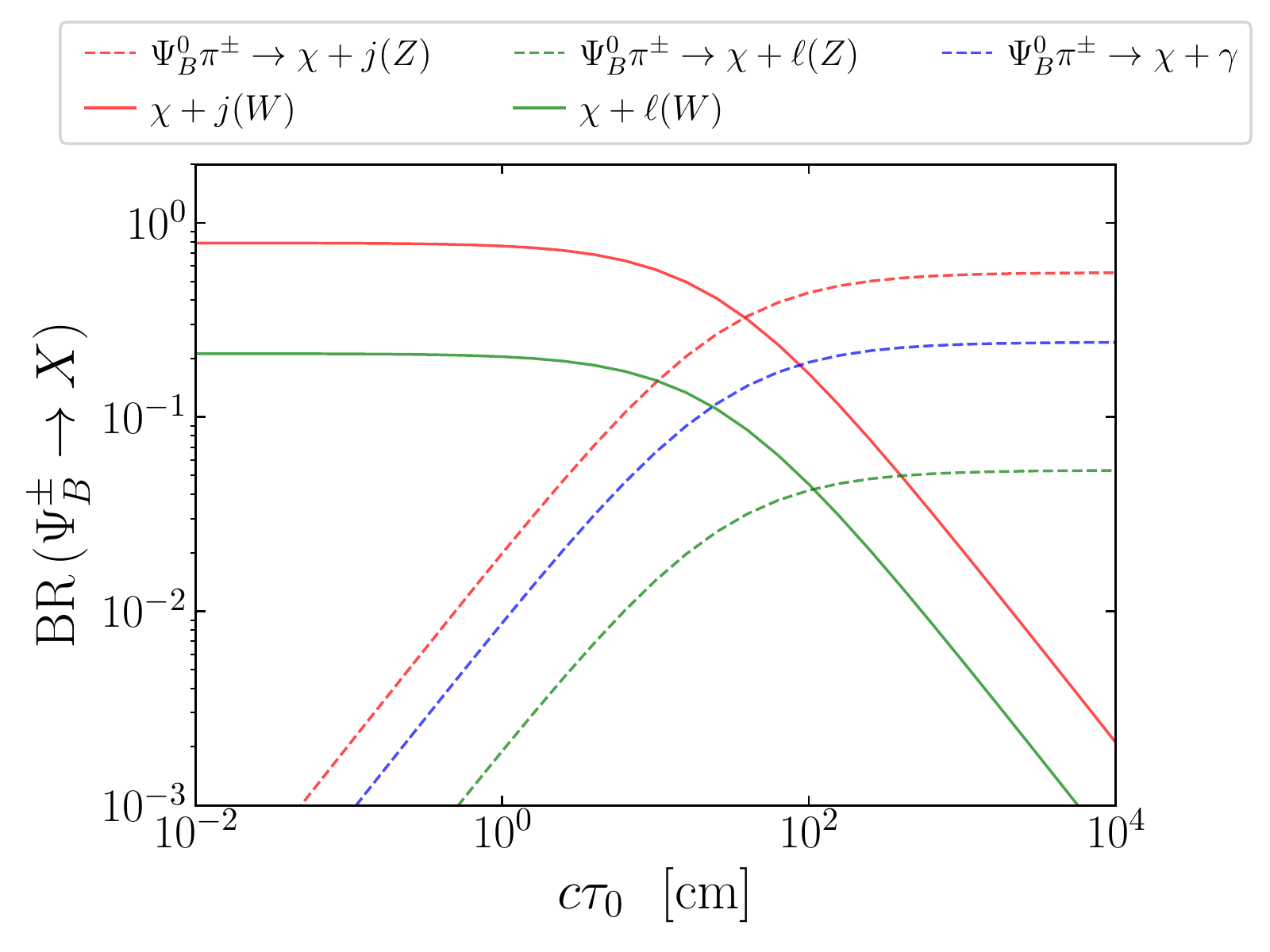}}
	\caption{Typical decay processes of the singlet-triplet model (left). Charged mediator branching ratio to the products listed in the legend as a function of the neutral mediator decay length $c\tau_0$ as a proxy for the scale $\Lambda$ for a mediator mass $m_T$ = 500 GeV (right).}
	\label{fig:spectrum}
\end{figure}

\paragraph{Collider constraints.} It is helpful to visualize the mass spectrum and the decay channels of the triplet components illustrated in Figure~\ref{fig:spectrum}. The neutral heavy fermion $\Psi_B^0$ can decay into DM emitting a $\gamma$ or a $Z$ boson through the dimension-five operator, and the decay width scaling as $\sim {m_T^3}/{\Lambda^2}$ is typically macroscopic and in the parameter space relevant to freeze-in. The heavy charged fermion can either decay into the neutral heavy fermion plus a soft pion through the exchange of a $W$ or directly into DM plus a $W$ boson. The first decay mode is purely due to gauge interactions, while the second one is due to the dimension-five operator, hence it is controlled by $\Lambda$. Different $\Psi_B^{\pm}$ decay products give different signatures at the LHC. In the right panel of Figure~\ref{fig:spectrum}, we illustrate how the different branching ratios (BR) of $\Psi_B^\pm$ decay products depend on the neutral mediator lifetime $c\tau_0$, which we take as a proxy for the scale $\Lambda$ (as $c\tau_0\propto \Lambda^2$). The $\Lambda$ parameter indeed drives the relative importance of the $\Psi^{\pm}_B$ decays induced by the dimension-five operator relative to the gauge-induced decays. The decay length of the $\Psi_B^{\pm}$ is at most few cm, when the decay is dominated by the gauge-induced interactions. Shorter lifetimes can be obtained only when $\Lambda$ is small such that the decay channels into DM dominate.

%%%%%%%%%%%%%
%%%%%%%%%%%%%%%%%

We distinguish in Figure~\ref{fig:spectrum} between the hadronic and leptonic subsequent decays of the $Z$ and $W$ bosons. For small $c\tau_0$ (small $\Lambda$), $\Psi_B^{\pm}$ principally decays into $\chi W^{\pm}$. For larger $c \tau_0$ (larger $\Lambda$), $\Psi_B^{\pm}$
mainly decay into a soft $\pi^\pm$ and a heavy neutral fermion $\Psi_B^0$ which in turn decays to $\chi$ + jets, leptons or $\gamma$. Let us also emphasize that, at the LHC, mediator pair production involves the production of charged fermions $\Psi_B^{+} \Psi_B^{-}$ or the associated production of the $\Psi_B^{\pm} \Psi_B^0$ states, in all cases through s-channel electroweak bosons exchange.\footnote{There is no vertex involving a SM gauge boson and a pair of neutral heavy fermions $\Psi_B^0$.}

We denote with colored areas in Figure~\ref{fig:ExclRegion13} the regions excluded by LLP searches at LHC. For large values of the $\Psi_B^0$ lifetime, the strongest bound come from searches for disappearing tracks (DT)~\cite{Aaboud:2017mpt,CMS:2018ail} excluding the purple region up to $m_C\simeq m_T \lesssim 480$~GeV. This topology is relevant when producing a charged $\Psi_B^\pm$ decaying to $\pi^\pm\Psi_B^0$ with the pion being too soft to be detected and
$\Psi_B^0$ too long-lived to decay inside the detector. If instead $\Psi_B^0$ decays inside the detector, into DM plus a $Z$ or a photon, DT searches can be no longer sensitive since extra hits could be recorded in other parts of the detector. In order to avoid this issue, we conservatively require that the decay of the neutral component $\Psi_B^0$ occurs outside the tracker for the DT search to be applicable (see Appendix~\ref{sec:def-DT} for details). This is the reason why DT searches appear to be sensitive only for $c\tau_0\gtrsim 1$~m.  In the plots of Figure~\ref{fig:ExclRegion13}, we show the DT exclusion region following from the ATLAS search. Indeed, ATLAS performs better than CMS for small decay lengths (see the discussion in Appendix~\ref{sec:def-DT}) which seems to be more relevant for the model under consideration.

%%%%%%%%%%%%%%
%%%%%%%%%%%%%%

Other signatures arise when the neutral component of the triplet decays inside the detector, or when the charged component decays directly to the singlet and a $W$ boson.\footnote{The charged component will only decay to the singlet when $\Lambda$ (equivalently, $c\tau_0$) is small enough that $\Gamma_{\Psi_B^\pm \rightarrow \chi W^\pm} > \Gamma_{\Psi_B^\pm \rightarrow \Psi_B^0 \pi^\pm}$. This predominantly happen inside the detector, see the right panel of
  Fig.~\ref{fig:spectrum}.} If the $Z$ or $W$ boson decay hadronically, a displaced vertex (DV) or delayed jet (DJ) + MET signature could be observed. The heavy neutral $\Psi_B^0$ is in some cases produced in the decay chain of charged $\Psi_B^\pm$, which is itself long lived (with a decay length of at most few cm), together with a soft pion (see Figure~\ref{fig:spectrum}). In this case, we assume that $\Psi_B^0$ is emitted in the same direction as $\Psi_B^\pm$, and hence we add the two displacements to define the total displacement of the resulting gauge boson. We consider all combinations of production and decay modes (with the corresponding branching ratios) to obtain the final states leading to displaced jets plus missing energy. Final states always contain two DM particles, consistently with $Z_2$ conservation, and two gauge bosons which can be either $W$ or $Z/\gamma$ (plus possibly soft pions). The relevant decay chains for these searches are the ones leading to jets (i.e.~the hadronic decays of $W$ or $Z$). The hadronic BR of $\Psi_B^{\pm}$, as shown in Figure~\ref{fig:spectrum}, depends on the neutral component decay length. The BR of $\Psi_B^0$ into hadronic final states is instead independent on the value of its decay length. In Figure~\ref{fig:ExclRegion13}, the regions excluded by the DV~\cite{Aaboud:2017iio} and DJ \cite{Sirunyan:2019gut} searches are shown in light and dark blue, respectively.  For moderate values of the lifetime ($c\tau_0 \lesssim 10$~m) they put strong bounds on the mass of the triplet, excluding up to $m_T \lesssim 1.3$~TeV.

The other LLP signatures shown in Table~\ref{tab:models-searches} arise in this model. Leptonic decays of the $Z$ and $W$ boson lead to a rich variety of topologies, including displaced vertices + $\mu$~\cite{Aad:2020srt}, displaced lepton vertices~\cite{Aad:2019tcc} and displaced leptons~\cite{Khachatryan:2014mea,CMS:2016isf,Aad:2020bay}. However, because of the comparatively small leptonic BR of the electroweak gauge boson, we expect these searches to be less constraining that the ones targeting hadronic final states plus missing energy.  From Figure \ref{fig:ExclRegion13} we observe for instance that the displaced jets+muon search is once again slightly less constraining than the DV search, as already happened in the topphilic model.  In Figure~\ref{fig:ExclRegion13} we show also some representative sensitivity curves for purely leptonic LHC searches.
In uniform green we display the reach of the recent DL ATLAS analysis~\cite{Aad:2020bay}. Note that this search targets same-flavour ($e^{\pm} e^{\mp}$ and $\mu^{\pm} \mu^{\mp}$) as well as different flavours ($e^{\pm} \mu^{\mp}$) final states,  which are all possible topologies for the singlet-triplet model (from leptonically decaying $W$ and/or $Z$). By combining all these channels we obtain the sensitivity in Figure~\ref{fig:ExclRegion13}. The reach of this search cannot overcome the search focussing on displaced jets+MET, but it provides nevertheless a  significant coverage of the parameter space of the model. For completeness we also
show the reach of the the displaced lepton vertex search~\cite{Aad:2019tcc} (black region denoted as DLV in Figure \ref{fig:ExclRegion13}), see Appendix~\ref{sec:def-DLV} for details. This search requires the two leptons to point to the same displaced vertex, hence effectively targets displaced $Z$ for our model. 

We have already discussed for the previous simplified models how the large luminosity of HL-LHC can further increase the mass reach of the LLP searches. Here we would like instead to emphasise the impact on this model
of future detectors targeting particles with large ($\mathcal{O}(10-100)$ m) decay lengths, such as MATHUSLA~\cite{Curtin:2018mvb} and CODEX-b~\cite{Aielli:2019ivi}. First, in the singlet-triplet model the long-lived particle is neutral and can naturally reach the far detector,
hence we expect that these facilities will significantly probe the large $c\tau$ region
(a dedicated investigation as the one performed for the singlet-doublet model in Ref.~\cite{No:2019gvl} is left for future works).
Second, the freeze-in DM production is UV sensitive. As a consequence, also for very small
couplings (corresponding to large $c{\tau}$) the value of $T_R$ is set by the relic abundance constraint, as it can be seen by the contours in the left panel of Figure~\ref{fig:ExclRegion13}, and thus one can infer an absolute upper bound on the allowed $T_R$.
For instance, as we can see from the figure, observing a signal of a $1$~TeV mediator with lifetime $c{\tau} \simeq 10^4$~cm would imply, in the singlet-triplet model, an upper bound on $T_R$ of approximately $T_R \lesssim 10^5$~GeV.
This would have relevant implications for well-motivated cosmological scenarios such as leptogenesis and for models of inflation.

Finally, comparing the results in Figure~\ref{fig:ExclRegion13} with those for
the other models in Figures~\ref{fig:lepto} and ~\ref{fig:top}, one
can see how the relative sensitivity of different LPP searches differs
in the three scenarios. This would provide an important handle to
discriminate among possible models, in the (lucky) case of the
observation of a pattern of signals from different LLP
searches. The problem of determining the theoretical
  model underlying the LLP production is quite generic. For
  significant progress in this direction, see
  e.g.~\cite{Barron:2020kfo}.

\section{Summary and conclusions}
\label{sec:conclusion}

Decades after its first convincing gravitational hints, we are still not aware about the microscopic particle identity of DM. This is a serious unresolved issue in particle physics and cosmology, and one of the most urgent questions to answer in fundamental interactions. When we look at a younger universe by detecting photons from distant objects, there is a limit on how much we can travel back in time. The universe was opaque to electromagnetic radiation before recombination, which happened when its age was approximately 380,000 years, and the CMB curtain hides whatever was behind. We can still look at earlier times by observing the BBN snapshot that features a radiation dominated universe only one second old. Looking at even earlier times is perhaps possible by observing spectral features in primordial gravitational waves, 
see e.g.~Refs.~\cite{Nakayama:2008wy,Kuroyanagi:2011fy,Cui:2018rwi,DEramo:2019tit,Bernal:2019lpc,Arias:2019uol}, but at the moment we have no information about the pre-BBN universe. The detection of displaced events at colliders with missing energy in the final state can provide hints about both these open questions. 

Focusing on the concrete framework illustrated in Figure \ref{fig:FI}, we have investigated the interplay between collider physics and early universe cosmology. We have considered DM particles with feeble couplings to the visible sector through the  three body interaction in Figure \ref{fig:FI} where $B$ is a new BSM particle in thermal equilibrium at early times with the primordial bath. These scenarios are not accessible at direct and indirect DM detection experiments, and they are quite a nightmare for experimentalists. Nevertheless, they can give rise to spectacular signals at colliders where the bath particle $B$ is pair produced and decays with macroscopic lifetime into DM. In the early universe, DM particles are produced through the freeze-in production mechanism (via decays or scattering). The decay length of the bath particle $B$, which one can measure from the displaced events at particle colliders, is directly linked to the DM relic abundance.

We have scrutinized this link with the aim to provide a deeper connection between the DM LLP signature and the cosmological history influencing the DM production. First, we have reviewed DM freeze-in production for a standard early universe with its energy budget dominated by a thermal bath of relativistic particles. The resulting prediction for the $B$ lifetime is typically too long to lead to displaced signatures at colliders unless the DM is significantly light (few keV). Then we have discussed how a non-standard early universe history can affect the predictions for DM abundance. We have focused in particular on an early MD era resulting from a late inflationary reheating of the universe. Low values of the reheating temperature $T_{R}$, as a consequence of the dilution of the DM relic density due to entropy release from inflaton decays, allow for higher values of the DM coupling to the SM and open the possibility to probe a larger part of the parameter space with LLP searches. We have improved previous studies in literature by going beyond the instantaneous reheating approximation and by considering both IR- and UV-dominated freeze-in. Remarkably, under certain generic assumptions, we have shown that DM LLP signatures can imply an upper bound on the reheating temperature of our Universe.

We have provided in Table~\ref{tab:classification} a systematic classification of simplified models giving rise to the cubic interaction displayed in Figure~\ref{fig:FI}. After reviewing the relevant LLP searches within this context, we provided an in-depth study of the viable parameter space of three selected models that cover a wide spectrum of collider signatures (see Table~\ref{tab:models-searches}). The first two involve renormalizable operators with DM preferably coupling to muons and top quarks respectively, while the last one features a non-renormalizable operator with preferred DM coupling to the EW gauge bosons. Processes driving freeze-in production are IR-dominated when renormalizable operators are involved and/or freeze-in mainly occurs via decays. Instead, when non-renormalizable operators are responsible for the three body interaction, freeze-in become UV-dominated for $T_{R}\gg T_{FI}$ with scatterings playing a major role. 

The interplay between DM relic abundance, affected by $T_{R}$, and LLP searches is shown in Figures~\ref{fig:lepto},~\ref{fig:top} and~\ref{fig:ExclRegion13}. Our study highlights the model discriminating power of the LLP searches that have been performed by LHC collaborations. While the minimal models of Table~\ref{tab:classification} should not be regarded as ``realistic'' SM extensions, they still serve to illustrate how the underlying new physics could be identified in presence of a pattern of observed exotic signatures (cf.~Table~\ref{tab:models-searches}). They also represent useful building blocks that a complete theory may need to incorporate to account for signals for LLP at colliders.

A full reconstruction of the displaced events kinematics is highly unlikely. Notwithstanding the fact that we will probably be unable to measure the DM mass from these events, we have shown how displaced missing energy can still provide hints about the early universe. By setting the DM mass to the lowest value allowed by the Lyman-$\alpha$ forest bounds and imposing the relic density constraint, we have argued that one can still derive a generic {\it upper bound} on $T_{R}$ if the $B$ mass and mediator lifetime could be measured at colliders by means of LLP searches. Colliders could thus indirectly provide a constrain on the early universe cosmology. 

While we have focused on an early period of MD following inflationary reheating, our strategy can be applied to other scenarios with different early cosmological histories. One motivated example is the one of a ``fast-expanding'' universe~\cite{DEramo:2017gpl} where the Hubble parameter has a stronger power-law temperature dependence, as it is the case for a kination phase~\cite{DEramo:2017ecx,Redmond:2017tja,Visinelli:2017qga,Biswas:2018iny}. 

Our investigation highlights the importance of DM LLP signatures at colliders as a powerful strategy to search for feebly interacting DM. Indeed, beyond their promising discovery prospects in view of the HL-LHC, they can also represent indirect probes of the cosmological history of our Universe.

\acknowledgments We would like to thank F.~Blekman, S.~Lowette and S.~Pagan Griso for clarifications on collider searches, B.~Zaldivar for valuable discussions on FI computations and Q.~Decant, A.~Lenoci, D.C.~Hooper and R.~Murgia for clarifications on Lyman-$\alpha$ constraints on FIMPs and cross-checks. LC~is partially supported by the National Natural Science Foundation of China under the grant No.~12035008. The work of FD~is supported by the research grants: ``The Dark Universe: A Synergic Multi-messenger Approach'' number 2017X7X85K under the program PRIN 2017 funded by the Ministero dell'Istruzione, Universit\`a e della Ricerca (MIUR); ``New Theoretical Tools for Axion Cosmology'' under the Supporting TAlent in ReSearch@University of Padova (STARS@UNIPD); ``New Theoretical Tools to Look at the Invisible Universe'' funded by the University of Padua. FD~is also supported by Istituto Nazionale di Fisica Nucleare (INFN) through the Theoretical Astroparticle Physics (TAsP) project. FD~acknowledges support from the European Union's Horizon 2020 research and innovation programme under the Marie Sk\l odowska-Curie grant agreement No 860881-HIDDeN. LLH is a Research associate of the Fonds de la Recherche Scientifique FRS-FNRS. LLH and SJ work are supported by the FNRS research grant number F.4520.19.  AM and SJ are supported by the Strategic Research Program High-Energy Physics and the Research Council of the Vrije Universiteit Brussel, and by the ``Excellence of Science - EOS'' - be.h project n.30820817.

%%%%%%%%%%%%%%%%%%%%%%%%%%%%%%%%%%%%%%%%%%%%%%%%%%%%%%%%%%%%%%%%%%%%%%
\appendix
%%%%%%%%%%%%%%%%%%%%%%%%%%%%%%%%%%%%%%%%%%%%%%%%%%%%%%%%%%%%%%%%%%%%%%

%%%%%%%%%%%%%%%%%%%%% APP COSMOLOGY %%%%%%%%%%%%%%%%%%%%%
\section{Cosmological Histories}
\label{app:cosmology}

We summarize in this appendix the key features of the two cosmological histories analyzed in this paper: a standard radiation dominated universe and inflationary reheating. The expanding universe is described by the Friedmann-Lemaitre-Robertson-Walker (FLRW) metric where physical length scales are proportional to the scale factor $a(t)$. The Hubble parameter $H(t)$ quantifies the expansion rate and it is set by the energy content via the Friedmann equation
\begin{equation}
H(t) \equiv \frac{\dot{a}(t)}{a(t)} = \frac{\sqrt{\rho_{\rm tot}}}{\sqrt{3} \, M_{\rm Pl}} \ .
\label{eq:Friedmann}
\end{equation}
Here, we employ the reduced Planck mass $M_{\rm Pl} = (8 \pi G)^{-1/2} = 2.4 \times 10^{18} \, {\rm GeV}$. The species contributing to the total energy density $\rho_{\rm tot}$ depend on the cosmological history under consideration. Consequently, the explicit functional form of the Hubble parameter in terms of the cosmic time $t$ also depends on the cosmological history. 

\subsection{Radiation dominated universe}

Once we consider a thermal bath of relativistic particles dominating the early universe, the energy density can be conveniently expressed as follows
\begin{equation}
\rho_R = \frac{\pi^2}{30} g_*(T) T^4 \ .
\label{eq:rhoRvsT}
\end{equation}
Here, $T$ is the bath temperature and $g_*(T)$ accounts for the effective number of relativistic degrees of freedom in the primordial plasma. All SM particles are relativistic for temperatures much higher than the weak scale, and this gives a contribution $g_*^{\textsc{sm}} = 106.75$. On the contrary, at temperature as low as the BBN epoch ( $T_{\rm BBN} \simeq {\rm fews \, MeV}$, for a recent reassessment see~\cite{Hasegawa:2019jsa}) we have $g_*^{\rm BBN} = 10.75$. Thus the effective number of relativistic degrees of freedom changes approximately by a factor of $10$ in the temperature range of interest. Another important quantity describing the thermal bath is its entropy density
\begin{equation}
s = \frac{2 \pi^2}{45} g_{*s}(T) T^3 \ ,
\end{equation}
where $g_{*s}(T)$ accounts for the effective number of degrees of freedom contributing to the entropy density. In our analysis we can safely take $g_*(T) \simeq g_{*s}(T)$.

The entropy in a comoving volume, $S = s a^3$, is a conserved quantity for such a radiation dominated (RD) universe. This is not the case anymore once we consider inflationary reheating because inflaton decays dump entropy in the radiation bath. The conservation of entropy, $dS / dt = 0$, provides us with the useful relation
\begin{equation}
\frac{d T}{d t} = - \frac{H \, T}{1 + \frac{1}{3} \frac{\partial \ln g_{*s}}{\partial \ln T}}  \ .
\label{eq:Tvst}
\end{equation}
It is important to remember that it is valid only if entropy is conserved.

\subsection{Inflationary reheating}
\label{app:reheating}

Once the inflationary phase is over, the inflaton field oscillates around its minimum and it decays to radiation, a process known as reheating. The intricate physics of reheating is accounted for by the compact set of coupled Boltzmann equations
\begin{align}
\label{eq:BEback1} \frac{d \rho_M}{dt} + 3 H \rho_M  = & \, - \Gamma_M \rho_M \ , \\ 
\label{eq:BEback2} \frac{d \rho_{\rm R}}{dt} + 4 H \rho_{\rm R} = & \, \Gamma_M \rho_M \ .
\end{align}
Here, $\rho_M$ is the inflaton energy density and $\Gamma_M$ its decay width. The factor of $3 H$ in Eq.~\eqref{eq:BEback1} shows how the energy density stored in the inflaton oscillations evolves as non-relativistic matter (as opposed the factor of $4 H$ in Eq.~\eqref{eq:BEback2}). Our universe undergoes an early matter dominated (MD) epoch. Strictly speaking, Eq.~\eqref{eq:BEback2} is only valid if $g_* = g_{*s} = {\rm const}$; for the values mentioned above, the largest error we can potentially make is approximately $10 \%$. As we discuss in our work for each case, the error due to this approximation is much smaller in our study and thus we safely employ it. 

We determine the cosmological background by solving numerically the Boltzmann equation system in our analysis. In this Appendix, we provide an approximate analytical solution that allows us to understand the behavior of the energy densities as a function of time. We start solving the system at the time $t_{\rm in}$ when the inflationary phase is terminated and the initial radiation energy density is vanishing, $\rho_R(t_{\rm in}) = 0$; even if there was any radiation after the Big Bang, it would have been inflated away. On the contrary, the initial inflaton energy density is equal to $\rho_M(t_{\rm in}) = E_I^4$ with $E_I$ the energy scale of inflation. 

If we consider our universe at times much earlier than the reheating lifetime, $t\ll \Gamma_M^{-1}$, the inflaton energy density just red-shifts as the inverse cubic power of the scale factor
\begin{equation}
\rho_M \simeq \rho_M(t_{\rm in}) \, \left(\frac{a_{\rm in}}{a} \right)^3 \ ,
\end{equation}
with $a_{\rm in}$ is the value of the FLRW scale factor at the initial time $t_{\rm in}$. Meanwhile, the radiation bath begins being populated by inflaton decays and the approximate solution for the its energy density during this initial reheating phase reads~\cite{Co:2015pka}
\begin{equation}
\rho_R \simeq \frac{2 \sqrt{3}}{5} M_{\rm Pl} \Gamma_M \sqrt{\rho_M(t_{\rm in})} \left[\left(\frac{a_{\rm in}}{a} \right)^{3/2} - \left(\frac{a_{\rm in}}{a} \right)^4 \right] \ .
\label{eq:rhoRapprox}
\end{equation}
The radiation temperature is connected to its energy density via Eq.~\eqref{eq:rhoRvsT}. 

Initially, there is not radiation at all and the initial growth is controlled by the second term in the square brackets of Eq.~\eqref{eq:rhoRapprox}. The radiation energy density and its temperature reach a maximum for a value of the scale factor $a_\textsc{max} = ( 8/3)^{2/5} \, a_{\rm in}$. Neglecting the $g_*$ dependence on the temperature, the associated maximum temperature results in
\begin{equation}
T_\textsc{max} \simeq \frac{\left(M_{\rm Pl} \Gamma_M \right)^{1/4} \rho_M(t_{\rm in})^{1/8}}{g_*^{1/4}} \simeq \frac{\left(M_{\rm Pl} \Gamma_M E_I^2 \right)^{1/4}}{g_*^{1/4}}  \ .
\end{equation}
Afterwards, the decrease of the radiation energy density and temperature is controlled by the first term in the square brackets of Eq.~\eqref{eq:rhoRapprox}. The radiation energy density features the peculiar dependence $\rho_R \propto a^{-3/2}$; this has to be compared with the adiabatic expansion for a RD universe where the energy density decrease is faster, $\rho_R \propto a^{-4}$. Likewise, we notice that the temperature behaves as $T \propto a^{-3/8}$ when the universe expands and cools down; for a RD universe the temperature decreases with the scale factor is faster, $T \propto a^{-1}$. 

The approximation that lead us to Eq.~\eqref{eq:rhoRapprox} breaks down when most inflatons decays and the energy budget starts being dominated by the radiation bath: we recover the RD universe that serves as a background for BBN. The transition to a RD universe happens when the temperature of the thermal bath has a value $T_R$ satisfying the condition 
\begin{equation}
H(T_R) \simeq \Gamma_M \simeq \frac{\rho_R(T_R)^{1/2}}{\sqrt{3} M_{\rm Pl}} \ .
\end{equation}
The so called reheating temperature reads
\begin{equation}
T_ R = \left(\frac{90}{\pi^2 g_*}\right)^{1/4}\sqrt{\Gamma_M M_{\rm Pl}} \ .
\label{eq:TRH}
\end{equation}
The radiation temperature spans a potentially large range from $T_\textsc{max}$ down to $T_R$, and they are connected via the relation $T_\textsc{max} \simeq \left(T_R E_I \right)^{1/2}$.

%%%%%%%%%%%%%%%%%%%%% APP BOLTZMANN %%%%%%%%%%%%%%%%%%%%%
\section{Boltzmann equation for DM production}
\label{app:FI}

The general form of the Boltzmann equation tracking the DM number density reads
\begin{equation}
\frac{d n_X}{d t} + 3 H n_X = \sum_\alpha \mathcal{C}_\alpha \ .
\label{eq:BE}
\end{equation}
Besides the dilution due to the expansion, an effect controlled by the Hubble parameter $H$, the number density $n_X$ can change because of processes altering the number of $X$ particles. This effect is accounted for by collision operator $\mathcal{C}_\alpha$, and the index $\alpha$ runs over all possible processes. Here, we provide the collision operators relevant to our analysis. 

Although our work focuses on $B$ decays to a 2-body final state, we present the collision operator for the more general case of n-body decays
\begin{equation}
B \; \rightarrow \; X \, B_2 \, \ldots \, B_n \ .
\end{equation}
Here, $B_i$ denotes a generic particle belonging to the primordial thermal bath (either a SM or a BSM degree of freedom). The associated collision operator reads
\begin{equation}
\mathcal{C}_{B \rightarrow  X  B_2  \ldots B_n} = n_B^{\rm eq} \; \Gamma_{B \rightarrow  X  B_2 \ldots B_n} \; \frac{K_1[m_B / T]}{K_2[m_B / T]} \ ,
\label{eq:Cdecay}
\end{equation}
where $\Gamma_{B \rightarrow  X  B_2 \ldots B_n}$ is the partial decay width and $K_{1,2}$ are modified Bessel functions. The equilibrium number density for $B$ can be obtained upon the phase space integration
\begin{equation}
n_B^{\rm eq} = g_B \int \frac{d^3 p}{(2 \pi)^3} f_B^{\rm eq} = \frac{g_B}{2 \pi^2} m_B^2 T K_2[m_B / T] \ ,
\end{equation}
with $g_B$ the internal degrees of freedom (spin, color, etc.) of the bath particle. The last equality holds for a Maxwell-Boltzmann statistics, $f_B^{\rm eq} = \exp[ - E_B / T]$, and it leads to an exponential suppression in the non-relativistic regime ($T \ll m_B$).

For a generic binary collision
\begin{equation}
B_1 \, B_2 \; \rightarrow \; B_3 \, X \ ,
\end{equation}
where as usual $B_i$ denotes particles in thermal equilibrium at early times, the collision operator reads
\begin{equation}
\mathcal{C}_{B_1 B_2 \rightarrow B_3 X} = \frac{g_{B_1} g_{B_2}}{32 \pi^4} T^4 \, 
\int_{s_{\rm min}}^\infty ds \frac{\lambda(s, m_{B_1}, m_{B_2})}{s^{1/2}} \, \sigma_{B_1 B_2 \rightarrow B_3 X}(s) \, K_1[\sqrt{s} / T] \ .
\label{eq:Cscattering}
\end{equation}
Here, $\sigma$ is the Lorentz invariant scattering cross section as a function of the (squared of the) energy $s$ in the center of mass frame. The lower integration extreme is
\begin{equation}
s_{\rm min} = (m_{B_1} + m_{B_2})^2 \ ,
\end{equation} 
and is set by a kinematical threshold, whereas the function $\lambda$ is defines as follows
\begin{equation}
\lambda(x,y,z) \equiv [x - (y + z)^2] [x - (y - z)^2] \ .
\end{equation}

\subsection{FIMP production during a RD era}

Besides decays and scattering, the $X$ number density gets diluted as $n_X \propto a^{-3}$ because the universe is expanding. At the same time, the entropy in a comoving volume, $S = s a^3$, is conserved for a RD universe. Thus in the absence of processes changing the number of $X$'s, the comoving number density 
\begin{equation}
Y_X \equiv \frac{n_X}{s} 
\end{equation}
is a conserved quantity. Such a variable is convenient because it scales out the effect of the expansion, and it changes over time only if there are number changing processes. 

For a RD universe, the Boltzmann equation in Eq.~\eqref{eq:BE} becomes 
\begin{equation}
\frac{d Y_X}{d t} = \frac{1}{s} \sum_\alpha \mathcal{C}_\alpha \ .
\end{equation}
We trade the time variable $t$ with the thermal bath temperature $T$ (see Eq.~\eqref{eq:Tvst})
\begin{equation}
\frac{d Y_X}{d \ln T} = - \left( 1 + \frac{1}{3} \frac{\partial \ln g_{*s}}{\partial \ln T} \right) \frac{1}{H \, s} \sum_\alpha \mathcal{C}_\alpha \ .
\end{equation}
It is convenient to introduce a dimensionless ``time variable'' $x = M / T$ to solve the equation numerically. The overall mass scale $M$ is purely conventional. For production via decays it is advantageous to set it to the decaying particle mass $m_B$, whereas for production via scattering it is practical to set it to the heaviest mass involved in the process. Regardless of the specific choice for $M$, the Boltzmann equation reads
\begin{equation}
\frac{d Y_X}{d \ln x} =  \left( 1 - \frac{1}{3} \frac{\partial \ln g_{*s}}{\partial \ln x} \right) \frac{1}{H \, s} \sum_\alpha \mathcal{C}_\alpha  \ .
\end{equation}
This expression, written in terms of only dimensionless quantities, is the most general equation describing DM production during RD. We solve it numerically in our analysis.

For a 2-body decay, with collision operator from Eq.~\eqref{eq:Cdecay}, the equation reads
\begin{equation}
\frac{d Y_X}{d \ln x} =  \left( 1 - \frac{1}{3} \frac{\partial \ln g_{*s}}{\partial \ln x} \right) 
\frac{\Gamma_{B \rightarrow   A_\textsc{sm} X}}{H} \, Y_B^{\rm eq}  \, \frac{K_1[x]}{K_2[x]}  \ .
\end{equation}
We set the appropriate initial condition, $Y_X(x=0) = 0$, and we solve this differential equation by just performing an integral. The asymptotic value of the comoving number density, which we take as its value for very large $x$ (i.e., small temperatures), results in
\begin{equation}
Y_X^\infty = \int_0^\infty \frac{d x}{x} \left( 1 - \frac{1}{3} \frac{\partial \ln g_{*s}}{\partial \ln x} \right) 
\frac{\Gamma_{B \rightarrow   A_\textsc{sm} X}}{H} Y_B^{\rm eq} \frac{K_1[x]}{K_2[x]} \ .
\end{equation}
The integral can be computed analytically if we set $g_* \simeq g_{*s} \simeq {\rm const}$ so that the number of relativistic degrees of freedom do not vary during the DM production era (see Refs.~\cite{Belanger:2018mqt,Belanger:2020npe} for relaxing this assumption). We set them at the time when freeze-in is mostly efficient, namely when $T\simeq m_B$ ($x \simeq 1$), and we find
\begin{equation}
Y_X^\infty \simeq \frac{405 \sqrt{5}}{4 \sqrt{2} \, \pi^4} \frac{g_B}{g_{*s} g_*^{1/2}} \frac{\Gamma_{B \rightarrow   A_\textsc{sm} X} M_{\rm Pl}}{m_B^2} \ .
\label{eq:YFIasymptotic}
\end{equation}

For the case of FIMP production via scattering, with collision operator given in Eq.~\eqref{eq:Cscattering}, we need to know how the cross section depends on the Mandelstam variable $s$.

Finally, we can evaluate the contribution to the $\Omega$ parameter defined as follows
\be
\Omega_X h^2 \equiv \frac{\rho_X}{\rho_{\rm cr} / h^2} = \frac{m_X Y_X^\infty s_0}{\rho_{\rm cr} / h^2} \ ,
\ee
where we define the critical density $\rho_{\rm cr} = 3 H_0^2 M_{\rm Pl}^2$ in terms of the current value of the Hubble parameter expressed in the conventional form $H_0 = 100 h \, {\rm km} \, {\rm s}^{-1} \, {\rm Mpc}^{-1}$. 

\subsection{FIMP production during an early MD era}

The optimal numerical technique to solve the Boltzmann equation is different if freeze-in happens during an early MD era. The starting point is still Eq.~\eqref{eq:BE}, but using the comoving number density is rather inconvenient. Most of FIMPs are produced around the freeze-in temperature $T_{FI} > T_R$, and their number density dilutes afterwards with the scale factor as $n_X \propto a^{-3}$. At the same time, as long as the thermal bath temperature is larger than $T_R$, inflaton decays dump entropy in the primordial bath and the entropy in a comoving volume $s a^3$ grows with time. As a consequence, the comoving number density $Y_\chi$ decreases with time when the bath temperature spans between $T_{FI}$ and $T_R$. 

This suggests a new variable accounting for the number density of DM particles
\begin{equation}
\mathcal{X} \equiv n_X a^3 \ .
\end{equation}
Once freeze-in productions runs its course, the quantity $\mathcal{X}$ remains constant until the present time. The Boltzmann equation tracking the FIMP number density becomes
\begin{equation}
\frac{d \mathcal{X}}{d t} = a^3 \sum_\alpha \mathcal{C}_\alpha \ .
\label{eq:BEMD}
\end{equation}
The cosmic time $t$ is not an ideal evolution variable, the scale factor is quite convenient instead. We trade time derivative with derivative with respect to $a$ by using the Friedmann equation in Eq.~\eqref{eq:Friedmann}, and we find
\begin{equation}
\frac{d \mathcal{X}}{d \ln a} = \frac{\sqrt{3} M_{\rm Pl} \, a^3}{\sqrt{\rho_M(a) + \rho_R(a)}} \sum_\alpha \mathcal{C}_\alpha(a) \ .
\end{equation}
The inflaton and radiation energy density as a function of the scale factor can be determined from the procedure described in Appendix~\ref{app:cosmology}. In particular, such a procedure provides us with a relation between the scale factor $a$ and the bath temperature $T$ which we can use to express the collision operators as a function of the scale factor. The asymptotic value of $\mathcal{X}$, once we set $\mathcal{X}(a_{\rm in}) = 0$ as initial condition, results in
\begin{equation}
\mathcal{X}^\infty = \sqrt{3} M_{\rm Pl} \int_0^\infty \frac{d a}{a} \frac{a^3}{\sqrt{\rho_M(a) + \rho_R(a)}} \sum_\alpha \mathcal{C}_\alpha(a) \ .
\end{equation}

At late enough times, once the thermal bath temperature drops below $T_R$ and the universe is dominated by radiation, we can trust entropy conservation again. In order to compute the FIMP contribution to the current energy density, we identify a late temperature $T_* < T_R$ and we evaluate $n_X(T_*)$ and the entropy density $s(T_*)$. The corresponding comoving number density:
\begin{equation}
Y_X(T_*) = \frac{n_X(T_*)}{s(T_*)} =   \frac{\mathcal{X}(T_*)}{S(T_*)} = \frac{\mathcal{X}^\infty}{S(T_*)}   \quad [T_*< T_R]
\end{equation}
which is constant in the subsequent evolution of the universe,
i.e.~$Y_X^\infty= Y_X(T_*)$. The energy density today thus 
reads $\rho_X(t_0) = m_X  Y_X^\infty\, s_0$ where $s_0$ is the current entropy
density.

\subsection{Comparison with previous studies}
\label{sec:comparison}

We conclude this appendix with a comparison between our analysis and previous studies about freeze-in via decays with subsequent dilution. As already explained above, we consider the full reheating process that happens after the inflationary phase of the early universe. In particular, we study the evolution of the thermal bath since its maximum temperature $T_\textsc{max}$ all the way down to $T_R$. Our typical situation is when there is the hierarchy $T_\textsc{max} \gg T_{FI} \gg T_R$, in such a way that freeze-in happens during an early MD era with the thermal bath filled of $B$ particles with a relativistic abundance. Later on, until we reach the reheating temperature $T_R$, DM particles produced via these decays get diluted.

An alternative methodology, followed e.g.~by Refs.~\cite{Belanger:2018mqt,Belanger:2018sti}, is to neglect the extension in time of the early MD era. Within this approach, the two temperatures $T_\textsc{max}$ and $T_R$ coincide and this is achieved if the reheating process is instantaneous. On one hand, the amount of DM predicted suffers from a Maxwell-Boltzmann suppression of the $B$ particles at temperature $T_R \ll m_B$. On the other hand, the dilution due to inflaton decays after DM production is not present within this methodology. 

\begin{figure}[t]
	\centering
	\includegraphics[width=0.55\textwidth]{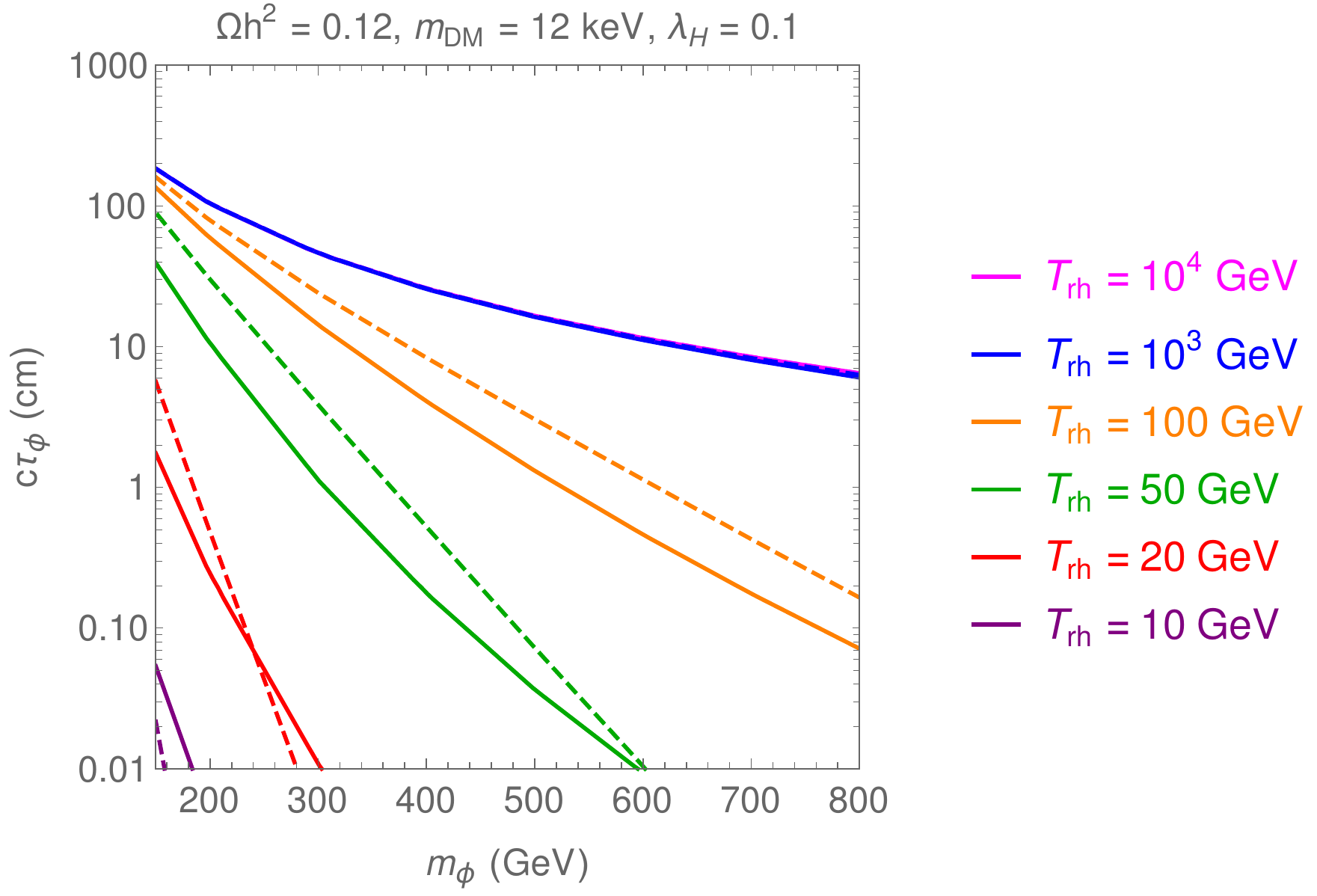}
		\caption{Comparison between our implementation of $T_{R}$ (continuous curve) and a cut in the time integration of the DM yield as in e.g.~\cite{Belanger:2018mqt,Belanger:2018sti}.}
	\label{fig:TRHcompare}
\end{figure}

We illustrate in Figure~\ref{fig:TRHcompare} how these two different approaches affect the estimation of the collider prospects. We plot with continuous curves the results obtained following our methodology within the context of our leptophilic scenario of Section~\ref{sec:leptophilic-scenario}. The dashed lines of the same color are obtained following instead the instantaneous reheating approximation.  For renormalizable operators, this changes by a factor $\sim 2$ at most the estimate of the mother particle lifetime. We expect thus that the analysis provided in Ref.~\cite{Belanger:2018sti} for scalar DM coupled to a vector-like heavy lepton mediator would give qualitatively equivalent results when following the methodology presented here.

%%%%%%%%%%%%
\section{LHC searches for LLPs: description and recasting}
\label{sec:Ap-LHC}
%%%%%%%%%%%

In this appendix we briefly describe the LLP LHC searches tabulated in
Table~\ref{tab:searches} and  provide more details on our recasting procedure for each of them.

\subsection{Heavy stable charged particle (HSCP) and R-hadrons (RH)}
\label{sec:def-HSCP}

\paragraph{Description.}
When a charged mediator has a long enough lifetime to cross
the detector completely ({$c\tau_B > \mathcal{O}(10)$~m), it can leave a
highly ionized track in the detector. This type of signature is usually
referred to as Heavy Stable Charged Particle (HSCP). When the charged
particle also carries QCD color, the long-lived particle is expected to hadronize, hence one refers to ``R-hadron'' (RH)
searches. Searches of this type have been performed by both
CMS~\cite{Khachatryan:2015lla,CMS:2016ybj} and
ATLAS~\cite{Aaboud:2019trc}. Both collaborations have publicly provided efficiency
tables for the cases of slepton-like HSCP and gluino- or squark-like
R-hadrons. The efficiency tables of the CMS search are gathered in the public code {\tt
  SModelS}~\cite{Kraml:2013mwa,Ambrogi:2018ujg}, which we use for
recasting this search, see~\cite{Junius:2019dci} for details. In order to reinterpret the ATLAS search, a public code is provided in the ``LLP recasting repository''~\cite{LLPrepos}, a repository on GitHub holding various example codes to recast existing LLP searches. Notice that we use both codes to constrain R-hadrons originating
from vectorlike quarks instead of squarks. In the latter case we
assume that the differences in efficiency due to the difference in
spin are minor.

As can be seen in Table~\ref{tab:searches}, one would expect the ATLAS search to be slightly more constraining since the employed luminosity is about three times higher. However, there is an important difference between ATLAS and CMS searches, which unfolds itself when the search strategy is applied to look for charged particles/R-hadrons with a mass of the order of 100-200 $\rm GeV$. The main difference between the two searches relies on how the signal regions (SR) are defined. Both make use of the reconstructed mass to define the SR, however the CMS SR are defined such that any heavy LLP/R-hadron will fit in at least one of the SR, while the ATLAS search only probes heavy LLPs/R-hadrons with a reconstructed mass larger than a certain threshold.\footnote{This limit depends on the type of particle/R-hadron considered.} Hence, the CMS search is more sensitive in the lower mass region. Details on our recasting of these searches are provided in the following.

\paragraph{Recasting.}
Searches for highly ionized tracks originating from heavy stable
charged particles (HSCP) and R-Hadrons (RH) have been both performed
by CMS~\cite{CMS:2016ybj} and ATLAS~\cite{Aaboud:2019trc}. As discussed above,
the main
differences between these two searches are the integrated luminosity
and the sensitivity to the low mass range of the HSCPs/RHs.

In order to reinterpret the search performed by CMS, we employed a
publicly available code named {\tt SModelS}~\cite{Kraml:2013mwa,Ambrogi:2018ujg}. 
This code makes use
of the available efficiency tables in order to find the upper limit on
the cross section in any given model. However, the efficiency tables
included in {\tt SModelS} are built by assuming that the
HSCPs/RHs are detector stable (i.e.~decay way outside the
detector). For HSCPs/RHs with an intermediate lifetime, it can happen that
only a fraction of the produced particles traverse the whole
detector. {\tt SModelS} takes this effect into account by multiplying
the efficiency of a stable HSCP/RH by the probability that the HSCP/RH
with a specified lifetime traverses the detector completely (for more
details on our treatment, see~\cite{Junius:2019dci}).

In order the reinterpret the ATLAS search, we again made use of a
publicly available code in the ``LLP recasting repository'' on
GitHub~\cite{LLPrepos}. This code does not rely on previously
released efficiency tables, but rather makes use of {\tt Pythia 8}~\cite{Sjostrand:2014zea} to
perform an event-by-event based analysis such that the dependence on
the lifetime can be estimated with greater accuracy. The recasting
strategy is based on the information published by the ATLAS
collaboration, see~\cite{AtlasRHrecast}. For the leptophilic
scenario described in Section~\ref{sec:leptophilic-scenario}, all
necessary ingredients to extract the constraints are present in both
codes. There is however one caveat for the reinterpretation of the RH
searches for the topphilic scenario
(Section~\ref{sec:topphilic-scenario}). Both of these codes are able to
reinterpret the existing searches only for squark- or gluino-like
RHs.\footnote{In {\tt SModelS}, only efficiency tables for squark- and
  gluino-like RHs are present. Also {\tt Pythia 8} is only able to
  consider the formation RHs originating from squarks or gluinos.} Since we
wanted to use these codes to constrain an RH originating from a
vectorlike top quark, we made some small modification such
that they would treat a vectorlike top RH as a stop-like RH. As
the only difference between these two cases is the spin of the
particle the RH is originating from, we do not expect that full implementation
of a vectorlike top would have a large impact on our results.

\subsection{Disappearing tracks (DT) and kinked tracks (KT)}
\label{sec:def-DT}
\paragraph{Description.}
Disappearing tracks (DT) can arise in models with a charged mediator
that has a proper decay length $c\tau_B \sim \mathcal{O}(10)$~cm,
i.e.~smaller than the size of the inner tracking system of ATLAS and CMS, and its decay products
cannot be reconstructed, either because they are neutral or they carry low momentum.  
Searches for disappearing tracks have been performed at both
ATLAS~\cite{Aaboud:2017mpt} and CMS~\cite{CMS:2018ail,Sirunyan:2020pjd} with 13 TeV
data. 

In order to recast DT searches, we make use of the publicly available
efficiency table provided by the CMS and ATLAS collaborations,
see~\cite{Junius:2019dci} for details. One important difference
between the CMS and ATLAS search is the range of lifetimes they
probe. The innermost tracking layers in the ATLAS detector have been
upgraded and can focus more on very short tracks (tracklets). In
practice~ATLAS can probe shorter lifetimes $\tau \sim
\mathcal{O}$(1)~ns while CMS is more sensitive to larger lifetimes
with $\tau \sim \mathcal{O}$(10)~ns.
Below, we reproduce the
existing limits on the simplified model with a wino LSP as presented
by CMS~\cite{Sirunyan:2020pjd} and ATLAS~\cite{Aaboud:2017mpt}, in
order to validate our re-interpretations.

Note that these searches could also have some sensitivity for models exhibiting kinked track (KT) topologies. 
Indeed, let's assume that there is a displaced decay producing a stable SM charged particle plus DM.
If the track of the charged decay product is
not reconstructed, this signature would resemble a disappearing track. 
Some effort has been already done to verify the sensitivity of the disappearing track
searches to a kinked track
signature~\cite{Jung:2015boa,Evans:2016zau,Belyaev:2020wok,Belanger:2018sti}.
We remind here that DT searches impose a lepton veto and thus they cannot be 
employed
straightforwardly for DM models involving a direct DM coupling to leptons
and leading to displaced hard leptons in the final states. 
We discuss how we handle this issue in Section~\ref{sec:leptophilic-scenario} to provide an estimate of the DT search reach on the KT signature
which is present in that model.

\paragraph{Recasting.}
Here we reinterpret the disappearing track (DT) searches by
ATLAS~\cite{Aaboud:2017mpt} and CMS~\cite{Sirunyan:2020pjd}, performed
for a supersymmetric model with long-lived winos. For that purpose, we
apply the technique that some of us developed in~\cite{Junius:2019dci} and
applied to the DT searches in~\cite{CMS:2018ail, Aaboud:2017mpt}. 
We validated our technique for the DT searches used
here~\cite{Aaboud:2017mpt,Sirunyan:2020pjd} by reanalysing the case of
the wino within our framework and comparing our results to the ones
presented in the CMS and ATLAS papers. Figure~\ref{fig:ValidationDT} illustrates that 
 we find a very good agreement.
\begin{figure}
	\centering
	\subfloat{\includegraphics[width=0.45\textwidth]{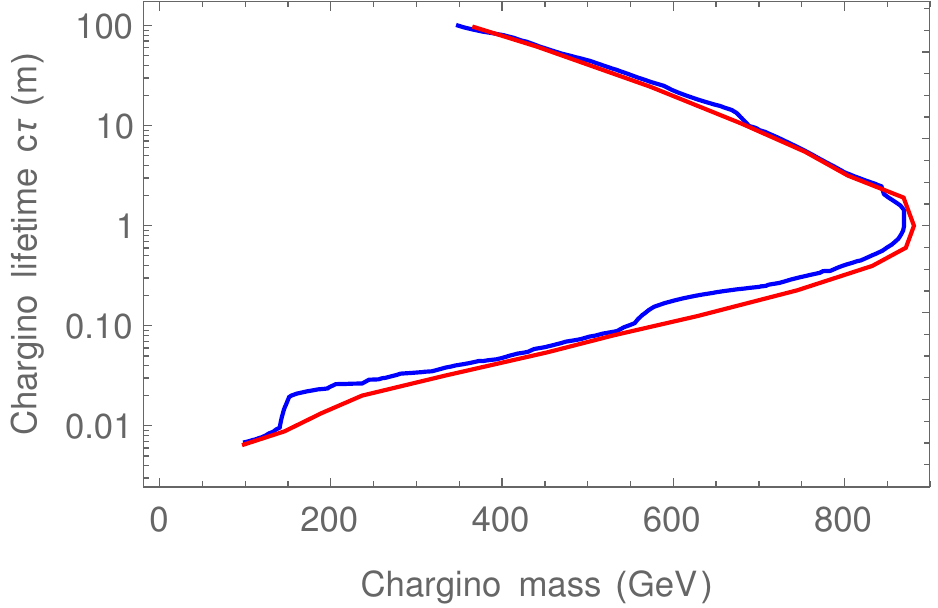}}
	\hfill	
	\subfloat{\includegraphics[width=0.46\textwidth]{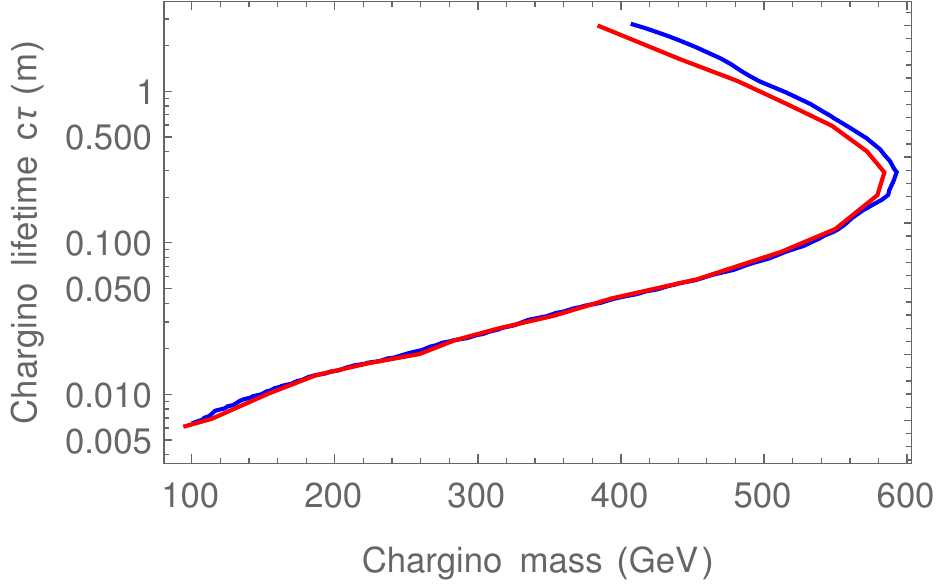}}
	\caption{Comparison between the experimental result (red) and our reinterpreted result (blue) for the latest DT searches performed by CMS~\cite{Sirunyan:2020pjd} (left) and ATLAS~\cite{Aaboud:2017mpt} (right).}
	\label{fig:ValidationDT}
\end{figure}

The DT signature is a smoking-gun signature for the singlet triplet
model studied in Section~\ref{sec:non-renorm-oper}, which is closely
related to the supersymmetric wino model, for which all DT searches are
originally performed. The DT originates from the charged component of
the triplet decaying into the neutral one in association with a soft pion that
evades detection: $\Psi_B^\pm\to \Psi_B^0 +\pi^\pm$.  For the
charged track to fulfill the disappearance requirement stated in the
CMS and ATLAS papers, no hits in the outer layers of the tracker have to
be associated to the track. If the lifetime of the
neutral component is short enough $\Psi_B^0$ for it to decay inside the
detector, hits in the outer layers of the tracker might
occur. Therefore, we require in our analysis that $\Psi_B^0$  to decay outside the tracker. 
In order to take this into account, we multiply the efficiency by the probability for
$\Psi_B^0$ to decay inside the tracker by taking $\langle
l_\text{outer}/\gamma\beta \rangle_\text{eff}$=0.36, as done in
\cite{Heisig:2018kfq}, where $l_{outer}$ is the outer radius of the inner tracker. Notice that the CMS search~\cite{Sirunyan:2020pjd} considered
here has been performed on a larger data set than the ATLAS search ($\mathcal{L}=140~\text{fb}^{-1}$ compared to
$36.1~\text{fb}^{-1}$). Naively, one would then expect CMS
to set the strongest constraints on our models. As mentioned above though, 
the ATLAS detector has been optimized
to study shorter disappearing tracks than the those studied by CMS, 
hence the CMS and ATLAS searches are complementary as one can see from
our reinterpretation of Figure~\ref{fig:ValidationDT}.

We have also estimated the efficiency of the CMS disappearing track search
in constraining a kinked-track signature for our leptophilic
model (see Section~\ref{sec:leptophilic-scenario}). For the DT
search to be sensitive to a kinked track, we require that the angle
between the mother and daughter particle tracks be large enough so as
to avoid them to be identified as collinear (and hence being
disentangled from one single track). We took the conservative limit on the angle
$\Delta R \equiv \sqrt{(\Delta \phi)^2+ (\Delta \eta )^2} > 0.1$. Additionally, 
 the CMS search imposes a cut on the amount of energy deposited in the calorimeters
within a cone of $\Delta R < 0.5$ about the disappearing track, which has to
be less than 10 GeV. When the charged daughter particle is a muon, it
will not leave a large energy deposit in the calorimeter, and hence,
we do not need any extra requirements. In contrast, when the charged
daughter particle is an electron or some hadronic final state, such an
extra cut on the energy deposit in the calorimeter has to be imposed
in order to estimate the sensitivity to the kinked track signature.

\subsection{Displaced leptons (DL)}
\label{sec:def-DL}
\paragraph{Description.}
In cases where the long-lived mediator decays to a lepton within the tracker, one would get displaced lepton signatures. 
Searches have been performed for events containing leptons with large impact parameters, i.e.~displaced
leptons, by CMS at both the 8 TeV~\cite{Khachatryan:2014mea} and  the 13
TeV~\cite{CMS:2016isf} run and by ATLAS at 13 TeV~\cite{Aad:2020bay}. Both CMS searches are maximally sensitive to particles with $c\tau_B \sim \mathcal{O}(1)$~cm while the ATLAS search is sensitive to slightly larger values of $c\tau_B \sim \mathcal{O}(10)$~cm. 
In our study we consider both 8 TeV and 13 TeV CMS searches since the 13 TeV analysis has been performed on a small
data set (corresponding to an integrated luminosity of $2.6~\text{fb}^{-1}$ compared to
the $19.7~\text{fb}^{-1}$ employed by the 8 TeV search) such that the 8 TeV search will be in
many cases more sensitive. 

Let us emphasize that only displaced $e^\pm \mu^\mp$ pairs (coming
from different decaying particles) are looked for by the CMS searches. In the cases where
the DM couples to one lepton flavor only, this specific search does
not a priori apply. In Section~\ref{sec:leptophilic-scenario} we nevertheless used the
available efficiency maps to provide a naive estimate of the possible
LHC sensitivity to displaced $\mu^+\mu^-$ pairs. 

Instead, the DL ATLAS search defines three signal regions depending on the lepton flavours, one where a displaced $e^\pm \mu^\mp$ pair is observed and two others where same flavor leptons are observed ($e^\pm e^\mp$ or $\mu^\pm \mu^\mp$). Hence, it is straightforwardly applicable also to cases where the DM couples to one lepton flavor only. More details on the
extracted sensitivities are provided below.

\paragraph{Recasting.}
The DL searches performed by
CMS~\cite{Khachatryan:2014mea,CMS:2016isf} look for events
containing a displaced electron-muon pair that do not necessarily
originate from the same vertex. The CMS collaboration has provided the
model-independent efficiency tables for detecting a single displaced
electron/muon as a function of its displacement and transverse
momentum. By using these tables and applying the cuts reported in
Table~\ref{tab:DLcuts} one can estimate the model-dependent efficiency
for the three signal regions (SR) that are defined for this
search. These are based on the transverse impact parameter of
both leptons ($|d_0|$) and they are referred to as SR1 for 0.05 cm >
$|d_0|$ > 0.02 cm, SR2 for 0.1cm > $|d_0|$ > 0.05 cm, and SR3 for $|d_0|$ >
0.1 cm. SR2 and SR3 are the most constraining as the SM background is
virtually zero in these signal regions.
\begin{table}
	\centering
	\begin{tabular}{|c|c|}
		\hline
		8 Tev & 13 TeV \\
		\hline
		\multirow{2}{4cm}{\centering $p_T$ > 25 GeV} & $p_T$ > 42 GeV (electrons)\\
		& $p_T$ > 40 GeV (muons) \\
		\hline
		$|\eta|$ < 2.5 & $|\eta|$ < 2.4\\
		\hline
		0.02 cm < $|d_0|$ < 2 cm & 0.02 cm < $|d_0|$ < 10 cm \\
		\hline
		$\Delta R$ > 0.5 & $\Delta R$ > 0.5 \\
		\hline 
	\end{tabular}
	\caption{DL searches at CMS: Cuts on the transverse momentum $p_T$, pseudorapidity $\eta$ and transverse impact parameter $d_0$ of the electron/muon and the angle $\Delta R = \sqrt{\Delta\phi^2+\Delta\eta^2}$ between the electron and muon for the 8 and 13 TeV searches.}
	\label{tab:DLcuts}
\end{table}

As mentioned above, we make use of the DL
searches to estimate the possible LHC sensitivity to a displaced
$\mu^+\mu^-$ pair. Since the experimental collaboration has published
the reconstruction efficiency for the electron and muon separately, we
obtained a quite accurate estimate for the total event efficiency of a
displaced $\mu^+\mu^-$ pair.  Notice that we have assumed zero background
events in SR2 and SR3, as is the case for the $e^\pm \mu^\mp$
topology. Performing a full background estimation for the $\mu^+\mu^-$
topology is beyond the scope of this work, but one should remind
that a more realistic background estimation can possibly reduce the
resulting sensitivity.\\

As mentioned above, the DL search performed by ATLAS~\cite{Aad:2020bay} defines three different signal regions (SR-$ee$, SR-$e\mu$ and SR-$\mu\mu$) and hence uses three different trigger strategies, namely single-photon, diphoton, and muon trigger. The photon triggers select events with an energy deposit in the electromagnetic calorimeter greater than 140 GeV (single-photon) and 50 GeV (diphoton) while the muon trigger select events with a signature in the muon spectrometer with $p_T>$60 GeV and $|\eta|<$1.05. For each event, two signal leptons are defined as the leptons with the highest transverse momentum. These leptons must further have an impact parameter $|d_0|$ between 3 and 300 mm, a transverse momenta $p_T > 65$~GeV and $|\eta|<2.5$. Apart from these cuts, there are two extra event requirements. First, there must be a clear separation between both leptons, $\Delta R>0.2$ and none of the muons can be cosmic tagged. The latter consists in requiring that the two signal muons cannot be produced back to back, a case in which they are considered of cosmic origin and are hence removed.
For events passing these selection cuts, the ATLAS collaboration also provides the model-independent reconstruction efficiencies for displaced electrons and muons~\cite{1831504}
to be further applied.
However, this information is only given for a benchmark with LLP mass of $400$~GeV and proper lifetime of $1$ ns.
Nevertheless, we use for concreteness the same efficiency maps also for benchmarks with different LLP mass and lifetime, so we expect some discrepancy with the experimental results in certain regions of the parameter space.

In Figure~\ref{fig:DL_validation} we show the validation of our recasting procedure for two representative models 
for which the full exclusion curves are reported in the auxiliary material of the ATLAS paper, that is, production of right-handed smuon and right-handed selectron.
\begin{figure}[t!]
	\centering
	\subfloat{\includegraphics[width=0.45\textwidth]{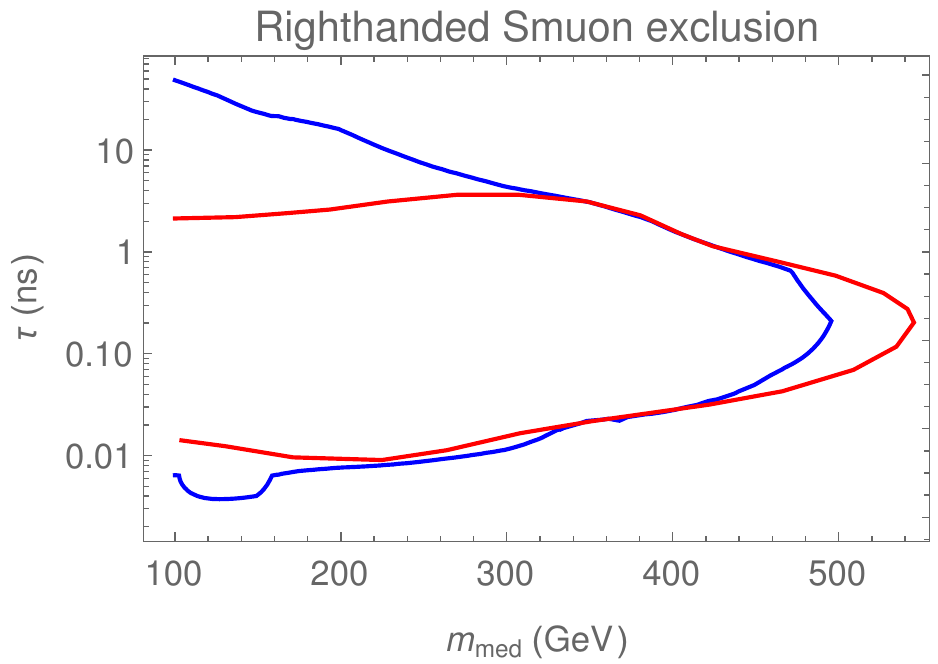}}
	\hfill	
	\subfloat{\includegraphics[width=0.45\textwidth]{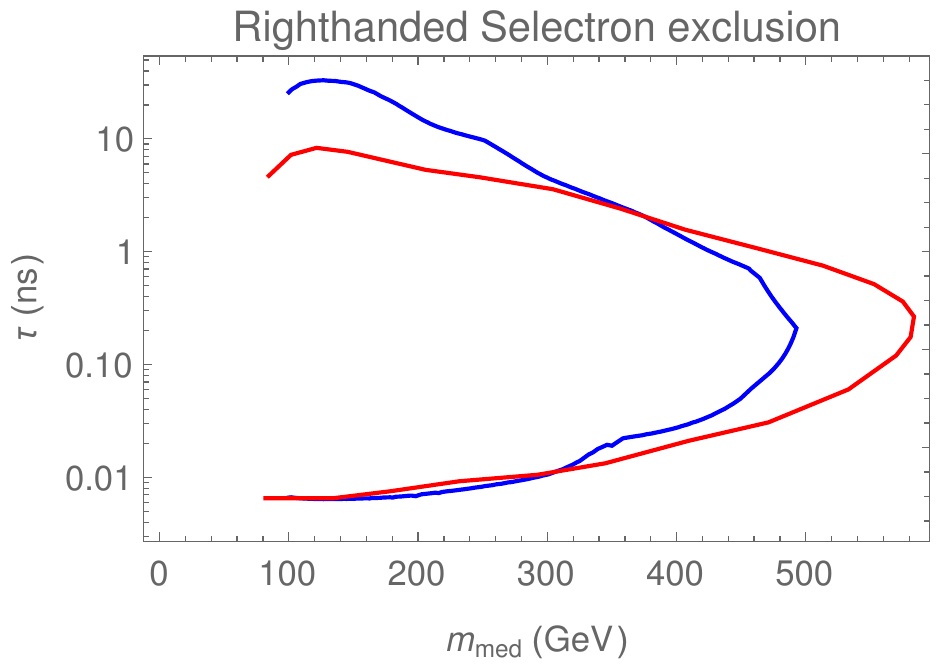}}
	\caption{Validation of the DL performed by ATLAS~\cite{Aad:2020bay}. Left: right-handed smuon production. Right: right-handed selectron production.}
	\label{fig:DL_validation}
\end{figure}
We see that our recasting procedure reproduces very well the experimental exclusion curves around the benchmark point for which detailed informations on the
electron and muon efficiency are provided by the ATLAS collaboration, as explained above.
For masses significantly smaller or larger than $400$~GeV the recasting is instead less precise. 
%We leave for future work an improvement of the recasting analysis that will be certainly possible when more information will be provided in the auxiliary material.
We nevertheless observe that, even if not quantitatively precise, our validation shows a qualitative agreement with the experimental results and hence
we employ it in the main text to estimate the expected sensitivity for the topphilic and the singlet-triplet model.

\subsection{Displaced vertices + MET (DV+MET)}
\label{sec:def-DV}
\paragraph{Description.}
We call displaced vertices + MET the case where the long-lived mediator
decays to DM and a color charged object before reaching the end of the
tracker. A jet will be produced but is not reconstructed using standard jet-clustering algorithms. 
The events will be rather analysed by looking at the individual tracks of the jet originating from a
displaced vertex, see~\cite{Aaboud:2017iio} for details.  In order to
recast this search, we made use of the information provided by the
ATLAS collaboration on the HEPdata page of the search~\cite{1630632}. We validated
our recasting techniques by applying them on the model studied by the
ATLAS collaboration and obtained very similar results, as discussed below.

\paragraph{Recasting.}
The ATLAS collaboration has released a search for events with
displaced vertices in combination with missing transverse momentum
in~\cite{Aaboud:2017iio}. They look for events with at least one
displaced vertex containing at least 5 tracks and missing transverse
energy (MET) larger or equal to 200 GeV. The efficiency tables are publicly
available on HEPData together with a procedure for recasting the
search to other models~\cite{1630632}. Following this procedure, we reproduced the
results of the ATLAS collaboration for long-lived gluinos decaying into
quarks and a neutralino by doing an event-by-event analysis. Our
results can be seen in Figure~\ref{fig:DV_validation}. A deviation
between the ATLAS limit and our recasting appears for small values of
the mass splitting between the gluino and the neutralino,
$\Delta m/m \lesssim 0.4$. This is because we have omitted a
cut, placed on the jets transverse momentum for 75\% of the events,
that only has a significant impact on the selection efficiency if
the mass splitting is small. In our study, we do not explore
compressed spectra and hence we can neglect this condition safely. 
We applied this recasting strategy to two of our models, the  topphilic and  singlet-triplet  scenarios of Sections~\ref{sec:topphilic-scenario} and~\ref{sec:non-renorm-oper}. Some special treatments were required for both of these models, as discussed in the following.
\begin{figure}[t!]
	\centering
	\subfloat{\includegraphics[width=0.45\textwidth]{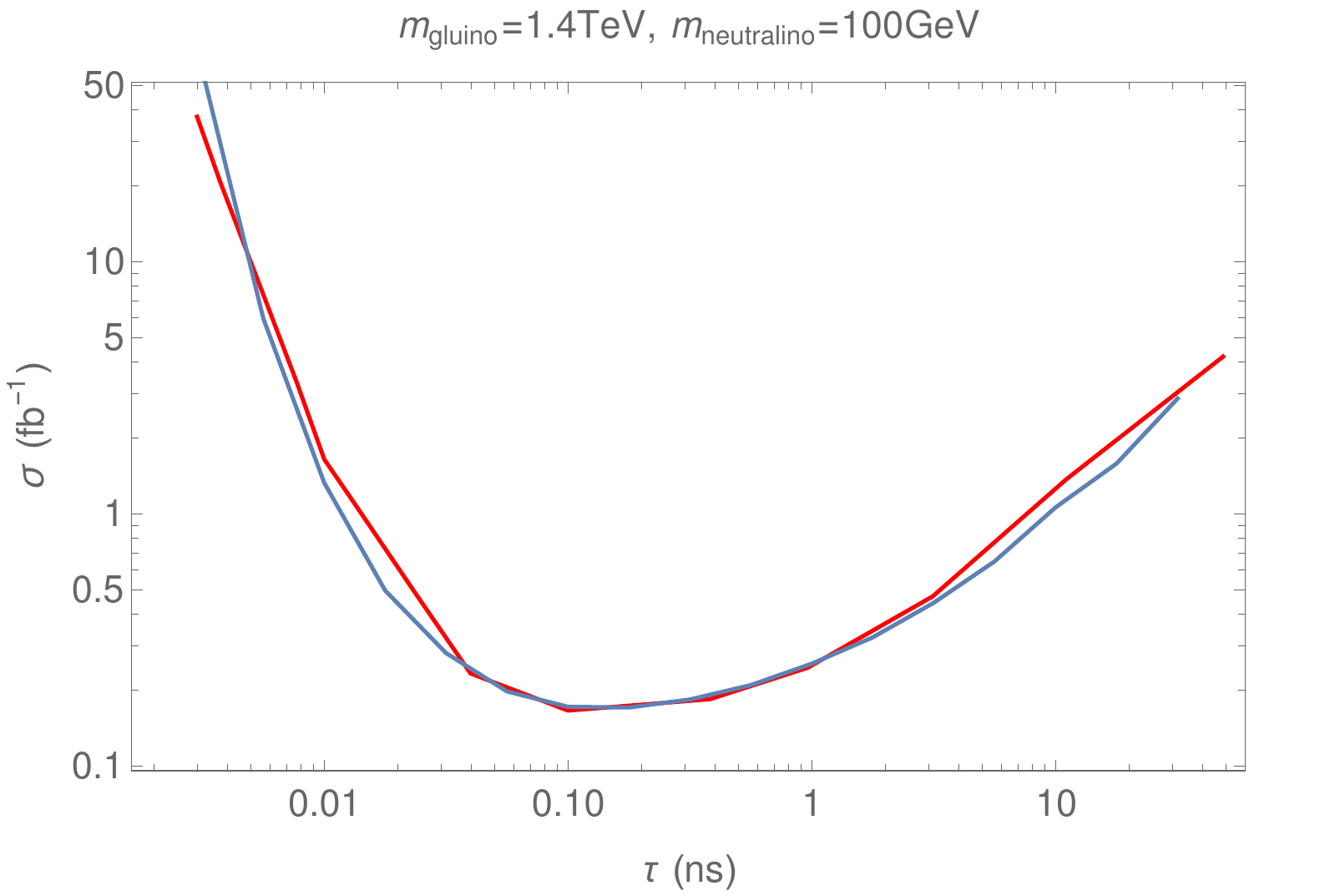}}
	\hfill	
	\subfloat{\includegraphics[width=0.45\textwidth]{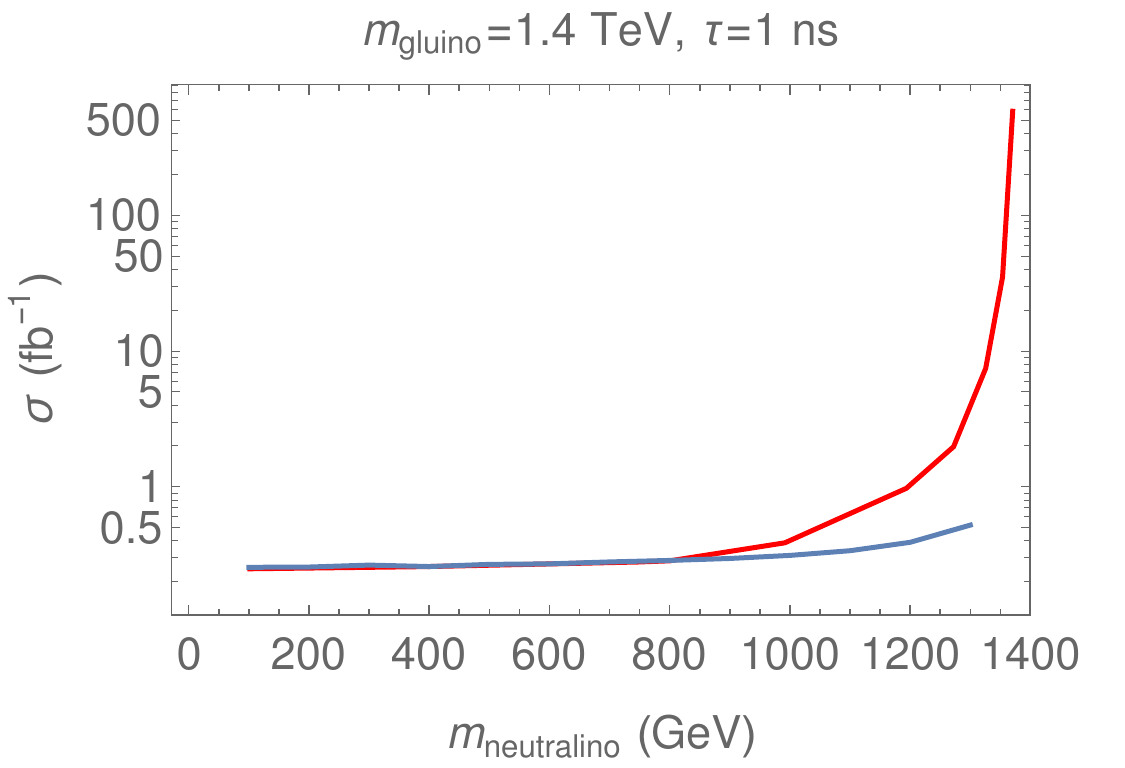}}
	\caption{Validation of the DV + MET search performed by ATLAS~\cite{Aaboud:2017iio} by comparing the ATLAS results (red) with the results from our recasting (blue). Left: Upper limit on the production cross section vs.~the gluino lifetime for a fixed gluino mass (1.4 TeV) and neutralino mass (100 GeV). Right: Upper limit on the production cross section vs~ the mass of the neutralino for a fixed gluino mass (1.4 TeV) and gluino lifetime (1 ns). Notice the discrepancy between the ATLAS limit and our result for a small gluino-neutralino mass splitting ($\Delta m/m \lesssim 0.4$) which is due to the omission of one of the selection criteria (see the text for details).}
	\label{fig:DV_validation}
\end{figure}

In the topphilic scenario, a displaced jet can arise from a long-lived
R-hadron decaying into a top quark and the DM scalar. The top quark
will predominantly decay to a bottom together with two other quarks
(through an intermediate $W$ boson). The quarks will promptly form a
jet. The bottom quark on the other hand eventually decays to a
slightly-displaced jet, with a displacement $\sim\mathcal{O}$(mm) due
to the non-negligible $b$ lifetime. For the DV+MET search,
reconstructed vertices that are within a distance of 1 mm of each
other are seen as one single vertex. Hence the tracks originating from
the bottom and from the two other quarks will not necessarily be
re-combined into one single vertex. Hence to be conservative, we
discard any track originating from decay of the bottom quark in the
recasting of the DV+ MET search (we also applied this conservative approach
in the DV+$\mu$ search discussed in Appendix~\ref{sec:def-DVmu}).

In the singlet-triplet model, a displaced jet can originate from the
decay of the charged and the neutral component of the triplet. The
neutral components can be produced in two ways, through its gauge
interactions or from the decay of the charged component. The decay of
the charged component to the neutral one (together with soft pions) is
slightly displaced, $\mathcal{O}$(cm), and this displacement has to
be taken into account in order to correctly interpret the efficiency
tables for the singlet-triplet model. Due to the small mass splitting
between the charged and the neutral component of the triplet, they
will be more or less collinear. Hence, to estimate the total
displacement of the displaced jet, we simply add the displacements of
the charged and neutral components of the triplet.

\subsection{Delayed jets + MET (DJ+MET)}
\label{sec:def-DJ}
\paragraph{Description.}
A delayed jet has been defined as a jet that is observed in the
calorimeter of the detector at a later time than one would expect from a
jet that is produced at the primary vertex. Such a time delay can be due
to a heavy, long-lived mediator that slowly crosses the detector
before decaying into a jet. This search relies on the timing
capabilities of the calorimeter and, as a result, there is no need to
reconstruct the displaced vertex. Here we use the search for delayed
jets + MET that has been performed by CMS in~\cite{Sirunyan:2019gut}.

Notice that the delayed jet + MET search is able to probe longer
lifetimes ($c\tau_B \sim 1-10$~m) than the displaced vertices
+ MET search discussed above (more sensitive to decaying particles
  with $c\tau_B \sim {\cal O} (1-10)$~cm) as the calorimeter lies further
from the centre than the tracker. The DV+MET search instead is more
sensitive to smaller lifetimes, and actually, the time delay will not
be enough to distinguish between a displaced or a prompt jet. Hence,
it is useful to consider both DV+MET and DJ+MET searches together given
their complementarity.

\paragraph{Recasting.}
In order to probe larger values of the lifetime of long-lived particles decaying into jets, 
the CMS collaboration made use of the timing capabilities of the
calorimeter to look for delayed jets~\cite{Sirunyan:2019gut}. They
studied explicitly the case of long-lived gluinos decaying into a gluon
and a gravitino. The efficiency tables for such a scenario are
publicly available, starting from gluino masses of 1 TeV. Since these
efficiency maps are blind to any jet activity arising from the primary
vertex, we can use them directly  for the scenarios studied here.

The gluino model studied in~\cite{Sirunyan:2019gut} always gives rise
to two delayed jets while in the models listed in
Table~\ref{tab:classification}, it can happen that only one delayed jet
arises.\footnote{Often, this is due to the fact that the jets
  originate from the hadronic decay of $W$ or $Z$ bosons, and these
  can also decay leptonically.} In order to address this difference,
we assume that the efficiency for an event with one delayed jet is
small enough ($\epsilon_{1jet}\ll1$) so as to approximate the
efficiency for an event with $N$ delayed jets with
$\epsilon_{Njets}=1-(1-\epsilon_{1jet})^N \approx N
\epsilon_{1jet}$. With this approximation we can derive the efficiency
for an event with one delayed jet, making use of the publicly
available efficiency tables for long-lived gluinos (involving events
with two delayed jets), and obtain the efficiencies for events with an
arbitrary number of delayed jets. By weighting the derived
efficiencies with the correct combination of branching ratios, we can
obtain model-dependent efficiency tables for the models studied in
this paper.

In the singlet-triplet model of Section~\ref{sec:non-renorm-oper}, we
have made extra assumptions in order to extract the corresponding
efficiencies. As already mention above, e.g.~in the discussion of
DV+MET searches, the neutral component of the triplet can get an extra
displacement from the decay of the charged component with $c\tau_C\sim
\cal O$(cm). Since the CMS delayed jet search probes relatively long
lifetimes of the neutral component ($c\tau_0\sim\mathcal{O}$(m)
compared to $c\tau_C\sim \cal O$(cm) ), we assume that we can ignore
the extra displacement arising from the decay of the charged triplet
component. We have also assumed the charged component decaying into
the neutral one and soft pions to one with a 100\% branching ratio,
i.e.~BR$(\Psi_B^{\pm}\to \Psi_B^0\pi^\pm)=1$. Indeed, even though
$\Psi_B^{\pm}$ can directly decay to DM, this decay width depends on
the new physics scale $\Lambda$, while the decay to $\Psi_B^{0}$ is
driven by gauge interactions, i.e.~independent of $\Lambda$. Since we
probe large lifetimes with the DJ+ MET search, we are always in a
regime where $\Lambda$ is large such that the direct decay of the
charged component to DM is suppressed compared to the decay
to the heavy neutral triplet component.

\subsection{Displaced vertices + muon (DV+$\mu$)}
\label{sec:def-DVmu}
\paragraph{Description.}
Another type of displaced vertex search is performed by ATLAS looking for events containing  displaced vertices together with a displaced muon track~\cite{Aad:2020srt,1788448}. The search defines two orthogonal signal regions, one containing events triggered by missing energy, the other one containing events triggered by a high $p_T$ muon. In the context of this paper, where we focus on simplified DM models, the MET signal region is expected to be the most efficient. The displaced vertex is reconstructed in a very similar way as for the DV+MET search discussed in Appendix~\ref{sec:def-DV}, so we expect these two searches to constrain similar ranges of lifetimes. The main differences between the two searches are the requirement of a muon and the higher luminosity used in the DV+$\mu$ search. 
%More details on the recasting of this search can be found in Appendix~\ref{sec:displ-vert-mu}.
%, where the detailed informations~\cite{Aad:2020srt} provided by the ATLAS collaboration are employed.

\paragraph{Recasting.}
We use the search for displaced vertices and a displaced muon
performed by ATLAS in~\cite{Aad:2020srt} (with $\sqrt{s} = 13$ TeV and
$\mathcal{L}=132~\text{fb}^{-1}$). It is particularly relevant for our simplified models as the
long-lived particles are always produced in pairs and $Z$ and
$W$ bosons potentially arising in their decays can partially decay to
leptons.

The ATLAS search defines two signal regions (SR): in the first SR
(SR1) the event is triggered by a large amount of missing energy
($E_T^{miss}>180$~GeV) while in SR2 the event is triggered by a track
in the muon spectrometer (with $p_T>62$~GeV, $|\eta|<1.05$). In
addition, in SR2, a cut of $E_T^{miss}<180$~GeV is imposed so as to
define two orthogonal SRs.  The experimental collaboration has
provided trigger efficiencies, as well as muon and displaced vertex
reconstruction efficiencies. We also make use of the cuts listed in
the Tables~1 and 2 of~\cite{Aad:2020srt} to reinterpret this search by
doing an event-by-event analysis. We validated our approach by
applying our recasting to the% model studied in the ATLAS paper. The
results of this validation can be found in
Figure~\ref{fig:DVmuon_validation}.
\begin{figure}[t!]
	\centering
	\includegraphics[scale=0.55]{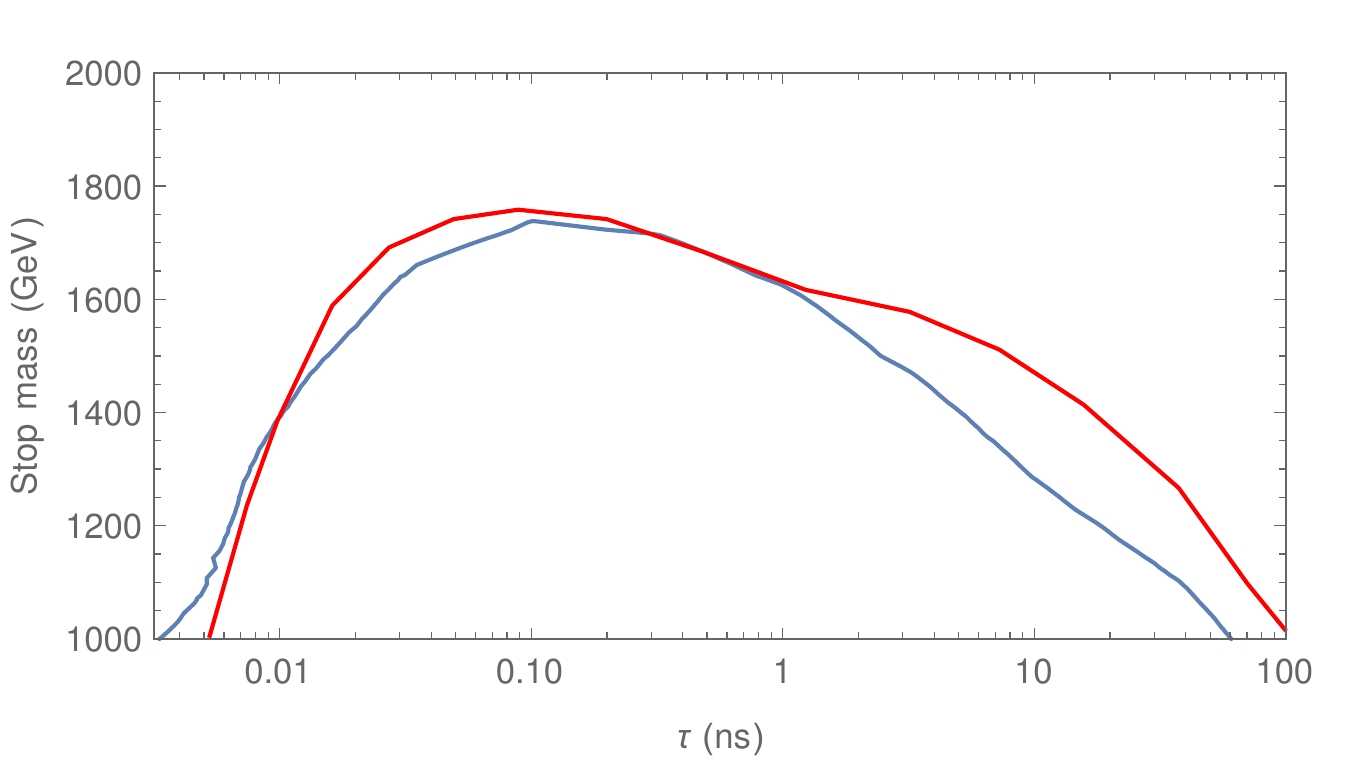}
	\caption{Validation of the DV + $\mu$ search by ATLAS. The red curve is the ATLAS result, the blue curve is our reinterpretation using the available efficiency tables.}
	\label{fig:DVmuon_validation}
\end{figure}

\subsection{Displaced dilepton vertex (DLV)} 
\label{sec:def-DLV}
\paragraph{Description.}
This kind of searches targets long-lived neutral particles decaying
into a lepton pair ($\mu^+\mu^-$, $e^+e^-$, or $\mu^\pm e^\mp$).  In
this work, we focus on simplified DM models where the connections
between dark and visible sector is governed by a three-particle
interaction (see Table~\ref{tab:classification}). This interaction
also governs the two body decay of the LLP in collider
experiments. Since one of the two daughter particles of the LLP is
always the DM, there will be only one SM particle involved in this
decay. Therefore, a displaced lepton vertex will occur in the
considered models only when the SM particle is a $Z$ boson which
decays to two leptons. The leptonic branching ratio of the $Z$ boson
is only about 10\%, while the hadronic branching ratio is about
70\%. Hence, we expect that the displaced lepton vertex search
performed by ATLAS~\cite{Aad:2019tcc} will be generically less
constraining than the DV+MET or DV+$\mu$ searches discussed in
Sections~\ref{sec:def-DV} and~\ref{sec:def-DVmu} above.  Nevertheless,
since the ATLAS collaboration provides useful informations to
reinterpret this search in a model independent manner~\cite{1745920},
we recast this analysis as discussed in the following and
we apply it when relevant for the considered models.

\paragraph{Recasting.}
Here we address the ATLAS search for a pair of oppositely charged
leptons originating from the same displaced vertex~\cite{Aad:2019tcc}.
The ATLAS collaboration has provided a document detailing how to
reinterpret this search for any model containing long-lived particles
decaying to oppositely charged leptons~\cite{1745920}. In order to
calculate the event acceptance, one has to apply some simple kinematic
cuts, among which the most relevant ones are that the invariant mass of the
lepton pair has to be above 12 GeV and their displacement must be
larger than 2 mm. The overall acceptance can be obtained by doing an
event-by-event analysis.

To obtain the detection efficiency of a displaced lepton pair, the
experimental collaboration provides two parameterisations. One for the
R-parity violation (RPV) SUSY model and one for a $Z'$ toy
model. There are two important differences between these
models. First, in the RPV SUSY model, the LLP is a bino-like
neutralino produced from the decay of a heavy squark. Due to the large
mass splitting between the heavy squarks and the LLP, the LLP will
have on average a much higher $p_T$ in the RPV model than in the $Z'$
model. As a result, the physical displacement in the detector will be
much larger (due to the larger boost) for the same proper lifetime and
the leptons arising in the decay will be also much more collimated in the
RPV case than in the $Z'$ case.  Another peculiarity of the RPV SUSY
model is that the LLP decays to a lepton pair and a neutralino, such
that the displacement vector pointing from the primary vertex to the
secondary vertex is not parallel to the momentum vector of the lepton
pair. This is also the case for the models studied in this work since
the LLP will always decay to a DM + SM (including leptons). As a
result, here we use the efficiency parametrisation provided for the
RPV SUSY model. We validated the procedure discussed
in~\cite{1745920} for the RPV SUSY model in Figure~\ref{fig:ValidationDLV}. Two different cases are illustrated, one where the LLP decays to either an electron pair or an electron-muon pair (left) and one where it decays to a muon pair or an electron-muon pair (right).
\begin{figure}[t!]
	\centering
	\subfloat{\includegraphics[width=0.45\textwidth]{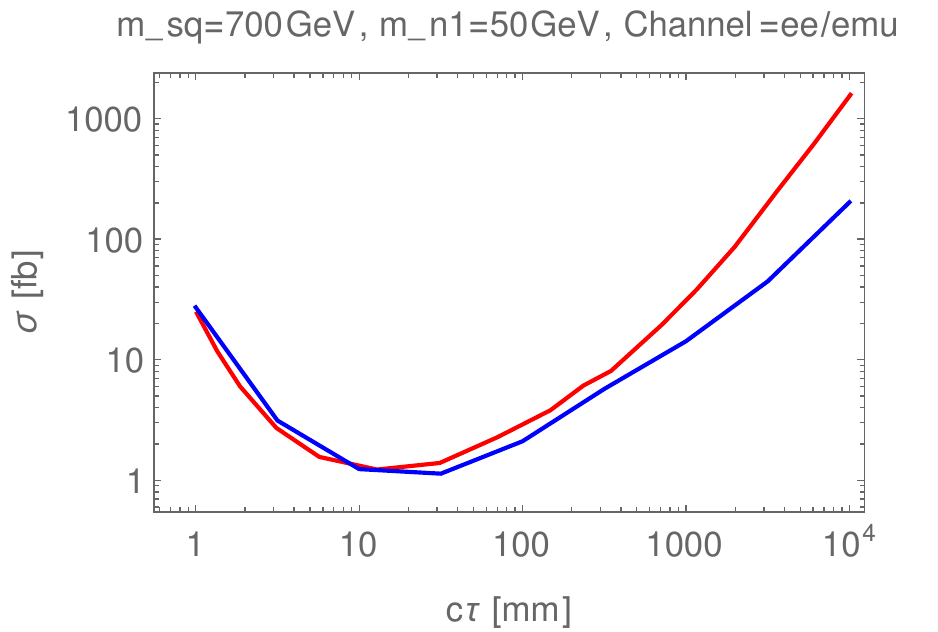}}
	\hfill
	\subfloat{\includegraphics[width=0.45\textwidth]{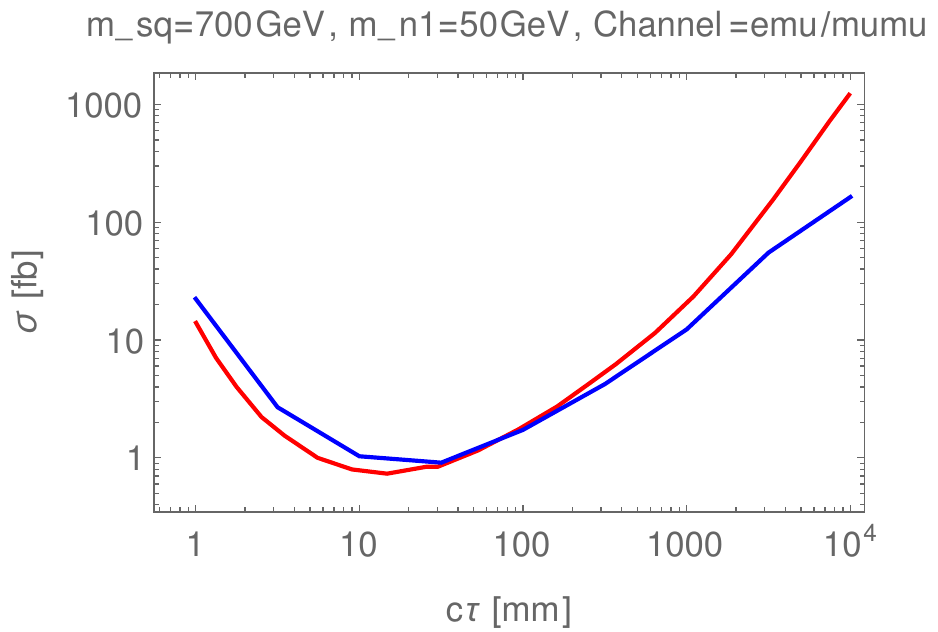}}
	\caption{Validation of the displaced lepton vertex search performed by ATLAS by comparing the upper limit on the cross section obtained by ATLAS (red) with the one from our recasting (blue) for the RPV SUSY model studied in~\cite{Aad:2019tcc}. }
	\label{fig:ValidationDLV}
\end{figure}

\subsection{Delayed photons (D$\gamma$)}
\label{sec:def-Dgam}
\paragraph{Description.}
If the long-lived mediator decays to a photon plus DM, the expected
LLP-type signature is displaced photons plus missing energy.  The CMS
analysis~\cite{Sirunyan:2019wau}, looking for delayed photons, covers
this topology.  When a photon is emitted from a secondary vertex, it
reaches the electromagnetic calorimeters (ECAL) at a later time
compared to a photon produced at the primary vertex. The reason for
the delay is twofold, firstly, there is a difference in trajectory,
and secondly, the long-lived mediator traverses the detector at a
lower speed. Another distinctive feature of a delayed photon is that
it enters the ECAL at non-normal impact angles. These two features are
used to distinguish a photon originating from a secondary vertex from
one produced at the primary vertex.

The models that we  investigate in Section~\ref{sec:results} include this
signature but with lower rate compared to other final states, so we do
not expect that this could lead to relevant constraints on the
parameter space of the models considered.  Hence, we chose not to
reinterpret this search. Moreover, little information for recasting is provided 
in~\cite{Sirunyan:2019wau}, hindering a trustful reinterpretation.

\subsection{Other searches}
\label{sec:def-other}
For completeness, 
we mention here other $\sqrt{s} = 13$ TeV LHC searches that could constrain 
our models~\cite{Sirunyan:2020cao,Aaboud:2018aqj,Aaboud:2019opc,Sirunyan:2018pwn,CMS:2020tzl}.
They all focus on displaced signatures involving jets, employing different search strategies and selection criteria.
Since they do not target specifically models of DM with large missing transverse energy,
we do not expect that they can 
test the parameter space of our models better than the DV+MET and DJ+MET analyses that we have re-interpreted
in the main body of the paper.
A possible interesting exception is given by the ATLAS analysis~\cite{Aaboud:2018aqj} which searches for displaced jets in the muon spectrometer and hence could possibly provide a slightly better coverage of the region with large $c \tau$. At present such ATLAS search employs only 36 fb$^{-1}$, so we can expect only a moderate impact on our parameter space, but it will be interesting in the future to consider possible updated analyses employing a similar search strategy.

Another search that could constrain our models but does not focus on signatures involving jets is the ATLAS analysis~\cite{Aaboud:2018hdl} searching for tracks with a high ionization energy loss. This search typically targets models with heavy charged LLPs similarly to the searches for HSCPs and R-hadrons. However, this search does not require the track to traverse the detector completely, so that it targets values of the order of few meters. In principle, the search is sensitive to our leptophilic and topphilic models. However, we checked that in the leptophilic model, the mass range we considered (a few hundred GeV) is too low for the track of the mediator to be efficiently distinguished from a track of a high $p_T$ SM particle. In the topphilic model, on the other hand, the search is mostly sensitive to the same $c\tau$ range as the DJ search~\cite{Sirunyan:2019gut}, but
 we expect the latter to have a higher reach in the mediator mass. This can be seen by comparing the maximal reach of the two searches for
 the model that they both employ to interpret their results (long-lived gluino production).

%%%%%%%%%%%%%%%%%%%%% APP SINGLET-TRIPLET %%%%%%%%%%%%%%%%%%%%%
\section{Details about the singlet-triplet model}
\label{sec:13}
In Section~\ref{sec:non-renorm-oper}, we have studied the
singlet-triplet model, denoted by ${\cal F}_{W\chi}$, which is an 
extension of the SM featuring a singlet and a triplet fermion, $\chi_S$ and
$\chi_T$.  The most general Lagrangian one can write including
interaction terms of dimension $\leq 5$ reads 
\begin{align}
\mathcal{L}_{\rm ST} =&  \frac{1}{2} \bar{\chi}_S i \cancel{\partial} \chi_S + \frac{1}{2} \text{Tr} \left[ \bar{\chi}_T i \cancel{D} \chi_T \right] - \frac{m_S}{2} \bar{\chi}_S \chi_S - \frac{m_T}{2} \text{Tr}\left[\bar{\chi}_T \chi_T \right] \nonumber \\
& + \frac{\kappa_{ST}}{\Lambda} \left( (H^{\dagger} \bar{\chi}_T H) \chi_S + \text{h.c.} \right) - \frac{\kappa_S}{\Lambda} H^\dagger H \bar{\chi}_S \chi_S - \frac{\kappa_T}{\Lambda} H^\dagger H \text{Tr} \left[\bar{\chi}_T\chi_T\right]  \nonumber \\
& + \frac{\kappa'_T}{\Lambda} \left(H^\dagger \bar{\chi}_T\chi_T H + \text{Tr} \left[ \bar{\chi}_T H^\dagger H \chi_T \right]\right) + \frac{\kappa}{\Lambda} (W^a_{\mu \nu} \bar{\chi}_S \sigma^{\mu \nu} \chi_T^a + \text{h.c.}),
\label{eq:LagrST}
\end{align}
where $\kappa,\kappa_{ST},\kappa_S,\kappa_T$ and $\kappa'_T$ are
dimensionless coefficients and $\Lambda$ is a common UV physics scale. 
In Section~\ref{sec:non-renorm-oper}, we have effectively
assumed that $\kappa\gg\kappa_{ST},\kappa_S,\kappa_T,\kappa'_T$ so as
to focus on the cubic interaction of Figure~\ref{fig:FI}. 
A complementary analysis focusing on the
Higgs portal to DM for a large range of portal couplings can be found in~\cite{Filimonova:2018qdc}.

\renewcommand{\arraystretch}{1.2}
\begin{table}[t]
	\centering
	\begin{tabular}{ | c | c \mycol c | c | }
		\hline
		\multicolumn{2}{|c\mycol}{Initial state} & \multicolumn{2}{|c|}{Final state}  \\ 
		\hline \hline
		\multicolumn{2}{|c\mycol}{$\Psi_B^\pm$} & $\chi$ & $W^\pm$ \\
		\hline
		\multicolumn{2}{|c\mycol}{$\Psi_B^0$} & $\chi$ & $Z,\gamma$ \\
		\hline
		\hline
		\multirow{5}{*}{$\Psi_B^\pm$} & $Z,\gamma,H$ & \multirow{5}{*}{$\chi$} & $W^\pm$  \\
		\cline{2-2}\cline{4-4}
		& $W^\pm$ & & $Z,\gamma,H$ \\
		\cline{2-2}\cline{4-4}
		& $l^\mp$ & & $\nu_l$ \\
		\cline{2-2}\cline{4-4}
		& $\bar{\nu_l}$ & & $l^\pm$ \\
		\cline{2-2}\cline{4-4}
		& $q$ & & $q'$ \\
		\hline
		\multirow{6}{*}{$\Psi^0_B$} & $W^\pm$ & \multirow{6}{*}{$\chi$} & $W^\pm$  \\
		\cline{2-2}\cline{4-4}
		& $Z$ & & $H$ \\
		\cline{2-2}\cline{4-4}
		& $H$ & & $Z$ \\
		\cline{2-2}\cline{4-4}
		& $l^\pm$ & & $l^\pm$ \\
		\cline{2-2}\cline{4-4}
		& $\nu_l$ & & $\nu_l$ \\
		\cline{2-2}\cline{4-4}
		& $q$ & & $q$ \\
		\hline
		$\Psi_B^+$ & $\Psi_B^-$ & \multirow{2}{*}{$\chi$} & $\Psi_B^0$ \\
		\cline{1-2}\cline{4-4}
		$\Psi_B^\pm$ & $\Psi^0_B$ & & $\Psi_B^\pm$ \\
		\hline
	\end{tabular}
	\caption{Decay and scattering processes, Initial state $\to$ Final state, contributing to the freeze-in production of DM for the singlet-triplet model.}
	\label{tab:processes}
\end{table}

If we neglect the terms that involve the Higgs field, i.e.~setting $\kappa_{ST}=\kappa_S=\kappa_T=\kappa'_T=0$, 
no mixing occurs between $\chi_S$ and $\chi_T^0$ and we can assume that the singlet field is the DM particle while
the triplet plays the role of the freeze-in mediator, that is $\chi\equiv \chi_S$, $\Psi_B^{0,\pm}\equiv\chi_T^{0,\pm}$, as we did in Section~\ref{sec:non-renorm-oper}. A mass splitting between $\Psi_B^0$ and $ \Psi_B^\pm$ always arises due to quantum corrections. As a
result, the mass of the charged component can be written $m_C = m_T +
\Delta m$, with~\cite{Cirelli:2009uv}
\begin{align}
\Delta m = \frac{\alpha_2 m_T}{4 \pi} \left[(s_w^2-1) f\left(\frac{m_Z}{m_T}\right)+f\left(\frac{m_W}{m_T}\right) \right],
\end{align}
where
\begin{align}
f(r) = r\left[2 r^3 \ln r - 2r + \sqrt{r^2-4}(r^2+2) \ln \left(\frac{r^2-2-r\sqrt{r^2-4}}{2}\right)\right].
\end{align}
Hence the mass splitting between the neutral and the charged components of the triplet
depends on the leading order mass $m_T$. However, such a 
dependence is soft and for $m_T$ in the 100 GeV$-$1~TeV range, $\Delta m
\approx 160$~MeV. As a consequence of this small mass splitting, the charged
component of the triplet can possibly decay into the neutral component
and a soft pion, with a decay width of~\cite{Cirelli:2009uv}
\begin{equation}
\Gamma(\Psi_B^\pm \rightarrow \Psi_B^0 \pi^\pm) = \frac{2 G_F^2 f_\pi^2 \Delta m^3}{\pi} \sqrt{1- \frac{m_\pi^2}{\Delta m^2}}.
\end{equation}

The gauge interactions of the triplet and the cubic-interaction term proportional to $\kappa$\footnote{Notice that in the discussion of Section~\ref{sec:non-renorm-oper}, we take $\kappa=1$, that is, 
we reabsorb the coupling in the definition of the scale $\Lambda$.} 
in Eq.~(\ref{eq:LagrST}) can be expanded as follows:
\begin{align}
\mathcal{L}_{int} =& \frac{e}{4 s_w} [ \bar\Psi_B^0 \cancel{W}^+\Psi_B^- +  \bar{\Psi}_B^- \cancel{W}^-\Psi_B^0 - \bar{\Psi}_B^+ \cancel{W}^+ \Psi_B^0 - \bar\Psi_B^0  \cancel{W}^- \Psi_B^+ \nonumber \\ 
& \qquad +  \bar{\Psi}^+_B\cancel{W}^0 \Psi_B^+ - \bar{\Psi}_B^- \cancel{W}^0 \Psi_B^-  ] \nonumber \\
& + \frac{\kappa}{\Lambda} [W_{\mu\nu}^- \bar\chi \sigma^{\mu\nu} \Psi_B^+ + W_{\mu\nu}^+ \bar \chi \sigma^{\mu\nu} \Psi_B^- +  W_{\mu\nu}^0 \bar \chi \sigma^{\mu\nu} \Psi_B^0 + \text{h.c.}
],
\end{align}
where $W_{\mu\nu}^\pm = \partial_\nu W^\pm_\mu - \partial_\mu
W^\pm_\nu + g (W^0_\mu W_\nu^\pm - W_\mu^\pm W_\nu^0)$ and
$W_{\mu\nu}^0 = \partial_\nu W^0_\mu - \partial_\mu W^0_\nu + g
(W^+_\mu W_\nu^- - W_\mu^- W_\nu^+)$. This Lagrangian captures all
possible interactions between the DM, the mediator and other SM
particles which give rise to decay and scattering processes that could
lead to freeze-in production of DM. The associated processes are listed in Table~\ref{tab:processes}.

\newpage

\bibliographystyle{JHEPmod}
\bibliography{DisplacedFI}

\end{document}